\documentclass[twocolumn,traditabstract,longauth]{aa}
\usepackage{graphicx,amsmath}
\usepackage{epsf}
\usepackage[breaklinks, colorlinks, citecolor=blue]{hyperref}
\usepackage{natbib}
\usepackage{lscape}
\usepackage{epsf}
\usepackage{txfonts}
\usepackage{upgreek}
\usepackage{array}
\usepackage{textcomp}

\bibpunct{(}{)}{;}{a}{}{,} 

\def\setsymbol#1#2{\expandafter\def\csname #1\endcsname{#2}}
\def\getsymbol#1{\csname #1\endcsname}

\def\Planck{{\it Planck\/}}

\newbox\tablebox    \newdimen\tablewidth
\def\leaderfil{\leaders\hbox to 5pt{\hss.\hss}\hfil}
\def\tablenote#1 #2\par{\begingroup \parindent=0.8em
    \abovedisplayshortskip=0pt\belowdisplayshortskip=0pt
    \noindent
    $$\hss\vbox{\hsize\tablewidth \hangindent=\parindent \hangafter=1 \noindent
    \hbox to \parindent{\sup{\rm #1}\hss}\strut#2\strut\par}\hss$$
    \endgroup}

\def\L2{\ifmmode L_2\else $L_2$\fi}

\def\DeltaT{\ifmmode \Delta T\else $\Delta T$\fi}
\def\deltat{\ifmmode \Delta t\else $\Delta t$\fi}
\def\fknee{\ifmmode f_{\rm knee}\else $f_{\rm knee}$\fi}
\def\Fmax{\ifmmode F_{\rm max}\else $F_{\rm max}$\fi}
\def\solar{\ifmmode{\rm M}_{\mathord\odot}\else${\rm M}_{\mathord\odot}$\fi}

\def\inv{\ifmmode^{-1}\else$^{-1}$\fi}
\def\mo{\ifmmode^{-1}\else$^{-1}$\fi}
\def\sup#1{\ifmmode ^{\rm #1}\else $^{\rm #1}$\fi}
\def\expo#1{\ifmmode \times 10^{#1}\else $\times 10^{#1}$\fi}
\def\,{\thinspace}
\def\lsim{\mathrel{\raise .4ex\hbox{\rlap{$<$}\lower 1.2ex\hbox{$\sim$}}}}
\def\gsim{\mathrel{\raise .4ex\hbox{\rlap{$>$}\lower 1.2ex\hbox{$\sim$}}}}

\def\simprop{\mathrel{\raise .4ex\hbox{\rlap{$\propto$}\lower 1.2ex\hbox{$\sim$}}}}
\def\deg{\ifmmode^\circ\else$^\circ$\fi}
\def\pdeg{\ifmmode $\setbox0=\hbox{$^{\circ}$}\rlap{\hskip.11\wd0 .}$^{\circ}
          \else \setbox0=\hbox{$^{\circ}$}\rlap{\hskip.11\wd0 .}$^{\circ}$\fi}
\def\arcs{\ifmmode {^{\scriptstyle\prime\prime}}
          \else $^{\scriptstyle\prime\prime}$\fi}
\def\arcm{\ifmmode {^{\scriptstyle\prime}}
          \else $^{\scriptstyle\prime}$\fi}
\newdimen\sa  \newdimen\sb
\def\parcs{\sa=.07em \sb=.03em
     \ifmmode \hbox{\rlap{.}}^{\scriptstyle\prime\kern -\sb\prime}\hbox{\kern -\sa}
     \else \rlap{.}$^{\scriptstyle\prime\kern -\sb\prime}$\kern -\sa\fi}
\def\parcm{\sa=.08em \sb=.03em
     \ifmmode \hbox{\rlap{.}\kern\sa}^{\scriptstyle\prime}\hbox{\kern-\sb}
     \else \rlap{.}\kern\sa$^{\scriptstyle\prime}$\kern-\sb\fi}
\def\ra[#1 #2 #3.#4]{#1\sup{h}#2\sup{m}#3\sup{s}\llap.#4}
\def\dec[#1 #2 #3.#4]{#1\deg#2\arcm#3\arcs\llap.#4}
\def\deco[#1 #2 #3]{#1\deg#2\arcm#3\arcs}
\def\rra[#1 #2]{#1\sup{h}#2\sup{m}}

\def\dots{\relax\ifmmode \ldots\else $\ldots$\fi}
\def\WHzsr{\ifmmode $W\,Hz\mo\,sr\mo$\else W\,Hz\mo\,sr\mo\fi}
\def\mHz{\ifmmode $\,mHz$\else \,mHz\fi}
\def\GHz{\ifmmode $\,GHz$\else \,GHz\fi}
\def\mKs{\ifmmode $\,mK\,s$^{1/2}\else \,mK\,s$^{1/2}$\fi}
\def\muKs{\ifmmode \,\mu$K\,s$^{1/2}\else \,$\mu$K\,s$^{1/2}$\fi}
\def\muKRJs{\ifmmode \,\mu$K$_{\rm RJ}$\,s$^{1/2}\else \,$\mu$K$_{\rm RJ}$\,s$^{1/2}$\fi}
\def\muKHz{\ifmmode \,\mu$K\,Hz$^{-1/2}\else \,$\mu$K\,Hz$^{-1/2}$\fi}
\def\MJysr{\ifmmode \,$MJy\,sr\mo$\else \,MJy\,sr\mo\fi}
\def\MJysrmK{\ifmmode \,$MJy\,sr\mo$\,mK$_{\rm CMB}\mo\else \,MJy\,sr\mo\,mK$_{\rm CMB}\mo$\fi}
\def\microns{\ifmmode \,\mu$m$\else \,$\mu$m\fi}

\def\muK{\ifmmode \,\mu$K$\else \,$\mu$\hbox{K}\fi}
\def\microK{\ifmmode \,\mu$K$\else \,$\mu$\hbox{K}\fi}
\def\muW{\ifmmode \,\mu$W$\else \,$\mu$\hbox{W}\fi}
\def\kms{\ifmmode $\,km\,s$^{-1}\else \,km\,s$^{-1}$\fi}
\def\kmsMpc{\ifmmode $\,\kms\,Mpc\mo$\else \,\kms\,Mpc\mo\fi}
\setsymbol{LFI:center:frequency:70GHz:units}{70.3\,GHz}
\setsymbol{LFI:center:frequency:44GHz:units}{44.1\,GHz}
\setsymbol{LFI:center:frequency:30GHz:units}{28.5\,GHz}

\setsymbol{LFI:center:frequency:70GHz}{70.3}
\setsymbol{LFI:center:frequency:44GHz}{44.1}
\setsymbol{LFI:center:frequency:30GHz}{28.5}

\setsymbol{LFI:center:frequency:LFI18:Rad:M:units}{71.7\GHz}
\setsymbol{LFI:center:frequency:LFI19:Rad:M:units}{67.5\GHz}
\setsymbol{LFI:center:frequency:LFI20:Rad:M:units}{69.2\GHz}
\setsymbol{LFI:center:frequency:LFI21:Rad:M:units}{70.4\GHz}
\setsymbol{LFI:center:frequency:LFI22:Rad:M:units}{71.5\GHz}
\setsymbol{LFI:center:frequency:LFI23:Rad:M:units}{70.8\GHz}
\setsymbol{LFI:center:frequency:LFI24:Rad:M:units}{44.4\GHz}
\setsymbol{LFI:center:frequency:LFI25:Rad:M:units}{44.0\GHz}
\setsymbol{LFI:center:frequency:LFI26:Rad:M:units}{43.9\GHz}
\setsymbol{LFI:center:frequency:LFI27:Rad:M:units}{28.3\GHz}
\setsymbol{LFI:center:frequency:LFI28:Rad:M:units}{28.8\GHz}
\setsymbol{LFI:center:frequency:LFI18:Rad:S:units}{70.1\GHz}
\setsymbol{LFI:center:frequency:LFI19:Rad:S:units}{69.6\GHz}
\setsymbol{LFI:center:frequency:LFI20:Rad:S:units}{69.5\GHz}
\setsymbol{LFI:center:frequency:LFI21:Rad:S:units}{69.5\GHz}
\setsymbol{LFI:center:frequency:LFI22:Rad:S:units}{72.8\GHz}
\setsymbol{LFI:center:frequency:LFI23:Rad:S:units}{71.3\GHz}
\setsymbol{LFI:center:frequency:LFI24:Rad:S:units}{44.1\GHz}
\setsymbol{LFI:center:frequency:LFI25:Rad:S:units}{44.1\GHz}
\setsymbol{LFI:center:frequency:LFI26:Rad:S:units}{44.1\GHz}
\setsymbol{LFI:center:frequency:LFI27:Rad:S:units}{28.5\GHz}
\setsymbol{LFI:center:frequency:LFI28:Rad:S:units}{28.2\GHz}

\setsymbol{LFI:center:frequency:LFI18:Rad:M}{71.7}
\setsymbol{LFI:center:frequency:LFI19:Rad:M}{67.5}
\setsymbol{LFI:center:frequency:LFI20:Rad:M}{69.2}
\setsymbol{LFI:center:frequency:LFI21:Rad:M}{70.4}
\setsymbol{LFI:center:frequency:LFI22:Rad:M}{71.5}
\setsymbol{LFI:center:frequency:LFI23:Rad:M}{70.8}
\setsymbol{LFI:center:frequency:LFI24:Rad:M}{44.4}
\setsymbol{LFI:center:frequency:LFI25:Rad:M}{44.0}
\setsymbol{LFI:center:frequency:LFI26:Rad:M}{43.9}
\setsymbol{LFI:center:frequency:LFI27:Rad:M}{28.3}
\setsymbol{LFI:center:frequency:LFI28:Rad:M}{28.8}
\setsymbol{LFI:center:frequency:LFI18:Rad:S}{70.1}
\setsymbol{LFI:center:frequency:LFI19:Rad:S}{69.6}
\setsymbol{LFI:center:frequency:LFI20:Rad:S}{69.5}
\setsymbol{LFI:center:frequency:LFI21:Rad:S}{69.5}
\setsymbol{LFI:center:frequency:LFI22:Rad:S}{72.8}
\setsymbol{LFI:center:frequency:LFI23:Rad:S}{71.3}
\setsymbol{LFI:center:frequency:LFI24:Rad:S}{44.1}
\setsymbol{LFI:center:frequency:LFI25:Rad:S}{44.1}
\setsymbol{LFI:center:frequency:LFI26:Rad:S}{44.1}
\setsymbol{LFI:center:frequency:LFI27:Rad:S}{28.5}
\setsymbol{LFI:center:frequency:LFI28:Rad:S}{28.2}

\setsymbol{LFI:white:noise:sensitivity:70GHz:units}{152.6\muKs}
\setsymbol{LFI:white:noise:sensitivity:44GHz:units}{173.1\muKs}
\setsymbol{LFI:white:noise:sensitivity:30GHz:units}{146.8\muKs}

\setsymbol{LFI:white:noise:sensitivity:70GHz}{152.6}
\setsymbol{LFI:white:noise:sensitivity:44GHz}{173.1}
\setsymbol{LFI:white:noise:sensitivity:30GHz}{146.8}

\setsymbol{LFI:white:noise:sensitivity:LFI18:Rad:M:units}{512.0\muKs}
\setsymbol{LFI:white:noise:sensitivity:LFI19:Rad:M:units}{581.4\muKs}
\setsymbol{LFI:white:noise:sensitivity:LFI20:Rad:M:units}{590.8\muKs}
\setsymbol{LFI:white:noise:sensitivity:LFI21:Rad:M:units}{455.2\muKs}
\setsymbol{LFI:white:noise:sensitivity:LFI22:Rad:M:units}{492.0\muKs}
\setsymbol{LFI:white:noise:sensitivity:LFI23:Rad:M:units}{507.7\muKs}
\setsymbol{LFI:white:noise:sensitivity:LFI24:Rad:M:units}{462.2\muKs}
\setsymbol{LFI:white:noise:sensitivity:LFI25:Rad:M:units}{413.6\muKs}
\setsymbol{LFI:white:noise:sensitivity:LFI26:Rad:M:units}{478.6\muKs}
\setsymbol{LFI:white:noise:sensitivity:LFI27:Rad:M:units}{277.7\muKs}
\setsymbol{LFI:white:noise:sensitivity:LFI28:Rad:M:units}{312.3\muKs}
\setsymbol{LFI:white:noise:sensitivity:LFI18:Rad:S:units}{465.7\muKs}
\setsymbol{LFI:white:noise:sensitivity:LFI19:Rad:S:units}{555.6\muKs}
\setsymbol{LFI:white:noise:sensitivity:LFI20:Rad:S:units}{623.2\muKs}
\setsymbol{LFI:white:noise:sensitivity:LFI21:Rad:S:units}{564.1\muKs}
\setsymbol{LFI:white:noise:sensitivity:LFI22:Rad:S:units}{534.4\muKs}
\setsymbol{LFI:white:noise:sensitivity:LFI23:Rad:S:units}{542.4\muKs}
\setsymbol{LFI:white:noise:sensitivity:LFI24:Rad:S:units}{399.2\muKs}
\setsymbol{LFI:white:noise:sensitivity:LFI25:Rad:S:units}{392.6\muKs}
\setsymbol{LFI:white:noise:sensitivity:LFI26:Rad:S:units}{418.6\muKs}
\setsymbol{LFI:white:noise:sensitivity:LFI27:Rad:S:units}{302.9\muKs}
\setsymbol{LFI:white:noise:sensitivity:LFI28:Rad:S:units}{285.3\muKs}

\setsymbol{LFI:white:noise:sensitivity:LFI18:Rad:M}{512.0}
\setsymbol{LFI:white:noise:sensitivity:LFI19:Rad:M}{581.4}
\setsymbol{LFI:white:noise:sensitivity:LFI20:Rad:M}{590.8}
\setsymbol{LFI:white:noise:sensitivity:LFI21:Rad:M}{455.2}
\setsymbol{LFI:white:noise:sensitivity:LFI22:Rad:M}{492.0}
\setsymbol{LFI:white:noise:sensitivity:LFI23:Rad:M}{507.7}
\setsymbol{LFI:white:noise:sensitivity:LFI24:Rad:M}{462.2}
\setsymbol{LFI:white:noise:sensitivity:LFI25:Rad:M}{413.6}
\setsymbol{LFI:white:noise:sensitivity:LFI26:Rad:M}{478.6}
\setsymbol{LFI:white:noise:sensitivity:LFI27:Rad:M}{277.7}
\setsymbol{LFI:white:noise:sensitivity:LFI28:Rad:M}{312.3}
\setsymbol{LFI:white:noise:sensitivity:LFI18:Rad:S}{465.7}
\setsymbol{LFI:white:noise:sensitivity:LFI19:Rad:S}{555.6}
\setsymbol{LFI:white:noise:sensitivity:LFI20:Rad:S}{623.2}
\setsymbol{LFI:white:noise:sensitivity:LFI21:Rad:S}{564.1}
\setsymbol{LFI:white:noise:sensitivity:LFI22:Rad:S}{534.4}
\setsymbol{LFI:white:noise:sensitivity:LFI23:Rad:S}{542.4}
\setsymbol{LFI:white:noise:sensitivity:LFI24:Rad:S}{399.2}
\setsymbol{LFI:white:noise:sensitivity:LFI25:Rad:S}{392.6}
\setsymbol{LFI:white:noise:sensitivity:LFI26:Rad:S}{418.6}
\setsymbol{LFI:white:noise:sensitivity:LFI27:Rad:S}{302.9}
\setsymbol{LFI:white:noise:sensitivity:LFI28:Rad:S}{285.3}

\setsymbol{LFI:knee:frequency:70GHz:units}{29.5\mHz}
\setsymbol{LFI:knee:frequency:44GHz:units}{56.2\mHz}
\setsymbol{LFI:knee:frequency:30GHz:units}{113.7\mHz}

\setsymbol{LFI:knee:frequency:70GHz}{29.5}
\setsymbol{LFI:knee:frequency:44GHz}{56.2}
\setsymbol{LFI:knee:frequency:30GHz}{113.7}

\setsymbol{LFI:knee:frequency:LFI18:Rad:M:units}{16.3\mHz}
\setsymbol{LFI:knee:frequency:LFI19:Rad:M:units}{15.1\mHz}
\setsymbol{LFI:knee:frequency:LFI20:Rad:M:units}{18.7\mHz}
\setsymbol{LFI:knee:frequency:LFI21:Rad:M:units}{37.2\mHz}
\setsymbol{LFI:knee:frequency:LFI22:Rad:M:units}{12.7\mHz}
\setsymbol{LFI:knee:frequency:LFI23:Rad:M:units}{34.6\mHz}
\setsymbol{LFI:knee:frequency:LFI24:Rad:M:units}{46.2\mHz}
\setsymbol{LFI:knee:frequency:LFI25:Rad:M:units}{24.9\mHz}
\setsymbol{LFI:knee:frequency:LFI26:Rad:M:units}{67.6\mHz}
\setsymbol{LFI:knee:frequency:LFI27:Rad:M:units}{187.4\mHz}
\setsymbol{LFI:knee:frequency:LFI28:Rad:M:units}{122.2\mHz}
\setsymbol{LFI:knee:frequency:LFI18:Rad:S:units}{17.7\mHz}
\setsymbol{LFI:knee:frequency:LFI19:Rad:S:units}{22.0\mHz}
\setsymbol{LFI:knee:frequency:LFI20:Rad:S:units}{8.7\mHz}
\setsymbol{LFI:knee:frequency:LFI21:Rad:S:units}{25.9\mHz}
\setsymbol{LFI:knee:frequency:LFI22:Rad:S:units}{15.8\mHz}
\setsymbol{LFI:knee:frequency:LFI23:Rad:S:units}{129.8\mHz}
\setsymbol{LFI:knee:frequency:LFI24:Rad:S:units}{100.9\mHz}
\setsymbol{LFI:knee:frequency:LFI25:Rad:S:units}{38.9\mHz}
\setsymbol{LFI:knee:frequency:LFI26:Rad:S:units}{58.9\mHz}
\setsymbol{LFI:knee:frequency:LFI27:Rad:S:units}{104.4\mHz}
\setsymbol{LFI:knee:frequency:LFI28:Rad:S:units}{40.7\mHz}

\setsymbol{LFI:knee:frequency:LFI18:Rad:M}{16.3}
\setsymbol{LFI:knee:frequency:LFI19:Rad:M}{15.1}
\setsymbol{LFI:knee:frequency:LFI20:Rad:M}{18.7}
\setsymbol{LFI:knee:frequency:LFI21:Rad:M}{37.2}
\setsymbol{LFI:knee:frequency:LFI22:Rad:M}{12.7}
\setsymbol{LFI:knee:frequency:LFI23:Rad:M}{34.6}
\setsymbol{LFI:knee:frequency:LFI24:Rad:M}{46.2}
\setsymbol{LFI:knee:frequency:LFI25:Rad:M}{24.9}
\setsymbol{LFI:knee:frequency:LFI26:Rad:M}{67.6}
\setsymbol{LFI:knee:frequency:LFI27:Rad:M}{187.4}
\setsymbol{LFI:knee:frequency:LFI28:Rad:M}{122.2}
\setsymbol{LFI:knee:frequency:LFI18:Rad:S}{17.7}
\setsymbol{LFI:knee:frequency:LFI19:Rad:S}{22.0}
\setsymbol{LFI:knee:frequency:LFI20:Rad:S}{8.7}
\setsymbol{LFI:knee:frequency:LFI21:Rad:S}{25.9}
\setsymbol{LFI:knee:frequency:LFI22:Rad:S}{15.8}
\setsymbol{LFI:knee:frequency:LFI23:Rad:S}{129.8}
\setsymbol{LFI:knee:frequency:LFI24:Rad:S}{100.9}
\setsymbol{LFI:knee:frequency:LFI25:Rad:S}{38.9}
\setsymbol{LFI:knee:frequency:LFI26:Rad:S}{58.9}
\setsymbol{LFI:knee:frequency:LFI27:Rad:S}{104.4}
\setsymbol{LFI:knee:frequency:LFI28:Rad:S}{40.7}

\setsymbol{LFI:slope:70GHz:units}{$-1.03$\mHz}
\setsymbol{LFI:slope:44GHz:units}{$-0.89$\mHz}
\setsymbol{LFI:slope:30GHz:units}{$-0.87$\mHz}

\setsymbol{LFI:slope:70GHz}{$-1.03$}
\setsymbol{LFI:slope:44GHz}{$-0.89$}
\setsymbol{LFI:slope:30GHz}{$-0.87$}

\setsymbol{LFI:slope:LFI18:Rad:M:units}{$-1.04$\mHz}
\setsymbol{LFI:slope:LFI19:Rad:M:units}{$-1.09$\mHz}
\setsymbol{LFI:slope:LFI20:Rad:M:units}{$-0.69$\mHz}
\setsymbol{LFI:slope:LFI21:Rad:M:units}{$-1.56$\mHz}
\setsymbol{LFI:slope:LFI22:Rad:M:units}{$-1.01$\mHz}
\setsymbol{LFI:slope:LFI23:Rad:M:units}{$-0.96$\mHz}
\setsymbol{LFI:slope:LFI24:Rad:M:units}{$-0.83$\mHz}
\setsymbol{LFI:slope:LFI25:Rad:M:units}{$-0.91$\mHz}
\setsymbol{LFI:slope:LFI26:Rad:M:units}{$-0.95$\mHz}
\setsymbol{LFI:slope:LFI27:Rad:M:units}{$-0.87$\mHz}
\setsymbol{LFI:slope:LFI28:Rad:M:units}{$-0.88$\mHz}
\setsymbol{LFI:slope:LFI18:Rad:S:units}{$-1.15$\mHz}
\setsymbol{LFI:slope:LFI19:Rad:S:units}{$-1.00$\mHz}
\setsymbol{LFI:slope:LFI20:Rad:S:units}{$-0.95$\mHz}
\setsymbol{LFI:slope:LFI21:Rad:S:units}{$-0.92$\mHz}
\setsymbol{LFI:slope:LFI22:Rad:S:units}{$-1.01$\mHz}
\setsymbol{LFI:slope:LFI23:Rad:S:units}{$-0.95$\mHz}
\setsymbol{LFI:slope:LFI24:Rad:S:units}{$-0.73$\mHz}
\setsymbol{LFI:slope:LFI25:Rad:S:units}{$-1.16$\mHz}
\setsymbol{LFI:slope:LFI26:Rad:S:units}{$-0.79$\mHz}
\setsymbol{LFI:slope:LFI27:Rad:S:units}{$-0.82$\mHz}
\setsymbol{LFI:slope:LFI28:Rad:S:units}{$-0.91$\mHz}

\setsymbol{LFI:slope:LFI18:Rad:M}{$-1.04$}
\setsymbol{LFI:slope:LFI19:Rad:M}{$-1.09$}
\setsymbol{LFI:slope:LFI20:Rad:M}{$-0.69$}
\setsymbol{LFI:slope:LFI21:Rad:M}{$-1.56$}
\setsymbol{LFI:slope:LFI22:Rad:M}{$-1.01$}
\setsymbol{LFI:slope:LFI23:Rad:M}{$-0.96$}
\setsymbol{LFI:slope:LFI24:Rad:M}{$-0.83$}
\setsymbol{LFI:slope:LFI25:Rad:M}{$-0.91$}
\setsymbol{LFI:slope:LFI26:Rad:M}{$-0.95$}
\setsymbol{LFI:slope:LFI27:Rad:M}{$-0.87$}
\setsymbol{LFI:slope:LFI28:Rad:M}{$-0.88$}
\setsymbol{LFI:slope:LFI18:Rad:S}{$-1.15$}
\setsymbol{LFI:slope:LFI19:Rad:S}{$-1.00$}
\setsymbol{LFI:slope:LFI20:Rad:S}{$-0.95$}
\setsymbol{LFI:slope:LFI21:Rad:S}{$-0.92$}
\setsymbol{LFI:slope:LFI22:Rad:S}{$-1.01$}
\setsymbol{LFI:slope:LFI23:Rad:S}{$-0.95$}
\setsymbol{LFI:slope:LFI24:Rad:S}{$-0.73$}
\setsymbol{LFI:slope:LFI25:Rad:S}{$-1.16$}
\setsymbol{LFI:slope:LFI26:Rad:S}{$-0.79$}
\setsymbol{LFI:slope:LFI27:Rad:S}{$-0.82$}
\setsymbol{LFI:slope:LFI28:Rad:S}{$-0.91$}

\setsymbol{LFI:FWHM:70GHz:units}{13\parcm01}
\setsymbol{LFI:FWHM:44GHz:units}{27\parcm92}
\setsymbol{LFI:FWHM:30GHz:units}{32\parcm65}

\setsymbol{LFI:FWHM:70GHz}{13.01}
\setsymbol{LFI:FWHM:44GHz}{27.92}
\setsymbol{LFI:FWHM:30GHz}{32.65}

\setsymbol{LFI:FWHM:LFI18:units}{13\parcm39}
\setsymbol{LFI:FWHM:LFI19:units}{13\parcm01}
\setsymbol{LFI:FWHM:LFI20:units}{12\parcm75}
\setsymbol{LFI:FWHM:LFI21:units}{12\parcm74}
\setsymbol{LFI:FWHM:LFI22:units}{12\parcm87}
\setsymbol{LFI:FWHM:LFI23:units}{13\parcm27}
\setsymbol{LFI:FWHM:LFI24:units}{22\parcm98}
\setsymbol{LFI:FWHM:LFI25:units}{30\parcm46}
\setsymbol{LFI:FWHM:LFI26:units}{30\parcm31}
\setsymbol{LFI:FWHM:LFI27:units}{32\parcm65}
\setsymbol{LFI:FWHM:LFI28:units}{32\parcm66}

\setsymbol{LFI:FWHM:LFI18}{13.39}
\setsymbol{LFI:FWHM:LFI19}{13.01}
\setsymbol{LFI:FWHM:LFI20}{12.75}
\setsymbol{LFI:FWHM:LFI21}{12.74}
\setsymbol{LFI:FWHM:LFI22}{12.87}
\setsymbol{LFI:FWHM:LFI23}{13.27}
\setsymbol{LFI:FWHM:LFI24}{22.98}
\setsymbol{LFI:FWHM:LFI25}{30.46}
\setsymbol{LFI:FWHM:LFI26}{30.31}
\setsymbol{LFI:FWHM:LFI27}{32.65}
\setsymbol{LFI:FWHM:LFI28}{32.66}

\setsymbol{LFI:FWHM:uncertainty:LFI18:units}{0.170\arcm}
\setsymbol{LFI:FWHM:uncertainty:LFI19:units}{0.174\arcm}
\setsymbol{LFI:FWHM:uncertainty:LFI20:units}{0.170\arcm}
\setsymbol{LFI:FWHM:uncertainty:LFI21:units}{0.156\arcm}
\setsymbol{LFI:FWHM:uncertainty:LFI22:units}{0.164\arcm}
\setsymbol{LFI:FWHM:uncertainty:LFI23:units}{0.171\arcm}
\setsymbol{LFI:FWHM:uncertainty:LFI24:units}{0.652\arcm}
\setsymbol{LFI:FWHM:uncertainty:LFI25:units}{1.075\arcm}
\setsymbol{LFI:FWHM:uncertainty:LFI26:units}{1.131\arcm}
\setsymbol{LFI:FWHM:uncertainty:LFI27:units}{1.266\arcm}
\setsymbol{LFI:FWHM:uncertainty:LFI28:units}{1.287\arcm}

\setsymbol{LFI:FWHM:uncertainty:LFI18}{0.170}
\setsymbol{LFI:FWHM:uncertainty:LFI19}{0.174}
\setsymbol{LFI:FWHM:uncertainty:LFI20}{0.170}
\setsymbol{LFI:FWHM:uncertainty:LFI21}{0.156}
\setsymbol{LFI:FWHM:uncertainty:LFI22}{0.164}
\setsymbol{LFI:FWHM:uncertainty:LFI23}{0.171}
\setsymbol{LFI:FWHM:uncertainty:LFI24}{0.652}
\setsymbol{LFI:FWHM:uncertainty:LFI25}{1.075}
\setsymbol{LFI:FWHM:uncertainty:LFI26}{1.131}
\setsymbol{LFI:FWHM:uncertainty:LFI27}{1.266}
\setsymbol{LFI:FWHM:uncertainty:LFI28}{1.287}

\setsymbol{HFI:center:frequency:100GHz:units}{100\,GHz}
\setsymbol{HFI:center:frequency:143GHz:units}{143\,GHz}
\setsymbol{HFI:center:frequency:217GHz:units}{217\,GHz}
\setsymbol{HFI:center:frequency:353GHz:units}{353\,GHz}
\setsymbol{HFI:center:frequency:545GHz:units}{545\,GHz}
\setsymbol{HFI:center:frequency:857GHz:units}{857\,GHz}

\setsymbol{HFI:center:frequency:100GHz}{100}
\setsymbol{HFI:center:frequency:143GHz}{143}
\setsymbol{HFI:center:frequency:217GHz}{217}
\setsymbol{HFI:center:frequency:353GHz}{353}
\setsymbol{HFI:center:frequency:545GHz}{545}
\setsymbol{HFI:center:frequency:857GHz}{857}

\setsymbol{HFI:Ndetectors:100GHz}{8}
\setsymbol{HFI:Ndetectors:143GHz}{11}
\setsymbol{HFI:Ndetectors:217GHz}{12}
\setsymbol{HFI:Ndetectors:353GHz}{12}
\setsymbol{HFI:Ndetectors:545GHz}{3}
\setsymbol{HFI:Ndetectors:857GHz}{4}

\setsymbol{HFI:FWHM:Maps:100GHz:units}{9\parcm88}
\setsymbol{HFI:FWHM:Maps:143GHz:units}{7\parcm18}
\setsymbol{HFI:FWHM:Maps:217GHz:units}{4\parcm87}
\setsymbol{HFI:FWHM:Maps:353GHz:units}{4\parcm65}
\setsymbol{HFI:FWHM:Maps:545GHz:units}{4\parcm72}
\setsymbol{HFI:FWHM:Maps:857GHz:units}{4\parcm39}
\setsymbol{HFI:FWHM:Maps:100GHz}{9.88}
\setsymbol{HFI:FWHM:Maps:143GHz}{7.18}
\setsymbol{HFI:FWHM:Maps:217GHz}{4.87}
\setsymbol{HFI:FWHM:Maps:353GHz}{4.65}
\setsymbol{HFI:FWHM:Maps:545GHz}{4.72}
\setsymbol{HFI:FWHM:Maps:857GHz}{4.39}

\setsymbol{HFI:beam:ellipticity:Maps:100GHz}{1.15}
\setsymbol{HFI:beam:ellipticity:Maps:143GHz}{1.01}
\setsymbol{HFI:beam:ellipticity:Maps:217GHz}{1.06}
\setsymbol{HFI:beam:ellipticity:Maps:353GHz}{1.05}
\setsymbol{HFI:beam:ellipticity:Maps:545GHz}{1.14}
\setsymbol{HFI:beam:ellipticity:Maps:857GHz}{1.19}

\setsymbol{HFI:FWHM:Mars:100GHz:units}{9\parcm37}
\setsymbol{HFI:FWHM:Mars:143GHz:units}{7\parcm04}
\setsymbol{HFI:FWHM:Mars:217GHz:units}{4\parcm68}
\setsymbol{HFI:FWHM:Mars:353GHz:units}{4\parcm43}
\setsymbol{HFI:FWHM:Mars:545GHz:units}{3\parcm80}
\setsymbol{HFI:FWHM:Mars:857GHz:units}{3\parcm67}

\setsymbol{HFI:FWHM:Mars:100GHz}{9.37}
\setsymbol{HFI:FWHM:Mars:143GHz}{7.04}
\setsymbol{HFI:FWHM:Mars:217GHz}{4.68}
\setsymbol{HFI:FWHM:Mars:353GHz}{4.43}
\setsymbol{HFI:FWHM:Mars:545GHz}{3.80}
\setsymbol{HFI:FWHM:Mars:857GHz}{3.67}

\setsymbol{HFI:beam:ellipticity:Mars:100GHz}{1.18}
\setsymbol{HFI:beam:ellipticity:Mars:143GHz}{1.03}
\setsymbol{HFI:beam:ellipticity:Mars:217GHz}{1.14}
\setsymbol{HFI:beam:ellipticity:Mars:353GHz}{1.09}
\setsymbol{HFI:beam:ellipticity:Mars:545GHz}{1.25}
\setsymbol{HFI:beam:ellipticity:Mars:857GHz}{1.03}

\setsymbol{HFI:CMB:relative:calibration:100GHz}{$\lsim 1\%$}
\setsymbol{HFI:CMB:relative:calibration:143GHz}{$\lsim 1\%$}
\setsymbol{HFI:CMB:relative:calibration:217GHz}{$\lsim 1\%$}
\setsymbol{HFI:CMB:relative:calibration:353GHz}{$\lsim 1\%$}
\setsymbol{HFI:CMB:relative:calibration:545GHz}{}
\setsymbol{HFI:CMB:relative:calibration:857GHz}{}

\setsymbol{HFI:CMB:absolute:calibration:100GHz}{$\lsim 2\%$}
\setsymbol{HFI:CMB:absolute:calibration:143GHz}{$\lsim 2\%$}
\setsymbol{HFI:CMB:absolute:calibration:217GHz}{$\lsim 2\%$}
\setsymbol{HFI:CMB:absolute:calibration:353GHz}{$\lsim 2\%$}
\setsymbol{HFI:CMB:absolute:calibration:545GHz}{}
\setsymbol{HFI:CMB:absolute:calibration:857GHz}{}

\setsymbol{HFI:FIRAS:gain:calibration:accuracy:statistical:100GHz}{}
\setsymbol{HFI:FIRAS:gain:calibration:accuracy:statistical:143GHz}{}
\setsymbol{HFI:FIRAS:gain:calibration:accuracy:statistical:217GHz}{}
\setsymbol{HFI:FIRAS:gain:calibration:accuracy:statistical:353GHz}{2.5\%}
\setsymbol{HFI:FIRAS:gain:calibration:accuracy:statistical:545GHz}{1\%}
\setsymbol{HFI:FIRAS:gain:calibration:accuracy:statistical:857GHz}{0.5\%}

\setsymbol{HFI:FIRAS:gain:calibration:accuracy:systematic:100GHz}{}
\setsymbol{HFI:FIRAS:gain:calibration:accuracy:systematic:143GHz}{}
\setsymbol{HFI:FIRAS:gain:calibration:accuracy:systematic:217GHz}{}
\setsymbol{HFI:FIRAS:gain:calibration:accuracy:systematic:353GHz}{}
\setsymbol{HFI:FIRAS:gain:calibration:accuracy:systematic:545GHz}{7\%}
\setsymbol{HFI:FIRAS:gain:calibration:accuracy:systematic:857GHz}{7\%}

\setsymbol{HFI:FIRAS:zero:point:accuracy:100GHz:units}{0.8\MJysr}
\setsymbol{HFI:FIRAS:zero:point:accuracy:143GHz:units}{}
\setsymbol{HFI:FIRAS:zero:point:accuracy:217GHz:units}{}
\setsymbol{HFI:FIRAS:zero:point:accuracy:353GHz:units}{1.4\MJysr}
\setsymbol{HFI:FIRAS:zero:point:accuracy:545GHz:units}{2.2\MJysr}
\setsymbol{HFI:FIRAS:zero:point:accuracy:857GHz:units}{1.7\MJysr}

\setsymbol{HFI:FIRAS:zero:point:accuracy:100GHz}{0.8}
\setsymbol{HFI:FIRAS:zero:point:accuracy:143GHz}{}
\setsymbol{HFI:FIRAS:zero:point:accuracy:217GHz}{}
\setsymbol{HFI:FIRAS:zero:point:accuracy:353GHz}{1.4}
\setsymbol{HFI:FIRAS:zero:point:accuracy:545GHz}{2.2}
\setsymbol{HFI:FIRAS:zero:point:accuracy:857GHz}{1.7}

\setsymbol{HFI:unit:conversion:100GHz:units}{0.2415\MJysrmK}
\setsymbol{HFI:unit:conversion:143GHz:units}{0.3694\MJysrmK}
\setsymbol{HFI:unit:conversion:217GHz:units}{0.4811\MJysrmK}
\setsymbol{HFI:unit:conversion:353GHz:units}{0.2883\MJysrmK}
\setsymbol{HFI:unit:conversion:545GHz:units}{0.05826\MJysrmK}
\setsymbol{HFI:unit:conversion:857GHz:units}{0.002238\MJysrmK}

\setsymbol{HFI:unit:conversion:100GHz}{0.2415}
\setsymbol{HFI:unit:conversion:143GHz}{0.3694}
\setsymbol{HFI:unit:conversion:217GHz}{0.4811}
\setsymbol{HFI:unit:conversion:353GHz}{0.2883}
\setsymbol{HFI:unit:conversion:545GHz}{0.05826}
\setsymbol{HFI:unit:conversion:857GHz}{0.002238}

\setsymbol{HFI:colour:correction:alpha=-2:V1.01:100GHz}{0.9893}
\setsymbol{HFI:colour:correction:alpha=-2:V1.01:143GHz}{0.9759}
\setsymbol{HFI:colour:correction:alpha=-2:V1.01:217GHz}{1.0007}
\setsymbol{HFI:colour:correction:alpha=-2:V1.01:353GHz}{1.0028}
\setsymbol{HFI:colour:correction:alpha=-2:V1.01:545GHz}{1.0019}
\setsymbol{HFI:colour:correction:alpha=-2:V1.01:857GHz}{0.9889}

\setsymbol{HFI:colour:correction:alpha=0:V1.01:100GHz}{1.0008}
\setsymbol{HFI:colour:correction:alpha=0:V1.01:143GHz}{1.0148}
\setsymbol{HFI:colour:correction:alpha=0:V1.01:217GHz}{0.9909}
\setsymbol{HFI:colour:correction:alpha=0:V1.01:353GHz}{0.9888}
\setsymbol{HFI:colour:correction:alpha=0:V1.01:545GHz}{0.9878}
\setsymbol{HFI:colour:correction:alpha=0:V1.01:857GHz}{1.0014}

\def\setsymbol#1#2{\expandafter\def\csname #1\endcsname{#2}}
\def\getsymbol#1{\csname #1\endcsname}
\def\lsim{\mathrel{\raise .4ex\hbox{\rlap{$<$}\lower 1.2ex\hbox{$\sim$}}}}
\def\gsim{\mathrel{\raise .4ex\hbox{\rlap{$>$}\lower 1.2ex\hbox{$\sim$}}}}

\def\simprop{\mathrel{\raise .4ex\hbox{\rlap{$\propto$}\lower 1.2ex\hbox{$\sim$}}}}

\def\arcm{\ifmmode {^{\scriptscriptstyle\prime}}
          \else $^{\scriptscriptstyle\prime}$\fi}

\def\amin{\ifmmode {^{\scriptscriptstyle\prime}}
          \else $^{\scriptscriptstyle\prime}$\fi}

\def\Planck{{\it Planck\/}\ {}}

\newcolumntype{B}{>{\centering\arraybackslash}p{78pt}}
\newcolumntype{S}{>{\centering\arraybackslash}p{66pt}}
\newcolumntype{L}{>{\raggedright\arraybackslash}p{156pt}}
\newcolumntype{M}{>{\raggedright\arraybackslash}p{48pt}}
\newcolumntype{N}{>{\raggedright\arraybackslash}p{18pt}}
\newcolumntype{C}{>{\centering\arraybackslash}p{42pt}}

\begin{document}
\title{\textit{Planck} early results XIV: ERCSC validation and extreme radio sources}
\author{\small
Planck Collaboration:
P.~A.~R.~Ade\inst{75}
\and
N.~Aghanim\inst{49}
\and
E.~Angelakis\inst{68}
\and
M.~Arnaud\inst{62}
\and
M.~Ashdown\inst{60, 4}
\and
J.~Aumont\inst{49}
\and
C.~Baccigalupi\inst{73}
\and
A.~Balbi\inst{28}
\and
A.~J.~Banday\inst{79, 8, 67}
\and
R.~B.~Barreiro\inst{56}
% ...
}
\institute{\small
Aalto University Mets\"{a}hovi Radio Observatory, Mets\"{a}hovintie 114, FIN-02540 Kylm\"{a}l\"{a}, Finland\\
\and
Agenzia Spaziale Italiana Science Data Center, c/o ESRIN, via Galileo Galilei, Frascati, Italy\\
\and
Astroparticule et Cosmologie, CNRS (UMR7164), Universit\'{e} Denis Diderot Paris 7, B\^{a}timent Condorcet, 10 rue A. Domon et L\'{e}onie Duquet, Paris, France\\
\and
Astrophysics Group, Cavendish Laboratory, University of Cambridge, J J Thomson Avenue, Cambridge CB3 0HE, U.K.\\
\and
Atacama Large Millimeter/submillimeter Array, ALMA Santiago Central Offices, Alonso de Cordova 3107, Vitacura, Casilla 763 0355, Santiago, Chile\\
\and
Australia Telescope National Facility, CSIRO, P.O. Box 76, Epping, NSW 1710, Australia\\
\and
CITA, University of Toronto, 60 St. George St., Toronto, ON M5S 3H8, Canada\\
\and
CNRS, IRAP, 9 Av. colonel Roche, BP 44346, F-31028 Toulouse cedex 4, France\\
\and
California Institute of Technology, Pasadena, California, U.S.A.\\
\and
DAMTP, University of Cambridge, Centre for Mathematical Sciences, Wilberforce Road, Cambridge CB3 0WA, U.K.\\
\and
DSM/Irfu/SPP, CEA-Saclay, F-91191 Gif-sur-Yvette Cedex, France\\
\and
DTU Space, National Space Institute, Juliane Mariesvej 30, Copenhagen, Denmark\\
\and
Departamento de F\'{\i}sica, Universidad de Oviedo, Avda. Calvo Sotelo s/n, Oviedo, Spain\\
\and
Department of Astronomy and Astrophysics, University of Toronto, 50 Saint George Street, Toronto, Ontario, Canada\\
\and
Department of Physics \& Astronomy, University of British Columbia, 6224 Agricultural Road, Vancouver, British Columbia, Canada\\
\and
Department of Physics and Astronomy, University of Southern California, Los Angeles, California, U.S.A.\\
\and
Department of Physics, Gustaf H\"{a}llstr\"{o}min katu 2a, University of Helsinki, Helsinki, Finland\\
\and
Department of Physics, Purdue University, 525 Northwestern Avenue, West Lafayette, Indiana, U.S.A.\\
\and
Department of Physics, University of California, Berkeley, California, U.S.A.\\
\and
Department of Physics, University of California, One Shields Avenue, Davis, California, U.S.A.\\
\and
Department of Physics, University of California, Santa Barbara, California, U.S.A.\\
\and
Department of Physics, University of Illinois at Urbana-Champaign, 1110 West Green Street, Urbana, Illinois, U.S.A.\\
\and
Dipartimento di Fisica G. Galilei, Universit\`{a} degli Studi di Padova, via Marzolo 8, 35131 Padova, Italy\\
\and
Dipartimento di Fisica, Universit\`{a} La Sapienza, P. le A. Moro 2, Roma, Italy\\
\and
Dipartimento di Fisica, Universit\`{a} degli Studi di Milano, Via Celoria, 16, Milano, Italy\\
\and
Dipartimento di Fisica, Universit\`{a} degli Studi di Trieste, via A. Valerio 2, Trieste, Italy\\
\and
Dipartimento di Fisica, Universit\`{a} di Ferrara, Via Saragat 1, 44122 Ferrara, Italy\\
\and
Dipartimento di Fisica, Universit\`{a} di Roma Tor Vergata, Via della Ricerca Scientifica, 1, Roma, Italy\\
\and
Discovery Center, Niels Bohr Institute, Blegdamsvej 17, Copenhagen, Denmark\\
\and
Dpto. Astrof\'{i}sica, Universidad de La Laguna (ULL), E-38206 La Laguna, Tenerife, Spain\\
\and
European Southern Observatory, ESO Vitacura, Alonso de Cordova 3107, Vitacura, Casilla 19001, Santiago, Chile\\
\and
European Space Agency, ESAC, Planck Science Office, Camino bajo del Castillo, s/n, Urbanizaci\'{o}n Villafranca del Castillo, Villanueva de la Ca\~{n}ada, Madrid, Spain\\
\and
European Space Agency, ESTEC, Keplerlaan 1, 2201 AZ Noordwijk, The Netherlands\\
\and
Finnish Centre for Astronomy with ESO (FINCA), University of Turku, V\"{a}is\"{a}l\"{a}ntie 20, FIN-21500, Piikki\"{o}, Finland\\
\and
Haverford College Astronomy Department, 370 Lancaster Avenue, Haverford, Pennsylvania, U.S.A.\\
\and
Helsinki Institute of Physics, Gustaf H\"{a}llstr\"{o}min katu 2, University of Helsinki, Helsinki, Finland\\
\and
INAF - Osservatorio Astrofisico di Catania, Via S. Sofia 78, Catania, Italy\\
\and
INAF - Osservatorio Astronomico di Padova, Vicolo dell'Osservatorio 5, Padova, Italy\\
\and
INAF - Osservatorio Astronomico di Roma, via di Frascati 33, Monte Porzio Catone, Italy\\
\and
INAF - Osservatorio Astronomico di Trieste, Via G.B. Tiepolo 11, Trieste, Italy\\
\and
INAF/IASF Bologna, Via Gobetti 101, Bologna, Italy\\
\and
INAF/IASF Milano, Via E. Bassini 15, Milano, Italy\\
\and
INRIA, Laboratoire de Recherche en Informatique, Universit\'{e} Paris-Sud 11, B\^{a}timent 490, 91405 Orsay Cedex, France\\
\and
IPAG: Institut de Plan\'{e}tologie et d'Astrophysique de Grenoble, Universit\'{e} Joseph Fourier, Grenoble 1 / CNRS-INSU, UMR 5274, Grenoble, F-38041, France\\
\and
ISDC Data Centre for Astrophysics, University of Geneva, ch. d'Ecogia 16, Versoix, Switzerland\\
\and
Imperial College London, Astrophysics group, Blackett Laboratory, Prince Consort Road, London, SW7 2AZ, U.K.\\
\and
Infrared Processing and Analysis Center, California Institute of Technology, Pasadena, CA 91125, U.S.A.\\
\and
Institut N\'{e}el, CNRS, Universit\'{e} Joseph Fourier Grenoble I, 25 rue des Martyrs, Grenoble, France\\
\and
Institut d'Astrophysique Spatiale, CNRS (UMR8617) Universit\'{e} Paris-Sud 11, B\^{a}timent 121, Orsay, France\\
\and
Institut d'Astrophysique de Paris, CNRS UMR7095, Universit\'{e} Pierre \& Marie Curie, 98 bis boulevard Arago, Paris, France\\
\and
Institut de Radioastronomie Millim\'{e}trique (IRAM), Avenida Divina Pastora 7, Local 20, 18012 Granada, Spain\\
\and
Institute of Astronomy and Astrophysics, Academia Sinica, Taipei, Taiwan\\
\and
Institute of Astronomy, University of Cambridge, Madingley Road, Cambridge CB3 0HA, U.K.\\
\and
Institute of Theoretical Astrophysics, University of Oslo, Blindern, Oslo, Norway\\
\and
Instituto de Astrof\'{\i}sica de Canarias, C/V\'{\i}a L\'{a}ctea s/n, La Laguna, Tenerife, Spain\\
\and
Instituto de F\'{\i}sica de Cantabria (CSIC-Universidad de Cantabria), Avda. de los Castros s/n, Santander, Spain\\
\and
Istituto di Fisica del Plasma, CNR-ENEA-EURATOM Association, Via R. Cozzi 53, Milano, Italy\\
\and
Jet Propulsion Laboratory, California Institute of Technology, 4800 Oak Grove Drive, Pasadena, California, U.S.A.\\
\and
Jodrell Bank Centre for Astrophysics, Alan Turing Building, School of Physics and Astronomy, The University of Manchester, Oxford Road, Manchester, M13 9PL, U.K.\\
\and
Kavli Institute for Cosmology Cambridge, Madingley Road, Cambridge, CB3 0HA, U.K.\\
\and
LERMA, CNRS, Observatoire de Paris, 61 Avenue de l'Observatoire, Paris, France\\
\and
Laboratoire AIM, IRFU/Service d'Astrophysique - CEA/DSM - CNRS - Universit\'{e} Paris Diderot, B\^{a}t. 709, CEA-Saclay, F-91191 Gif-sur-Yvette Cedex, France\\
\and
Laboratoire Traitement et Communication de l'Information, CNRS (UMR 5141) and T\'{e}l\'{e}com ParisTech, 46 rue Barrault F-75634 Paris Cedex 13, France\\
\and
Laboratoire de Physique Subatomique et de Cosmologie, CNRS/IN2P3, Universit\'{e} Joseph Fourier Grenoble I, Institut National Polytechnique de Grenoble, 53 rue des Martyrs, 38026 Grenoble cedex, France\\
\and
Laboratoire de l'Acc\'{e}l\'{e}rateur Lin\'{e}aire, Universit\'{e} Paris-Sud 11, CNRS/IN2P3, Orsay, France\\
\and
Lawrence Berkeley National Laboratory, Berkeley, California, U.S.A.\\
\and
Max-Planck-Institut f\"{u}r Astrophysik, Karl-Schwarzschild-Str. 1, 85741 Garching, Germany\\
\and
Max-Planck-Institut f\"{u}r Radioastronomie, Auf dem H\"{u}gel 69, 53121 Bonn, Germany\\
\and
MilliLab, VTT Technical Research Centre of Finland, Tietotie 3, Espoo, Finland\\
\and
National University of Ireland, Department of Experimental Physics, Maynooth, Co. Kildare, Ireland\\
\and
Niels Bohr Institute, Blegdamsvej 17, Copenhagen, Denmark\\
\and
Observational Cosmology, Mail Stop 367-17, California Institute of Technology, Pasadena, CA, 91125, U.S.A.\\
\and
SISSA, Astrophysics Sector, via Bonomea 265, 34136, Trieste, Italy\\
\and
SUPA, Institute for Astronomy, University of Edinburgh, Royal Observatory, Blackford Hill, Edinburgh EH9 3HJ, U.K.\\
\and
School of Physics and Astronomy, Cardiff University, Queens Buildings, The Parade, Cardiff, CF24 3AA, U.K.\\
\and
Space Sciences Laboratory, University of California, Berkeley, California, U.S.A.\\
\and
Special Astrophysical Observatory, Russian Academy of Sciences, Nizhnij Arkhyz, Zelenchukskiy region, Karachai-Cherkessian Republic, 369167, Russia\\
\and
Stanford University, Dept of Physics, Varian Physics Bldg, 382 Via Pueblo Mall, Stanford, California, U.S.A.\\
\and
Universit\'{e} de Toulouse, UPS-OMP, IRAP, F-31028 Toulouse cedex 4, France\\
% ...
}

 \abstract
 {\Planck\!\!'s all-sky surveys at 30--857 GHz provide an unprecedented opportunity to follow the radio spectra of a large sample of extragalactic sources to frequencies 2--20 times higher than allowed by past, large-area, ground-based surveys. We combine the results of the \Planck Early Release Compact Source Catalog (ERCSC) with quasi-simultaneous ground-based observations as well as archival data at frequencies below or overlapping \Planck frequency bands, to validate the astrometry and photometry of the ERCSC radio sources and study the spectral features shown in this new frequency window opened by \Planck\!\!. The ERCSC source positions and flux density scales are found to be consistent with the ground-based observations. We present and discuss the spectral energy distributions (SEDs) of a sample of ``extreme'' radio sources, to illustrate the richness of the ERCSC for the study of extragalactic radio sources.  Variability is found to play a role in the unusual spectral features of some of these sources.}


\authorrunning{Planck Collaboration}
\titlerunning{ERCSC validation and extreme radio sources}
 \maketitle

\section{Introduction}\label{sec:introduction}

This paper is one of a series based on observations of compact sources by the \Planck\!\!\footnote{\Planck (\url{http://www.esa.int/Planck}) is a project of the European Space Agency (ESA) with instruments provided by two scientific consortia funded by ESA member states (in particular the lead countries France and Italy), with contributions from NASA (USA) and telescope reflectors provided by a collaboration between ESA and a scientific consortium led and funded by Denmark.}~satellite that are included in the ERCSC \citep{Planck2011-1.10}. Among these ``Early Results''~\Planck papers there are three that address the extragalactic radio source population. \citet{Planck2011-6.1} examines statistical properties such as number counts and spectral index distributions, but only at frequencies $\ge$ 30~GHz.~\citet{Planck2011-6.3a} incorporates \Planck measurements and supporting ground-based and satellite observations to refine models for the physical properties of a sample of $\sim$100 bright blazars. Here we address the observed diverse, sometimes peculiar, spectral properties of sources in the ERCSC, which include peaked-spectrum, flat-spectrum, upturn-spectrum and multicomponent-spectrum sources. We combine archival data as well as new, ground-based, radio observations with the \Planck data to construct SEDs from $\sim$3 to $\sim$200~GHz (the exact frequency coverage varies case by case), to validate the astrometry and photometry of radio sources in the ERCSC. The \Planck data are also valuable for studying, for instance, high frequency peaked-spectrum sources that were previously underrepresented in radio source populations due to the lack of observations in the sub-millimeter regime. On the other hand, the ground-based data complement the \Planck data, and are crucial in defining the spectral shape of some sources by extending the observed SEDs. We present a sample of sources with near-simultaneous \Planck and ground-based observations (primarily employing the VLA, Effelsberg, IRAM and Mets\"ahovi telescopes in the northern hemisphere, and ATCA in the southern hemisphere), to control for variability. We also investigate a small number of ERCSC sources at 30 to 70~GHz without clear identification in existing radio surveys.

\subsection{The \Planck mission}

\Planck \citep{tauber2010a, Planck2011-1.1} is the third generation space mission to measure the anisotropy of the cosmic microwave background (CMB).  It observes the sky in nine frequency bands covering 30 to 857~GHz with high sensitivity and angular resolution from 32$'$ to 5$'$.  The Low Frequency Instrument (LFI; \citealt{Mandolesi2010, Bersanelli2010, Planck2011-1.4}) covers the 30, 44 and 70\,GHz bands with amplifiers cooled to 20~\hbox{K}.  The High Frequency Instrument (HFI; \citealt{Lamarre2010, Planck2011-1.5}) covers the 100, 143, 217, 353, 545 and 857\,GHz bands with bolometers cooled to 0.1~\hbox{K}.  Polarisation is measured in all but the highest two bands \citep{Leahy2010, Rosset2010}.  A combination of radiative cooling and three mechanical coolers produces the temperatures needed for the detectors and optics \citep{Planck2011-1.3}.  Two Data Processing Centers (DPCs) check and calibrate the data and make maps of the sky \citep{Planck2011-1.7, Planck2011-1.6}.  \Planck\!\!'s sensitivity, angular resolution and frequency coverage make it a powerful instrument for Galactic and extragalactic astrophysics as well as cosmology.  

\begin{figure}
   \centering
   \includegraphics[scale=0.20]{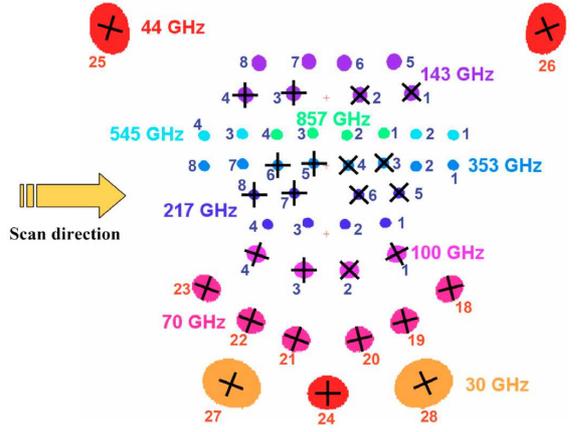}
   \caption{Focal plane, showing spacing of the \Planck receivers. Each day, the pointing is adjusted by $\sim$1$\degr$ in the vertical direction.  Note that the wide separation of the three 44\,GHz horns (two at the top, one at the bottom) causes the 44\,GHz observations of a given source to take place at two times separated by 7--10 days for each scan.}
              \label{fig_focalplane}%
    \end{figure}

The scan strategy employed in the \Planck mission is described in \citet{Planck2011-1.1}.  As the satellite spins, sources are swept over the focal plane, as indicated schematically in Figure~\ref{fig_focalplane}. In the course of each day, the pointing axis of the telescope is adjusted by $\sim$1$\degr$, so a given source will follow a slightly different track across the focal plane; thus its flux density is the average of many such scans. In addition, since the data included in the ERCSC amount to $\sim$1.6 full-sky surveys, some sources have been covered twice with a time separation of $\sim$6 months. Finally, sources near the ecliptic poles, where the scan circles intersect, are often covered multiple times. It is thus important to keep in mind that the flux densities cited in this paper (and indeed in the ERCSC as a whole) are {\it averaged}. Figure~\ref{fig_focalplane} illustrates that flux measurements at 44~GHz are particularly susceptible to time-dependent effects, because of the wide spacing of the 44\,GHz horns in the focal plane (see further discussion in \S\,\ref{sec:artifacts}).

 \subsection{The ERCSC}

The \Planck Early Release Compact Source Catalog \citep{Planck2011-1.10} provides lists of positions and flux densities of compact sources at each of the nine \Planck frequencies. For frequencies from 30 to 143~GHz (those mostly cited in this paper), sources were detected using Powell Snakes techniques \citep{carvalho2009}. In the four highest frequency channels, sources were located using the SExtractor method \citep{bertin1996}. A set of selection criteria was further applied to select sources that are included in the ERCSC. The primary criterion was a Monte Carlo assessment designed to ensure that $\ge\,$90\,\% of the sources in the catalogue are reliable and have a flux density accuracy better than 30\,\%. External validation (also discussed in \citealt{Planck2011-1.10}) shows that ERCSC met its reliability criterion, and we show evidence in \S\,\ref{sec:validation} that the flux density scale of ERCSC is accurate. Secondary selection criteria, including the elimination of extended sources, were also applied. Virtually all of the sources discussed in this paper, and the vast majority of extragalactic ERCSC sources, were unresolved by \Planck and in many cases even at the much higher angular resolution of the VLA or other ground-based instruments. 

The flux densities in the ERCSC are calculated using aperture photometry. The effective band centers or corresponding colour corrections depend to some degree on the spectrum of the source being observed. This relatively small dependence is discussed in the LFI and HFI instrument papers \citep{Planck2011-1.4, Planck2011-1.5, Planck2011-1.6, Planck2011-1.7}. We adopt here (and list in Table~\ref{table_bands}) the effective central frequencies defined in those papers, and the colour corrections defined for a source with a spectral index $\alpha = -0.5$ (using the convention $S_{\nu} \propto \nu^{\alpha}$). To obtain the correct flux density for an assumed narrow band measurement, we divide the tabulated ERCSC flux densities by the colour correction factor at the corresponding central frequency.  After the release of the ERCSC, it was found that the correction for aberration introduced by the motion of the satellite had not been correctly made.  This introduces small ($<$\,0.35$'$) errors in the catalogued positions of sources; these can in turn produce very small errors ($<$\,1\,\%) in flux densities.  We have not corrected the ERCSC flux densities for this small effect.

In the next section, we describe briefly some general properties of the extragalactic radio sources in the ERCSC. We describe in \S\,\ref{sec:observation} related ground-based observations. In \S\,\ref{sec:validation}, we discuss the identification of sources, positional accuracy and flux density scales, comparing those obtained from \Planck with ground-based measurements. This parallels the validation work discussed in \cite{Planck2011-1.10}. Variability of source luminosity and the issue of different resolutions are also discussed. We present and discuss in \S\,\ref{sec:ers} several examples of interesting classes of extragalactic radio sources. We conclude in the final section and point towards further research on these radio sources and many others contained in the ERCSC.

\begin{table}
\caption{Parameters of \Planck bands employed in this paper.}             
\label{table_bands}      
\centering                          
\begin{tabular}{c c c c }    
\hline\hline                
Planck & Central Frequency & Colour & Beam FWHM \\   

Band & [GHz] & Correction & [arcmin] \\
\hline\\                        
          30 & 28.5 & 1.037 & 32.6 \\
           44 & 44.1 & 1.018 & 27.0 \\
           70 & 70.3 & 1.031 & 13.0 \\
         100 & 100 & 0.999 & 9.94 \\
	143 & 143 & 1.006 & 7.04 \\
 	217 & 217 & 0.993 & 4.66 \\ 
	353 & 353 & 0.990 & 4.41 \\
\hline                        
\end{tabular}
\end{table}

\section{Radio sources in the ERCSC}

The ERCSC contains hundreds of extragalactic radio sources at frequencies up to 143~GHz. At frequencies $>$\,100~GHz, the \Planck surveys are unique. At 30 and 70~GHz, the higher sensitivity and resolution offered by \Planck allow us to detect more sources in a single sky survey than in seven years of survey by the WMAP satellite \citep{gold2010}.

A major finding from the analysis of ERCSC is that many bright radio sources have relatively flat ($\alpha$\,$>$\,$-0.5$) radio spectra extending up to and sometimes beyond 143~GHz. The vast majority of the extragalactic sources detected at 100, 143 and 217~GHz are synchrotron-dominated radio sources and not dusty galaxies. The statistical properties of radio sources in the ERCSC are discussed in much greater detail in the companion paper, \citet{Planck2011-6.1}. 

A second finding is the absence of compelling evidence for any new class of extragalactic radio sources.  At 30 to 70~GHz, more than 90\,\% of the extragalactic sources were reliably associated with radio sources in other large area surveys conducted at 8--20~GHz \citep{Planck2011-1.10}.  Still others have plausible identifications in lower frequency radio catalogues. We discuss those ERCSC sources without clear identification in \S\,\ref{sec:unmatched} since they may potentially contain new types of sources.  

Many of the identified radio sources are blazars, some of them clearly variable (see a detailed discussion in \citealt{Planck2011-6.3a}).  In \S\,\ref{sec:ers}, we discuss a small number of extreme or unusual radio sources as additional examples of the scientific richness the ERCSC provides for the study of extragalactic radio sources.

The \Planck data used here are drawn entirely from the ERCSC. These data are supplemented by ground-based observations at frequencies below and overlapping the \Planck frequency bands, which were generally made quasi-simultaneously with the \Planck observations of a given source (within 7--10 days, typically). This step was taken to monitor and control for variability.

\section{Ground-based observations}\label{sec:observation}

In planning the \Planck mission, it was recognised that the science yield of \Planck\!\!'s millimeter and FIR sky surveys would be increased if accompanying ground-based observations could be made. These include, but are not limited to, radio observations at frequencies below and overlapping \Planck frequency bands, optical observations for source identification and both ground- and satellite-based X- and $\gamma$-ray observations. This paper discusses only the supporting radio observations; for a discussion of the approximately simultaneous X- and $\gamma$-ray observations, see \citet{Planck2011-6.3a}.  Table 2 provides information on the radio observatories that are involved and produced data used in this paper. Most of the radio observations were made {\it preemptively}---that is, we observed sources expected to be detected by \Planck at about the same time they passed through the \Planck beam. The POFF software \citep{massardi2010} was used to predict which sources would be seen by \Planck in a given week. Obviously, only previously known sources can be observed preemptively. We later conducted a small number of follow-up observations on ERCSC sources.

In the southern hemisphere, a substantial amount of time was obtained at the Australia Telescope Compact Array (ATCA) to make preemptive observations at a wide range of frequencies up to and overlapping with the {\it Planck}. The Planck-ATCA Co-eval Observations (PACO) project \citep{massardi2011}  consists of several-epoch observations  of a compilation of sources selected from  the Australia Telescope 20\,GHz survey (AT20G, \citealt{murphy2010}) with $|\,b\,|$\,$>$\,$5\degr$. Observations were made with ATCA in the frequency range between 4.5 and 40 GHz at epochs close in time to the \Planck observations over the interval from July 2009 to August 2010. The PACO sample includes a complete flux density limited and spectrally selected sample over the whole Southern sky; 147 PACO point-like sources have at least one observation within 10 days of the \Planck observations.

In the northern hemisphere, quasi-simultaneous cm/mm radio spectra for a large number of \Planck blazars have been obtained within the framework of a {\sl Fermi}-GST related monitoring program of $\gamma$-ray blazars (F-GAMMA program, \citealt{fuhrmann2007, angelakis2008}). The frequency range spans from 2.64~GHz to 142~GHz using the Effelsberg 100\,m and IRAM 30\,m telescopes. The Effelsberg measurements were conducted with the secondary focus heterodyne receivers at 2.64, 4.85, 8.35,10.45, 14.60, 23.05, 32.00 and 43.00~GHz. The observations were performed quasi-simultaneously with cross-scans, that is slewing over the source position in azimuth and elevation direction with the number of sub-scans adjusted to reach the desired sensitivity (for details, see \citealt{fuhrmann2008, angelakis2008}). Subsequently, pointing off-set correction, gain correction, atmospheric opacity correction and sensitivity correction were applied to the data. The IRAM 30\,m observations were carried out with calibrated cross-scans using the EMIR horizontal and vertical polarisation receivers operating at 86.2 and 142.3 GHz. The opacity corrected intensities were converted into the standard temperature scale and finally corrected for small remaining pointing offsets and systematic gain-elevation effects. The conversion to the standard flux density scale was done using the instantaneous conversion factors derived from frequently observed primary (Mars, Uranus) and secondary(W3(OH), K3-50A, NGC7027) calibrators.

In the northern hemisphere, some data also come from the Mets\"ahovi telescope operating at 37 GHz. The observations were made with the 13.7\,m Mets\"ahovi radio telescope, which is a radome-enclosed, paraboloid antenna situated in Finland. The measurements were made with a 1\,GHz-band dual-beam receiver centered at 36.8~GHz. The observations are ON--ON, alternating the source and the sky in each feed horn. A typical integration time to obtain one flux density data point is 1200--1400~s. The detection limit of the telescope at 37~GHz is of the order of 0.2~Jy under optimal conditions. Data points with a signal-to-noise ratio $<$\,4 are treated as non-detections.  The flux density scale is set by observations of DR21. Sources NGC7027, 3C274 and 3C84 are used as secondary  calibrators. A detailed description of the data reduction and analysis is given in \citep{teraesranta1998}. The error estimate in the flux density includes the contribution from the measurement rms and the uncertainty of the absolute calibration.

\begin{table}
\label{table_followup} 
\centering
\caption{Ground-based radio observations.}   
\begin{tabular}{S B c }        
\hline\hline                 
Observatory & Project Leaders & Frequencies [GHz] \\   
\hline\\
ATCA, Australia \smallskip & Massardi \smallskip & 4.5--40 \smallskip \\
Effelsberg, Germany \smallskip & Fuhrmann, Angelakis and Rachen \smallskip & 2.6--43 \smallskip \\
IRAM, Spain & Fuhrmann, Ungerechts and Rachen & \smallskip 86.2, 142.3 \\
Mets\"ahovi, Finland & L\"ahteenm\"aki and Tornikoski & \smallskip 37 \\ 
VLA/EVLA, USA & Partridge and Sajina & \smallskip 5, 8, 22, 43 \\
\hline
\end{tabular} 
\end{table}

In addition, small amounts of time, scattered throughout the first 17 months of the \Planck mission, were obtained at the Very Large Array (VLA) of the National Radio Astronomy Observatory (NRAO). Observations at the VLA began 24~July 2009, slightly before the beginning of \Planck\!\!'s first sky survey. The first set of observations on 24~July 2009 is the only set not approximately simultaneous with \Planck observations. VLA/EVLA measurements continued at irregular intervals until November 2010, with a substantial gap in the spring of 2010 when the VLA was converted to the EVLA. Most of the VLA and EVLA runs were brief 1--2 hour chunks of time; 5--8 \Planck sources were typically observed per hour, besides flux calibrators and phase calibrators. In many cases, VLA flux and phase calibrators were of interest themselves because they were bright enough to be detected by {\it Planck}. The integration times were typically 30 seconds at 4.86~GHz and 8.46~GHz, 100 seconds at 22.46~GHz, and 120 seconds at 43.34~GHz. All u-v data were flagged, calibrated and imaged using standard NRAO software: AIPS or CASA. The flux density measurements were calibrated against one or both of the primary calibrators used by NRAO: 3C48 or 3C286. 

The positions and flux densities of sources observed by the VLA are listed in Table~\ref{table_sources_obs}. Since the VLA flux densities are accurate to a few percent for these bright sources, individual uncertainties are not included. Typical 1\,$\sigma$  errors at 22 GHz ranged from 2~mJy (for the fainter sources) to 15 mJy (for brighter sources); and from 3--15 mJy at 43 GHz; both are small compared to the uncertainties in Planck flux densities. The VLA (EVLA) was in different configurations at different times; hence the angular resolution of the array was changing. Additionally, for a given configuration, the resolution was much finer at higher frequencies. We thus flagged sources that showed signs of resolution in any configuration at any frequency. In general, the VLA observations were timed to occur during the same week as \Planck was expected to see a given source, but that was not always possible. Therefore, each observation is tagged with a date in Table~\ref{table_sources_obs}, to be compared with the dates of observation tabulated in the ERCSC. Given that the typical timescale for variability of radio sources is roughly weeks to months (see e.g., \citealt{hovatta2008, nieppola2009}), we were able to control for variability to some degree. Nevertheless, it must be borne in mind that the ground-based observations were not always taken at the same time \Planck was observing a source. The \Planck beams at different frequencies also swept over a source at different times due to the extended layout of its focal plane. We use this set of data in \S\,\ref{sec:validation} to check the positional and photometric accuracy of ERCSC sources.

\begin{table*}
\centering
\label{table_sources_obs}      
\caption{VLA observations on a set of bright ERCSC sources.}
\scalebox{0.87}{
\begin{tabular}{l c r c c c c l L}        
\hline\hline
\footnotesize
ERCSC name & VLA RA & {\hfill VLA Dec\hfill} & $S_{4.86~GHz}$ & $S_{8.46~GHz}$ & $S_{22.46~GHz}$ & $S_{43.34~GHz}$ & VLA obs. date & \Planck obs. date\tablefootmark{a} \\
 & [J2000] & {\hfill [J2000]\hfill} & [Jy] & [Jy] & [Jy] & [Jy] & [yymmdd] & [yymmdd]  \\ 
\hline\\
G052.38-36.49  & 21:34:10.3 & --01:53:17 & ... & ... & 1.698 & 1.218 & 091103 & 091102, 091103, 091108--091110, 100509--100511, 100517  \\
G056.70+80.65\tablefootmark{b} & 13:31:08.3 & 30:30:33 & 7.485 & 5.202 & 2.516 & 1.444  & 091022, 100103 & 091230, 091231 \\
G075.68-29.62 & 22:03:26.9 & 17:25:48 & 0.987 & 1.055 & 1.035 & 1.073 & 091211 & 091121, 091128--091202, 100524--100527, 100602, 100603 \\
G077.45-38.54 & 22:32:36.4 & 11:43:51 & 5.594 & 5.827 & 4.407 & 3.606 &  091211 & 091127, 091203--091207, 100531--100603 \\
 &  &  &  &  &  &  &  & 100506--100510, 100519, 100520 \\
G085.72+26.08 & 18:24:07.1 & 56:51:01  & 1.403 & 1.876 & 1.771  & 1.258 & 090724 & 090827--090829, 100417--100422, 100503,100504 \\
G085.86+83.31 & 13:10:28.7 & 32:20:44 & 1.427 & 2.133 & 2.856 & 1.977  &  100103 & 091221--091225, 100102, 100103 \\ 
G085.95-18.77\tablefootmark{c} & 22:03:15.0 & 31:45:38  & 2.165 & 2.495 & 2.576 & 2.841& 091211 & 091202, 091203, 091211--091215, 100531--100603 \\
G086.12-38.18\tablefootmark{c} & 22:53:57.8 & 16:08:54  & 9.463 & 8.176 & 11.834 & 21.570 & 091211 & 091206, 091207, 091213--091216 \\
G090.11--25.64 & 22:36:22.5 & 28:28:57 & 1.285 & 1.273 & 1.412 & 1.547 & 091211 & 091210, 091211, 091218--091221 \\
G090.14+09.66 & 20:23:55.8 & 54:27:36 & 0.952 & 1.132 & 0.804 & 0.502 & 090724 & 091222, 091223, 100517, 100518 \\ 
G092.62--10.44 & 22:02:43.3 & 42:16:39 & 3.994 & 3.996 & 3.440 & 3.085 & 091211 & 091213, 091214, 091222--091226, 100606, 100607 \\ 
G096.08+13.77 & 20:22:06.7 & 61:36:59 & 3.110 & 3.800 & 1.946 & 0.836 & 090724 & 091220--091222, 100103--100108, 100526--100531, 100602 \\ 
G097.46+25.04 & 18:49:16.1 & 67:05:42 & 1.284 & 2.431 & 3.553 & 2.956 & 090724 & 091012--091014, 091017--091022, 091112-091114, 091225--091227, \\
 &  &  &  &  &  &  &  & 100112--100119, 100512--100514, 100516--100521 \\
G098.28+58.29 & 14:19:46.6 & 54:23:14 & 1.178 & 1.249 & 1.231 & 1.176 & 100103 & 091206, 091207, 091209--091212, 091226, 091227 \\
G098.51+25.79 & 18:42:33.6 & 68:09:25 & 0.765 & 0.892 & 0.676 & 0.461 & 090724 & 091013--091015, 091110--091112, 091231, \\
 &  &  &  &  &  &  &  & 100101, 100116--100119, 100121, 100122, 100514--100516, 100518--100521, 100523, 100524 \\
G110.05+29.07 & 18:00:45.7 & 78:28:04 & 2.098 & 2.850 & 2.598 & 1.936 & 090724 & 091011--091016, 091026, 091027, 100202, 100203, 100214--100219 \\
G126.44--64.29\tablefootmark{c} & 00:57:34.9 & --01:23:28 & 0.338  & 0.097 & 0.060 &  0.070 & 091103 & 100107, 100108 \\
G128.95+11.97 & 02:17:30.8 & 73:49:33 & 4.168 & 4.090 & 2.623 & 1.709 & 090827 & 090911--090914 ,090920, 090921, 100211, 100212, 100217--100220 \\
G133.94--28.63\tablefootmark{b} & 01:37:41.3 & 33:09:35 & 5.426 & 3.140 & 1.098 & 0.501 & 090827, 091211 & 100130, 100131 \\
G143.55+34.41 & 08:41:24.4 & 70:53:42 & 1.762 & 1.485 & 2.267 & 2.900 & 091022 & 091012--091014, 091019, 100316, 100317, 100324--100327 \\
G145.58+64.96 & 11:53:24.5 & 49:31:09 & 1.167 & 0.949 & 0.885 & 0.798 & 100103 & 091116--091119, 091125, 091126, 100511, 100512, 100522--100526 \\
G174.47+69.79 & 11:30:53.3 & 38:15:19 & 1.334 & 1.356 & 1.019 & 0.680 & 100103 & 091121, 100527 \\
G195.26--33.13 & 04:23:15.8 & --01:20:33 & ... & ... & 7.500 & 7.289 & 091103 & 090828--090830, 090903, 100221, 100225--100227  \\
G199.42+78.39 & 11:59:31.8 & 29:14:44 & 1.609 & 1.530 & 1.416 & 1.141 & 100103 & 091204--091207, 091214, 100531, 100601 \\
G200.06+31.89 & 08:30:52.1 & 24:11:00 & 1.269 & 1.246 & 1.050 & 0.825 & 091022 & 091021--091023, 091027, 100409, 100410, 100414--100416 \\
G208.18+18.75 & 07:50:52.0 & 12:31:05 & 3.685 & 4.486 & 4.400 & 3.939 &  091022 & 091014--091016, 091020, 100404, 100408--100410 \\
G211.33+19.05 & 07:57:06.6 & 09:56:35 & 1.143 & 1.530 & 1.863 & 1.846 & 091022 & 091015--091018, 091022, 100406, 100410--100412 \\
G216.97+11.36 & 07:39:18.0 & 01:37:04 & 1.143 & 1.260 & 1.514 & 1.553 & 091022 & 091012--091014, 100408--100410 \\
G221.26+22.36 & 08:25:50.3 & 03:09:24 & 0.777 & 0.737 & 0.668 & 0.648 & 091022 & 091024, 091025, 100419, 100420 \\
G251.59+52.70 & 10:58:29.6 & 01:33:59 & 3.193 & 4.186 & 5.317 & 5.078 & 100103 & 091207--091210, 091216,  100524, 100530--100602 \\
G255.00+81.65 & 12:24:54.5 & 21:22:47 & 1.233 & 1.267 & 1.412 & 1.184 & 100103 & 091218--091221 \\
G283.75+74.54\tablefootmark{c}  & 12:30:49.4 & 12:23:28 & 62.000 & 39.000 & 12.000 & ... & 100103 & 091226--091229, 100104, 100628, 100629 \\
\hline                                   
\end{tabular}
}
\tablefoot{
\tablefoottext{a}{\Planck observation dates record the days that the source is observed by any of the 30, 44 and 70 GHz channels.}
\tablefoottext{b}{Standard VLA flux calibrator, observed twice. Reported flux densities are the average of the two observations.}
\tablefoottext{c}{Resolved source in the VLA.}
}
\end{table*}

\section{Identification and validation}\label{sec:validation}

To construct SEDs that include measurements from both \Planck and ground-based telescopes, we need to ensure that the identification of a source in the ERCSC has been correctly made, and that the flux density scales are consistent. Both properties were extensively checked as part of the validation work of the ERCSC \citep{Planck2011-1.10}. We present here the comparison of quasi-simultaneous VLA and \Planck observations of a set of bright radio sources.

\subsection{Source identification}\label{sec:identification}

\begin{figure*}
   \centering
   \hspace{0.2 in}
 \includegraphics[scale=0.3]{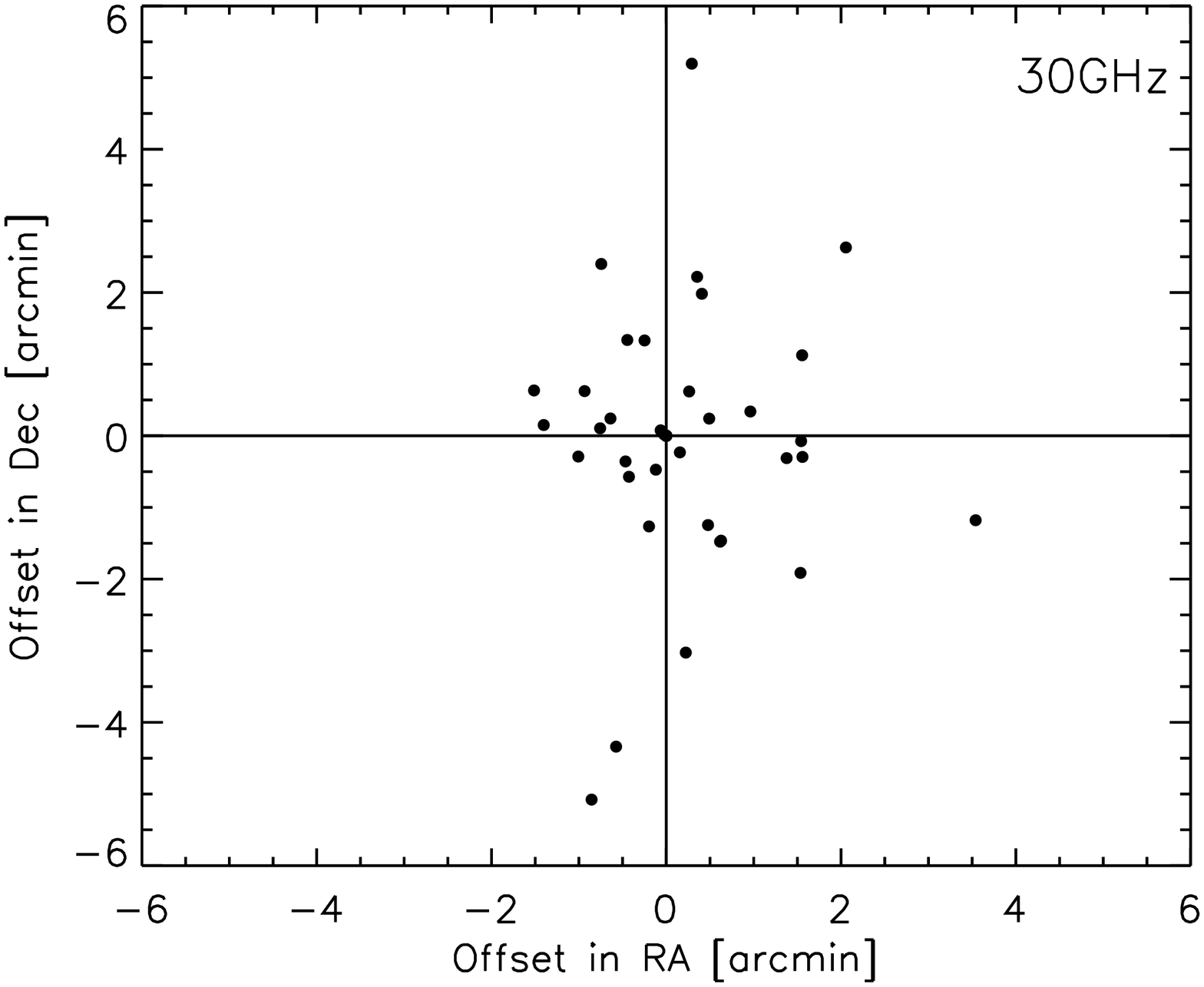} 
 \hspace{0.3 in}
 \includegraphics[scale=0.3]{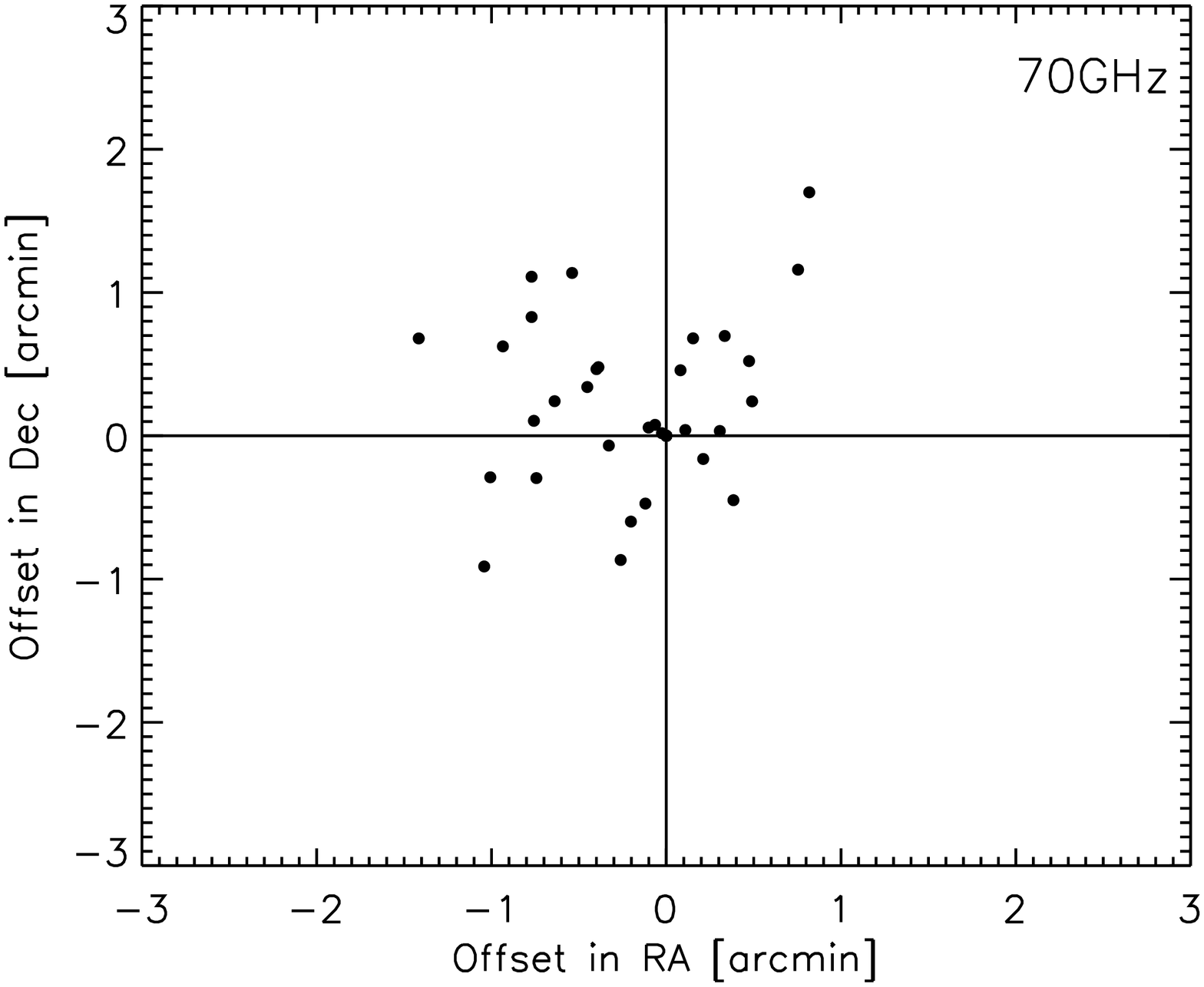}
    \caption{Comparison of the VLA measurements made at $\sim$22 GHz with the ERCSC source positions at 30 GHz (left) and 70 GHz (right).}
              \label{fig_posits}
    \end{figure*}

Identifying a source for supporting ground-based observations requires an understanding of the astrometric accuracy of the ERCSC. In Figure~\ref{fig_posits}, we show a comparison of ERCSC source positions at 30 and 70~GHz with those measured for the presumed identification by the VLA at 22~GHz. The VLA positions for such bright sources are typically accurate to  a few arcseconds even for the most compact configurations. Figure~\ref{fig_posits} indicates that (a) the ERCSC meets its specification of having a positional accuracy good to FWHM/5 (or $\sim$6.5$'$~at 30~GHz); (b) a search radius of 1/2 FWHM of the beam around each ERCSC source position is sufficient to locate any related source; and (c) any bright radio source within a few arcmin of the ERCSC position is likely the correct low frequency counterpart. This last point can be made more quantitative using the AT20G survey, which covers the entire southern hemisphere to $\sim$40 mJy at 20~GHz, and is essentially complete above 100~mJy. The density of sources with $S$\,$>$\,100\,mJy is $\sim$0.25 per square degree, and the integral counts of sources have a slope of $-$1.15. Thus we expect on average 0.07 sources$\cdot$deg$^{-2}$ above 300~mJy, the threshold we applied when cross-correlating AT20G sources with the ERCSC. At 30~GHz, the probability of a random AT20G source appearing within 16.25$'$~of a given ERCSC source is $\sim$1.6\,\%. The ERCSC contains several hundred extragalactic sources, depending on frequency; only 5--8 of these might be falsely identified or confused. Additionally, all the sources described in \S\,\ref{sec:ers} fall within a radius of 5$'$~of the ERCSC position, further reducing the chance coincidence rate by a factor of $\sim$10. The ERCSC source positions were also independently compared to those recorded for several hundred bright quasars. The median scatters in offset are 2.0$'$, 1.7$'$, 1.1$'$, 0.8$'$, 0.7$'$ and 0.3$'$ for  the \Planck frequency bands 30 to 217~GHz \citep{Planck2011-1.10}. This is consistent with our observations shown in Figure~\ref{fig_posits}.

\begin{figure*}
\centering
 \includegraphics[scale=0.3]{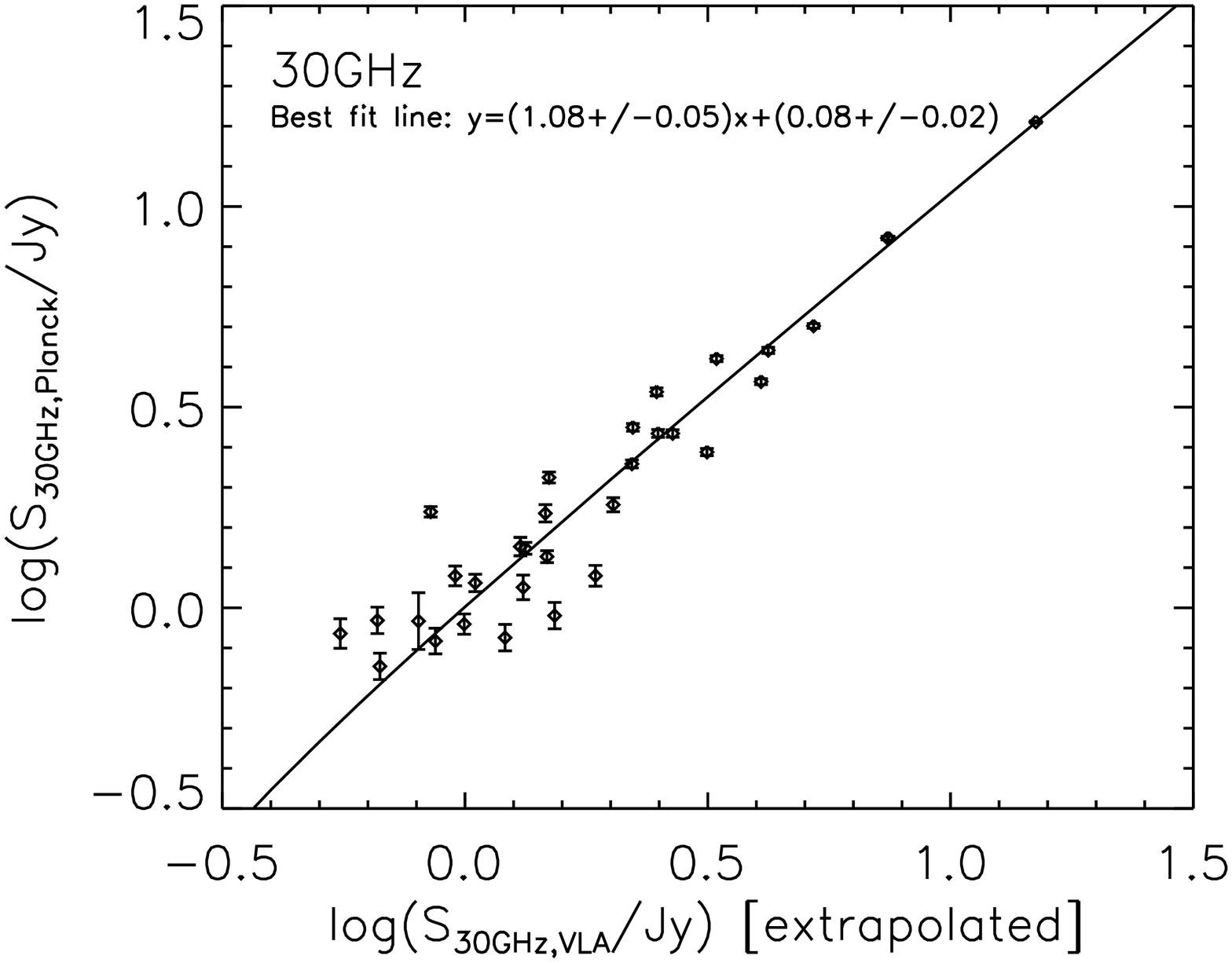}  
 \includegraphics[scale=0.3]{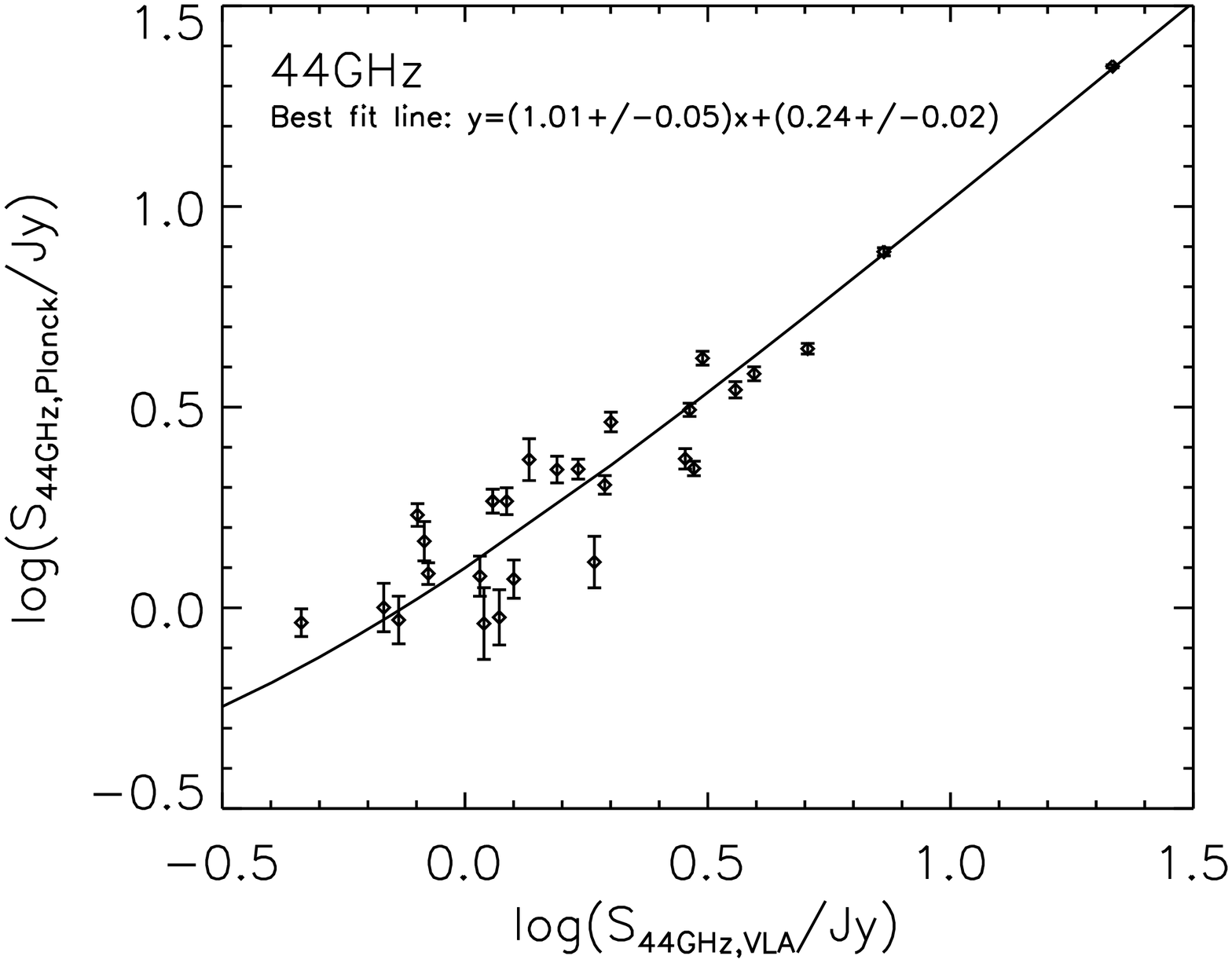} 
 \includegraphics[scale=0.3]{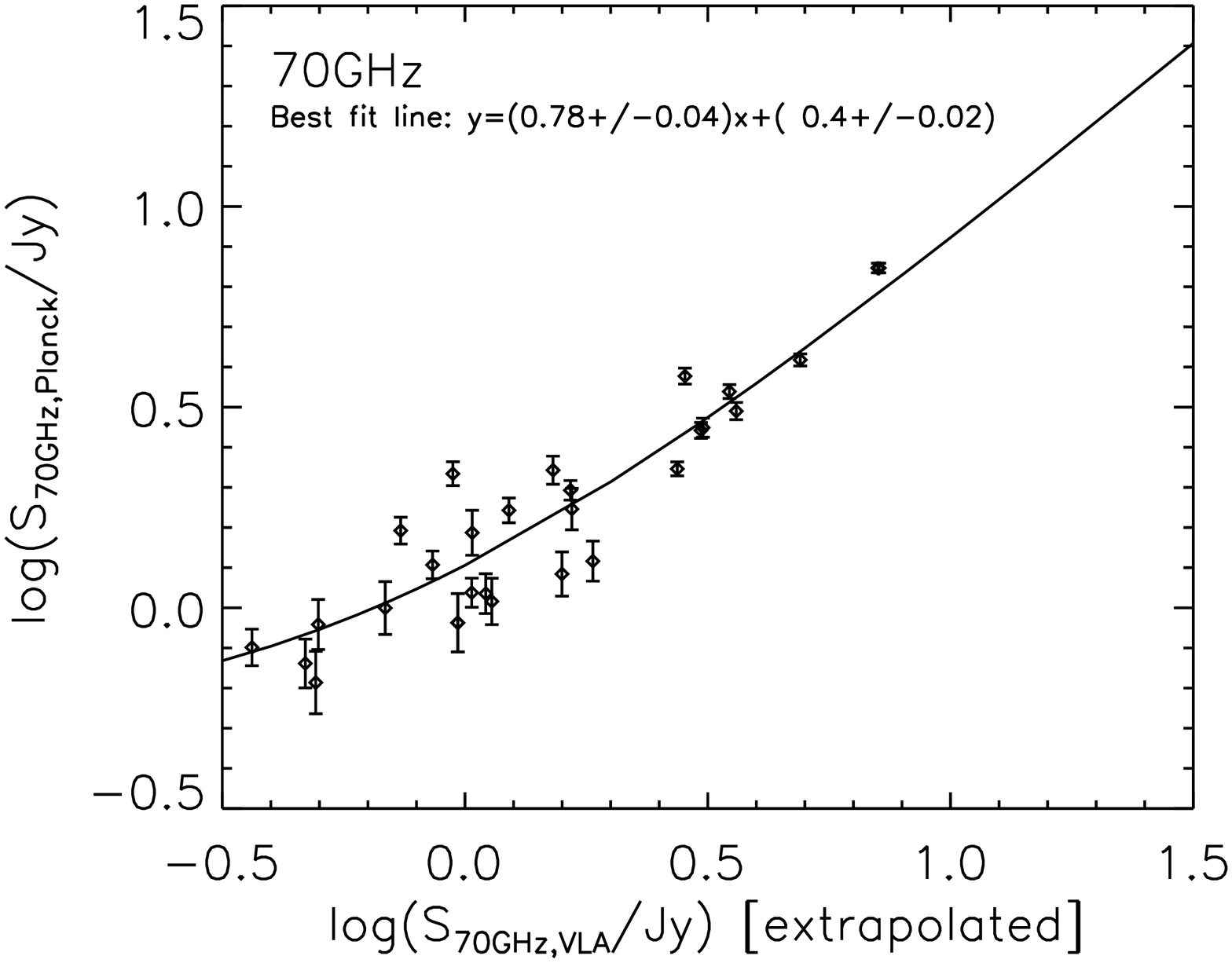}  

 \caption{{\it Upper left:} Comparison of \Planck measurements in the ``30\,GHz'' band with interpolated values from the VLA. {\it Upper right:} \Planck vs. VLA measurements at 44~GHz. {\it Lower center:} \Planck measurements at 70 GHz compared to the {\it extrapolated\/} VLA measurements. Note that although the log-log plot is presented for clarity, the fits were done on the {\it linear} data.}

    \label{fig_30GHz}
    \end{figure*}

\subsection{Comparison of flux density scales}

The flux density scale of \Planck is ultimately tied to the amplitude of the CMB dipole, and careful measurements of the angular resolution of \Planck and the ground-based instruments are required in order to convert from temperature units to flux density \citep{Planck2011-1.10}. Since the calibration standard was the very large angular scale dipole signal, it is important to confirm that flux density measurements of compact, unresolved, sources are accurate. 

A potential complication in comparing flux densities is the greatly different angular resolution of  \Planck and the ground-based instruments. As discussed above, depending on the frequency and configuration, the angular resolution of the VLA can be as small as a fraction of an arcsecond, whereas the \Planck beams are typically many arcminutes in size. Thus the VLA can ``miss'' flux included in the \Planck beam. To first order, we take account of this by excluding from the flux density comparison any source seen to be resolved at any frequency by the VLA (see Table~\ref{table_sources_obs}). A clear example of such a source is 3C274 (J1230+1223).

We then compare in Figure \ref{fig_30GHz} the VLA measurements with the ERCSC flux densities at 30, 44 and 70 GHz. The VLA 22 and 43\,GHz measurements were used to interpolate the VLA flux densities to the center frequency of \Planck 30\,GHz band. In the case of 44 GHz, the VLA measurements centered at 43.34 GHz and the \Planck measurements centered at 44.1 GHz were directly compared without any correction. We performed least squares fits to the data in the linear regime. The agreement at 30 GHz (28.5 GHz to be precise) is very good, with the measured slope equal to 1.08 and an intercept of 0.08. At 44 GHz, the agreement is also reasonably good, with a measured slope of 1.01 and an intercept of 0.24. 

At 70 GHz, we {\it extrapolate} the VLA measurements from 43 GHz using the 22--43\,GHz spectral index. As shown in Figure~\ref{fig_30GHz}, the extrapolated VLA values differ from the \Planck values by $\sim$20\,\%. 
There is strong evidence, presented in the companion paper \citep{Planck2011-6.1}, that the spectral index of radio sources detected by \Planck steepens at frequencies above
70 GHz or perhaps 44 GHz for some sources. Thus some of our extrapolated VLA 70\,GHz measurements may be biased high. Green Bank Telescope (GBT) 90\,GHz measurements of a sample of VLA sources \citep{sajina2011}, show that spectral curvature indeed plays a role. Among our VLA sample, 2 sources have 90\,GHz measurements. Replacing the extrapolated 70\,GHz values with interpolated ones from VLA 44 GHz and GBT 90 GHz changes the slope of the best fit line to be closer to unity, from 0.78 to 0.86. We emphasise that the 90GHz observations were not made simultaneously with the \Planck measurements, so they are less well controlled for variability than the quasi-simultaneous VLA (or EVLA) observations.
 
We also examined the median ratios of the \Planck measurements and the extrapolated VLA flux densities.  These are respectively 1.08\,$\pm$\,0.04, 1.12\,$\pm$\,0.07 and 0.99\,$\pm$\,0.06 for the 30, 44 and 70\,GHz bands. These median values have the advantage over the linear fits in that they effectively force a zero intercept and mitigate the effect of outliers. The median ratios confirm the good agreement at 30 and 44 GHz and show that the agreement between our VLA and \Planck flux densities at 70 GHz is better than implied by the linear fit. There is still, however, substantial scatter with the standard deviations of these ratios being 0.22, 0.38 and 0.29 for the 30, 44 and 70\,GHz bands, respectively. The scatter is likely dominated by the intrinsic variability of our sources.

\section{Extreme radio sources}\label{sec:ers}

In this section, we present several examples of sources illustrating the broad range of high frequency spectral behavior in bright, extragalactic radio sources as observed by \Planck\!\!. This is not intended as an exhaustive list of extreme sources or an extensive discussion to explore the phenomenology of these sources, but rather a sample from the rich data set provided in the ERCSC. Here we employ both \Planck and ground-based observations, using the latter to extend the source SEDs to lower frequencies. In some cases,  the ground-based observations were made at roughly the same time as \Planck observed the source, as mentioned in \S\,\ref{sec:observation}. Archival data were used when such nearly simultaneous observations were not available.

\subsection{Peaked spectrum sources}\label{sec:peaked}

The term Gigahertz-Peaked-Spectrum (GPS) sources, in principle, refers to a morphological type of the radio spectra, namely, to a spectral index $\alpha>0$ for $\nu<\nu_{\rm p}$, and $\alpha<0$ for $\nu>\nu_{\rm p}$ with a peak frequency $\nu_{\rm p}$ in the GHz regime. The GPS phenomenon was originally thought to be associated with compact, putatively very young radio sources \citep{odea1998}, and in fact many examples for this association have been found \citep{conway2002}. Recent research, however, has shown that a large fraction of sources with GPS features are not of this type, but associated with compact, beamed jet sources, commonly identified as blazars. These two classes produce slightly different spectra: the former shows a narrow peak, while the latter is typically broadly peaked. To date, however,  the only secure method to distinguish between them is to employ VLBI observations, revealing the source morphology \citep{bolton2006,vollmer2008}. In this paper, we use the term ``GPS sources'' purely phenomenologically, and discuss in the following sections the different, known, classes of sources for which we have examples in the ERCSC and also a set of ERCSC sources that are candidate GPS sources.

\begin{figure}
\centering
\includegraphics[scale=0.5]{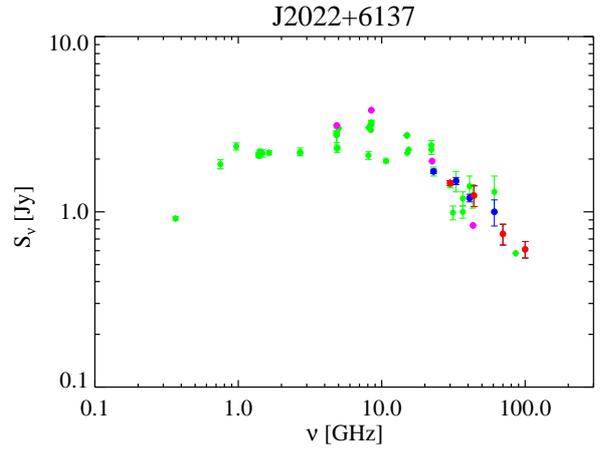}
\caption{SED of a known CSO source, J2022+6137. \Planck measurements are shown in red, {\it WMAP}-7yr flux densities in blue, our new VLA measurements in pink and archival data obtained from NED in green.}
              \label{J2022+6137}
    \end{figure}

\begin{figure*}
\centering
\begin{tabular}{cc}
\includegraphics[scale=0.45]{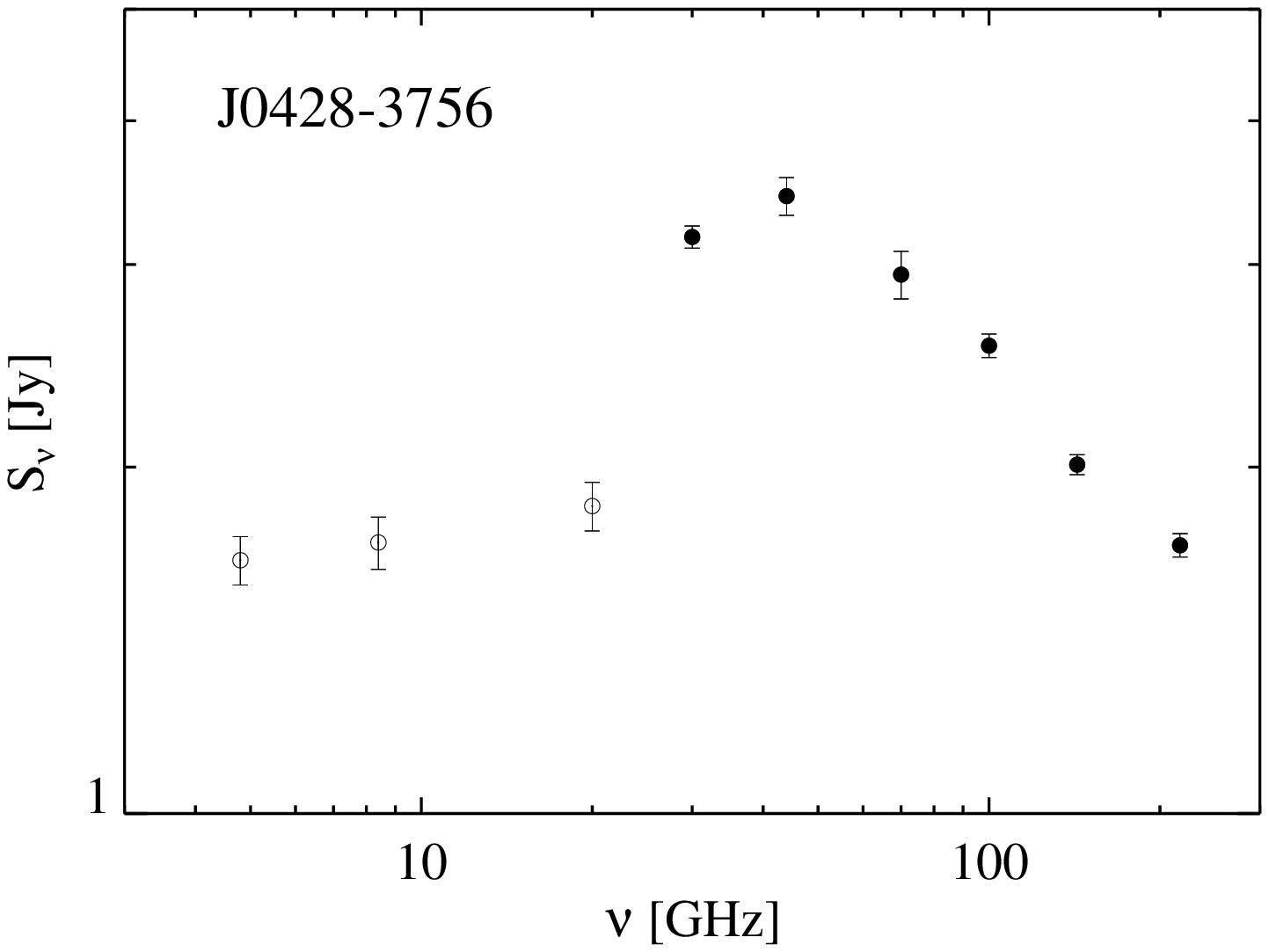} &
\includegraphics[scale=0.45]{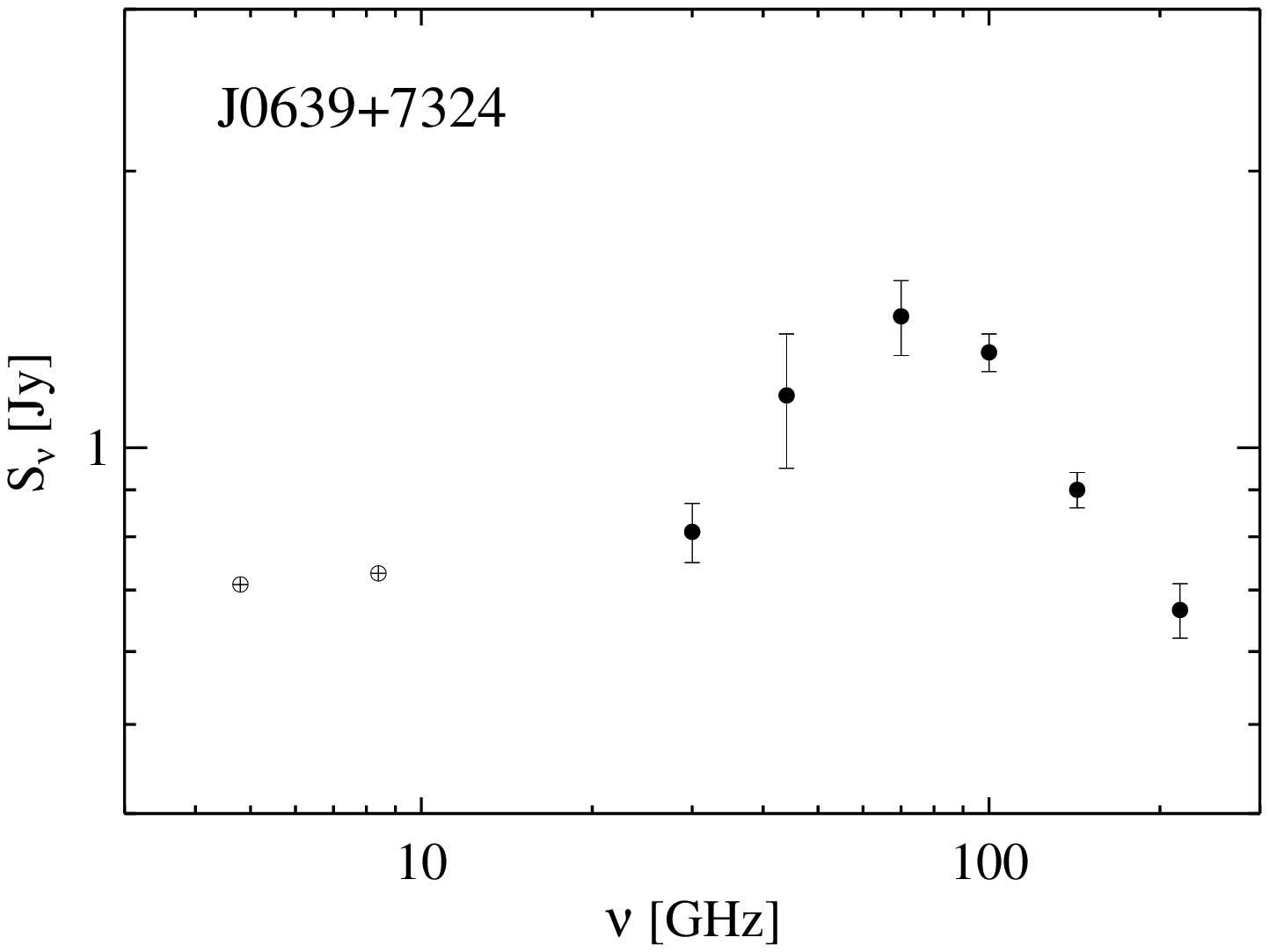} \\
\includegraphics[scale=0.45]{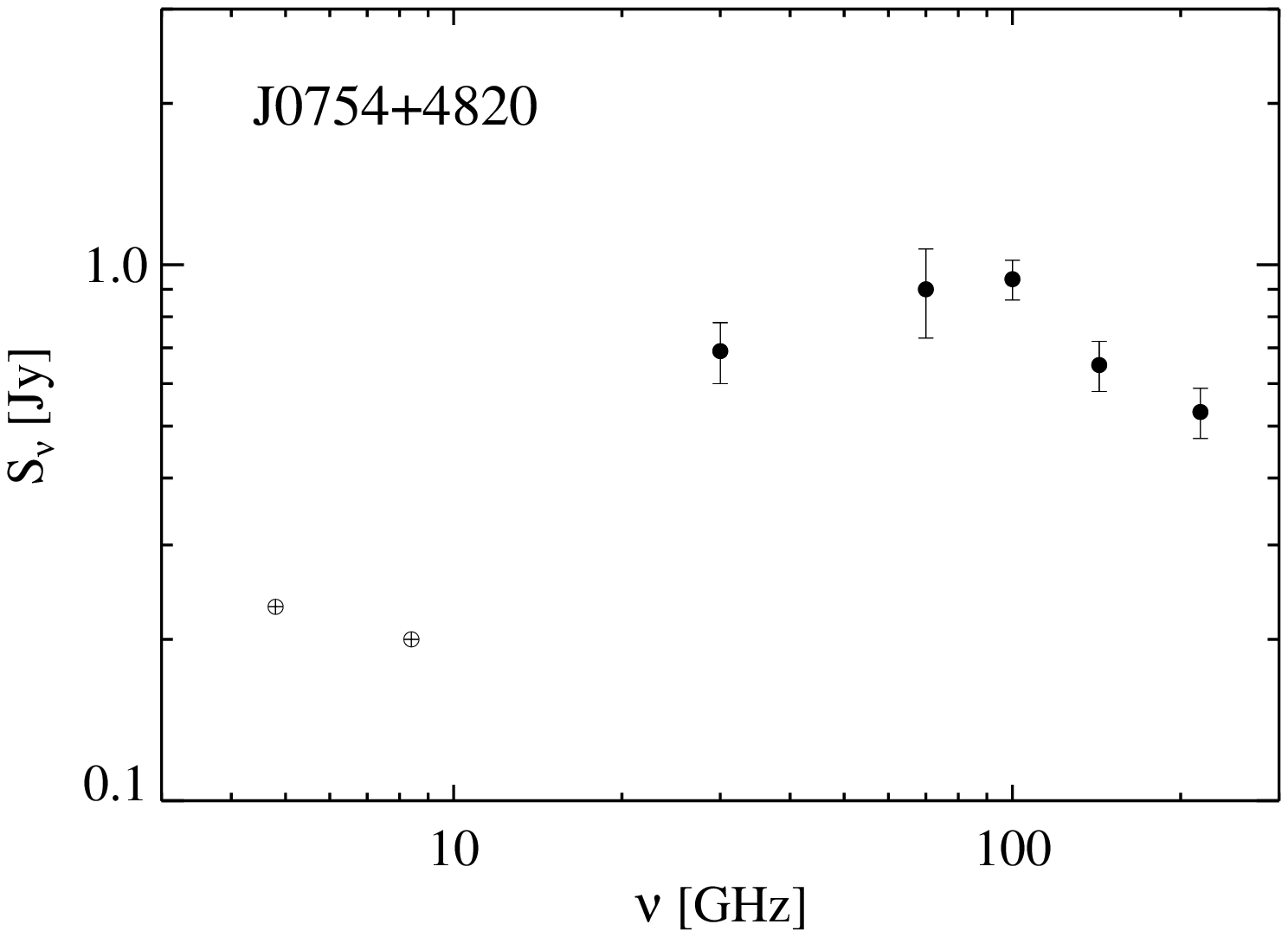} &
\includegraphics[scale=0.45]{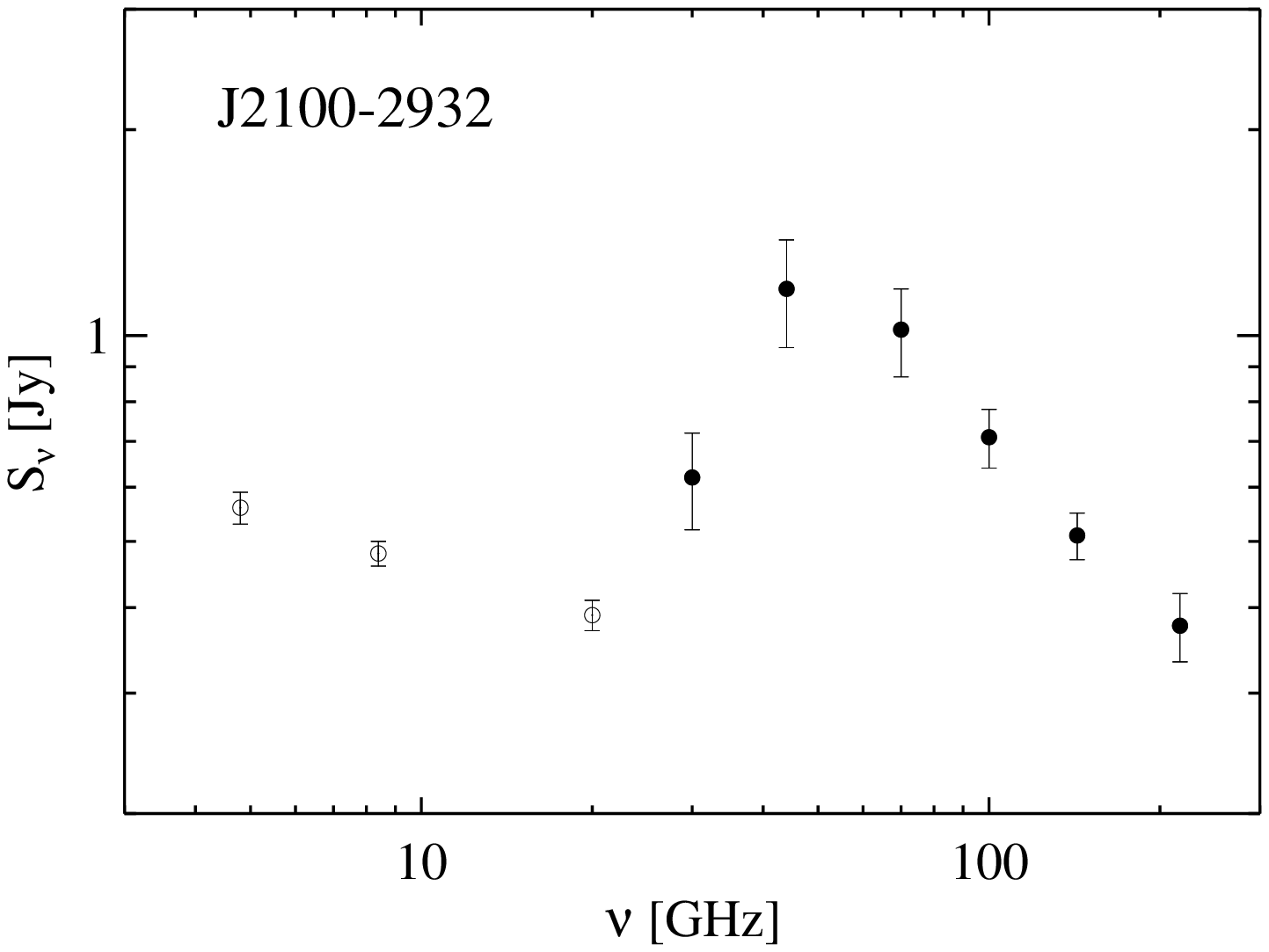} 
\end{tabular}
\caption{SEDs of sources that show spectral peak in the \Planck bands.  ERCSC data are shown in filled circles and low frequency archival data are shown in open circles.}
              \label{fig_hfp}%
    \end{figure*}

\subsubsection{ERCSC spectra of known Compact Symmetric Objects}

One class of GPS sources is Compact Symmetric Objects (CSO), thought to be either very young or very recently activated radio galaxies \citep{owsianik1998}. Mapped with VLBI resolution, these sources show a typical, symmetric radio-double morphology, but with linear extensions of 10 pc or less. They produce GPS type spectra by essentially single-zone synchrotron emission with synchrotron self-absorption causing a spectral turn-over at $\nu > 1$ GHz \citep{odea1998}. Unlike blazars, their emission is most likely not Doppler boosted, and they show no hint of fast variability, although their interpretation as young radio galaxies suggests that a spectral evolution over time scales of decades may be possible.  There is also a well-established connection between the peak frequency of a CSO-type GPS source and its linear extent:  $d\sim 100\nu^{-1}_{\rm p}\,$pc, with $\nu_{\rm p}$ in GHz. Thus sources with peak frequencies in the \Planck regime would point to objects at most a few parsecs in linear size. Previously, populations of sources peaking above an observed frequency of 10 GHz (in the observer's frame) might have been strongly underrepresented because they are relatively faint at the low frequencies where most large surveys have been made. \Planck opens the possibility of detecting such ``extreme GPS'' sources, or ``high frequency peakers" in the nomenclature of \citet{dallacasa2000}.  In addition, \Planck allows the examination of the spectral decline of GPS sources at frequencies far above the peak frequency where optically thin synchrotron radiation is thought to dominate the emission. In Figure \ref{J2022+6137}, we show the spectrum of one known CSO source, J2022+6137 (B2021+614 from \citealt{conway2002}). \Planck data have been combined with archival data as CSO sources are not expected to be variable.

\subsubsection{Additional high frequency peakers in ERCSC}

As mentioned above, \Planck observations make possible the detection and confirmation of rare, bright, GPS sources with very high peak frequencies. We provide in Table~\ref{table_candi_gps} a list of potential GPS sources that show convex SEDs constructed using the 30 to 143~GHz ERCSC data, along with the archival data at 20, 8.6 and 4.8 GHz \citep{murphy2010, healey2007} if available. We restricted our search to $|\,b\,|$\,$>$\,10$\degr$ sources. While all of the tabulated sources display spectral peaks in the \Planck frequency range, there is no guarantee that they are particularly young and compact CSO sources. Indeed, some of these bright sources have been previously studied, and most of them appear to be variable flat spectrum sources (see the notes in Table~\ref{table_candi_gps} and \citealt{torniainen2005}). We discuss sources of this kind in more detail in \S\,\ref{M-C-spec}.

\begin{figure}
\centering
\includegraphics[scale=0.45]{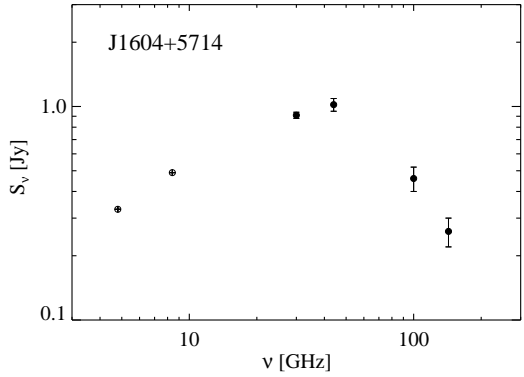}
\caption{With the additional \Planck data, source J1604+5714 is a newly disclosed GPS source candidate that previously showed an inverted spectrum at low frequencies. ERCSC data are shown in filled circles and low frequency archival data are shown in open circles. }
              \label{fig:J1604+5714}
    \end{figure}

Figure~\ref{fig_hfp} shows four sources that are quite undistinguished in low frequency catalogues, but reveal peaks in the \Planck frequency range. Note that sources J0639+7324, J0754+4820 and J2100--2932 all present peaks at 70~GHz or higher. The recorded redshift of J0639+7324 is 1.85; thus the rest frame frequency of the emission peak is extremely high at $\sim$200~GHz. In addition, \Planck data also helped to disclose the GPS-like spectrum for some sources that show inverted spectra at lower frequencies, as in the case of J1604+5714 (Figure \ref{fig:J1604+5714}).

\begin{figure}
   \centering
   \begin{tabular}{c}
\includegraphics[scale=0.5]{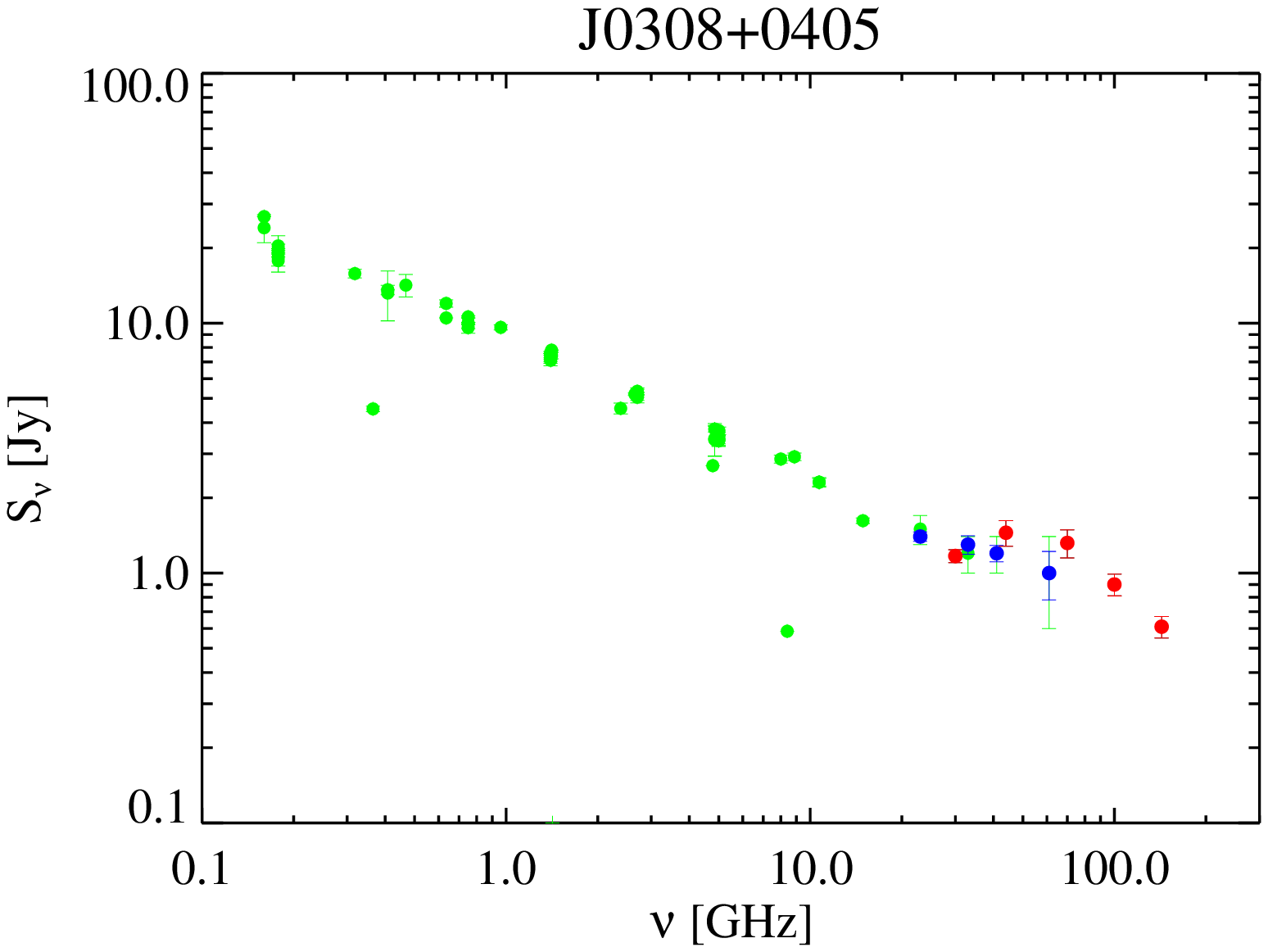}\\
\includegraphics[scale=0.5]{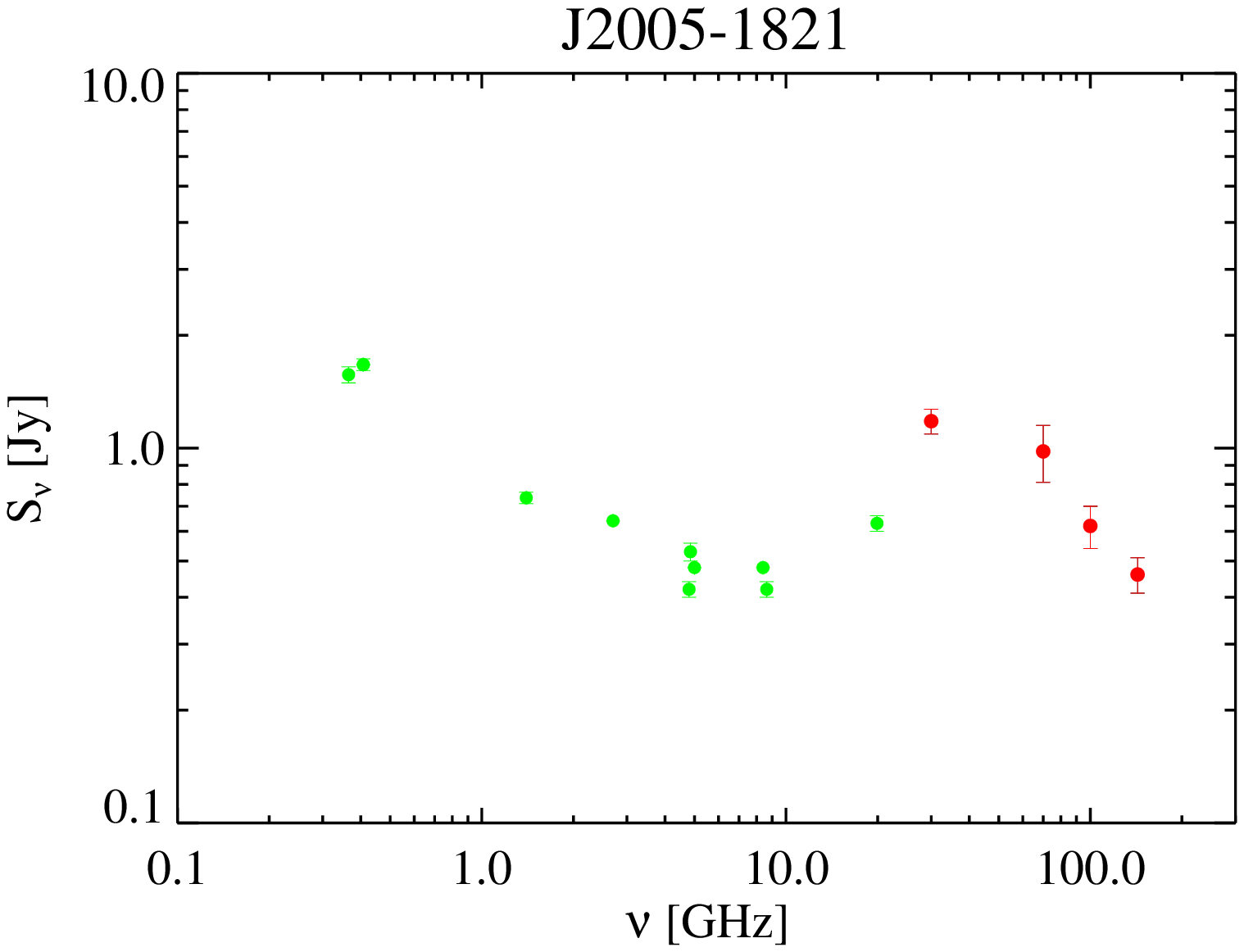}
\end{tabular}
   \caption{SED plots of J0308+0405 (upper) and J2005--1821 (lower). \Planck data are in red, {\it WMAP}-7yr data (if available) are in blue, archival data obtained from NED are in green.}
 \label{fig:2except}
    \end{figure}

As already mentioned, it is not possible from spectral information alone to identify high frequency peakers in the ERCSC with a new, very compact, radio source population. In fact, all but 5 sources in Table~\ref{table_candi_gps} can be identified with known blazars in the Roma-BZCAT blazar catalogue \citep{massaro2008}. Out of the five exceptions, 3 are further associated with blazars in the CGRaBS blazar catalogue \citep{healey2008}. The last two sources are identified as NGC1218 (J0308+0405) and HB89 2002--185 (J2005--1821) in NED. The SEDs of these two sources are shown in Figure \ref{fig:2except}. NGC1218 has a typical steep spectrum of radio galaxy up to  20 GHz, then a bump is seen at $\sim$40 GHz. Since WMAP (in the 7 year co-added map) did not catch this feature whereas \Planck did (during its 1.6 sky surveys), this peak is clearly caused by variability, suggesting some blazar like features in the galaxy, likely from close to the center. The spectral feature of HB89 2002-185 is evidently similar to that of source J2100--2932 in Figure \ref{fig_hfp} which has been identified as a flaring blazar. Therefore we suggest that this source could very well be a blazar. Further observations are needed to confirm this assumption.   Our finding that the peaked spectra in many bright sources is the result of flaring is consistent with earlier discussions by \citet{tornikoski2001,torniainen2005, bolton2006}.

\begin{landscape}
\begin{table}
\caption{GPS source candidates in the ERCSC.}           
\label{table_candi_gps}
\begin{tabular}{M CrC CCC CCC CCN}        
\hline\hline                 
\footnotesize
Name  & RA & {\hfill Dec\hfill}  & $S_{143~GHz}$ & $S_{100~GHz}$  & $S_{70~GHz}$ & $S_{44~GHz}$ & $S_{30~GHz}$ & $S_{20~GHz}$ & $S_{8.6~GHz}$ & $S_{4.8~GHz}$    & Redshift & Notes \\
 &  &  & [Jy] & [Jy]  & [Jy] &  [Jy] & [Jy]  & [Jy] & [Jy] & [Jy]  &  & \\
\hline\\

 J0010+1058  &   00:10:35.5  &   10:58:34  &        1.40$\pm$0.05  &        1.85$\pm$0.09  &        2.15$\pm$0.20  &        2.77$\pm$0.18  &        2.44$\pm$0.08  &        ...  &        0.25  &        0.44  & 0.089338 &  B   \\
 J0051--0651  &    00:51:14.9  &  --06:51:00  &        1.07$\pm$0.05  &        1.30$\pm$0.09  &        1.55$\pm$0.20  &        1.59$\pm$0.21  &        1.30$\pm$0.09  &        1.28$\pm$0.06  &        0.76  &        0.84  & 1.975000  & B \\
 J0118--2140  &    01:18:54.2  &  --21:40:55  &        0.76$\pm$0.04  &        0.86$\pm$0.07  &        1.34$\pm$0.16  &        ...  &        1.16$\pm$0.07  &        0.88$\pm$0.06  &        0.91$\pm$0.05  &        0.88$\pm$0.04   & 1.161000 & B \\
 J0121+1149  &   01:21:45.6  &   11:49:37  &        0.82$\pm$0.07  &        1.09$\pm$0.10  &        0.99$\pm$0.21  &        1.84$\pm$0.24  &        1.92$\pm$0.09  &        ...  &        1.87  &        1.13  & 0.570000 & B \\
 J0137--2430  &    01:37:41.8  &  --24:30:07  &        1.81$\pm$0.06  &        2.13$\pm$0.08  &        2.89$\pm$0.17  &        2.54$\pm$0.16  &        2.49$\pm$0.08  &        1.21$\pm$0.08  &        1.41$\pm$0.07  &        1.56$\pm$0.08  & 0.837000 & B  \\
 J0253--5441  &    02:53:32.6  &  --54:41:17  &        1.32$\pm$0.05  &        1.78$\pm$0.07  &        1.69$\pm$0.15  &        2.68$\pm$0.15  &        2.38$\pm$0.07  &        1.93$\pm$0.10  &        1.65$\pm$0.08  &        1.43$\pm$0.07 & 0.539000 & B, V  \\
 J0308+0405  &   03:08:22.8  &   04:05:53  &        0.61$\pm$0.06  &        0.90$\pm$0.09  &        1.32$\pm$0.17  &        1.45$\pm$0.17  &        1.17$\pm$0.07  &        ...  &        ...  &        ... & 0.028653 &  \\
 J0334--4008  &    03:34:20.6  &  --40:08:53  &        1.41$\pm$0.04  &        1.72$\pm$0.06  &        1.89$\pm$0.13  &        2.36$\pm$0.11  &        2.35$\pm$0.06  &        1.27$\pm$0.06  &        1.12$\pm$0.06  &        0.94$\pm$0.05  & 1.445000 & B, V \\
 J0423--0120  &    04:23:16.3  &  --01:20:31  &        5.27$\pm$0.07  &        6.51$\pm$0.08  &        7.24$\pm$0.19  &        7.85$\pm$0.17  &        8.65$\pm$0.09  &        6.00$\pm$0.29  &        2.41  &        4.36  & 0.914000 & B, V \\
 J0428--3756  &    04:28:40.8  &  --37:56:13  &        2.01$\pm$0.04  &        2.55$\pm$0.06  &        2.94$\pm$0.14  &        3.44$\pm$0.13  &        3.17$\pm$0.07  &        1.85$\pm$0.09  &        1.72$\pm$0.09  &        1.66$\pm$0.08 & 1.110000 & B \\
 J0450--8101  &    04:50:40.1  &  --81:01:05  &        0.95$\pm$0.04  &        1.13$\pm$0.05  &        1.29$\pm$0.09  &        1.73$\pm$0.10  &        1.75$\pm$0.05  &        1.45$\pm$0.16  &        1.20$\pm$0.06  &        1.07$\pm$0.05  & 0.444000 & B \\
 J0455--4616  &    04:55:53.0  &  --46:16:30  &        1.02$\pm$0.07  &        1.31$\pm$0.09  &        1.79$\pm$0.15  &        2.01$\pm$0.12  &        2.63$\pm$0.06  &        4.16$\pm$0.21  &        3.61$\pm$0.18  &        2.61$\pm$0.13  & 0.852800 & B \\
 J0457+0640  &   04:57:14.6  &   06:40:41  &        0.55$\pm$0.05  &        0.83$\pm$0.08  &        1.09$\pm$0.18  &        ...  &        0.80$\pm$0.10  &        ...  &        0.43  &        0.62  & 0.405000 & B \\
 J0525--2336  &    05:25:04.8  &  --23:36:40  &        0.42$\pm$0.03  &        0.65$\pm$0.07  &        ...  &        1.37$\pm$0.16  &        1.13$\pm$0.06  &        0.79$\pm$0.05  &        0.89$\pm$0.05  &        0.81$\pm$0.04  & 3.100000 & B \\
 J0526--4831  &    05:26:00.7  &  --48:31:08  &        0.39$\pm$0.06  &        0.65$\pm$0.07  &        ...  &        1.96$\pm$0.12  &        1.30$\pm$0.06  &        0.29$\pm$0.01  &        0.34$\pm$0.02  &        0.37$\pm$0.02  & ... & B \\
 J0534--6107  &    05:34:36.7  &  --61:07:26  &        ...  &        0.52$\pm$0.03  &        ...  &        0.83$\pm$0.10  &        0.79$\pm$0.04  &        0.44$\pm$0.02  &        0.45$\pm$0.02  &        0.44$\pm$0.02 & 1.997000 & B \\
 J0538--4405  &    05:38:54.7  &  --44:05:13  &        6.67$\pm$0.07  &        8.45$\pm$0.06  &        9.32$\pm$0.13  &        8.46$\pm$0.13  &        9.00$\pm$0.07  &        5.29$\pm$0.25  &        4.23$\pm$0.21  &        3.80$\pm$0.19  & 0.894000 & B, V \\
 J0540--5418  &    05:40:57.1  &  --54:18:29  &        0.43$\pm$0.03  &        0.72$\pm$0.05  &        0.97$\pm$0.10  &        1.10$\pm$0.08  &        1.06$\pm$0.05  &        1.13$\pm$0.06  &        0.99$\pm$0.05  &        0.71$\pm$0.04  & 1.185000 & B \\
 J0550--5732  &    05:50:07.0  &  --57:32:06  &        0.60$\pm$0.02  &        0.74$\pm$0.04  &        1.01$\pm$0.07  &        1.54$\pm$0.08  &        1.45$\pm$0.05  &        1.00$\pm$0.05  &        1.03$\pm$0.05  &        0.93$\pm$0.05  & 2.001000 & B \\
 J0622--6435  &    06:22:55.9  &  --64:35:17  &        0.53$\pm$0.03  &        0.76$\pm$0.03  &        0.86$\pm$0.06  &        1.07$\pm$0.07  &        0.99$\pm$0.03  &        0.72$\pm$0.04  &        0.84$\pm$0.04  &        0.87$\pm$0.04  & 0.128889 & B \\
 J0639+7324  &   06:39:10.1 &    73:24:29  &        0.90$\pm$0.04  &        1.27$\pm$0.06  &        1.39$\pm$0.13  &        1.14$\pm$0.19  &        0.81$\pm$0.06  &        ...  &        0.73  &        0.71  & 1.850000 & B \\
 J0646+4451  &   06:46:31.7  &   44:51:58  &        0.98$\pm$0.06  &        1.27$\pm$0.09  &        1.86$\pm$0.19  &        1.75$\pm$0.18  &        2.62$\pm$0.08  &        ...  &        2.26  &        1.22  & 3.396000 & B, V \\
 J0701--4633  &    07:01:44.2  &  --46:33:29  &        0.66$\pm$0.03  &        0.86$\pm$0.04  &        1.02$\pm$0.09  &        1.34$\pm$0.08  &        1.05$\pm$0.05  &        1.07$\pm$0.05  &        0.86$\pm$0.04  &        0.55$\pm$0.03  & 0.822000 & B \\
 J0717+4539  &   07:17:50.2  &   45:39:47  &        0.38$\pm$0.06  &        ...  &        0.98$\pm$0.14  &        ...  &        0.95$\pm$0.08  &        ...  &        0.56  &        0.47   & 0.940000 & B \\
 J0750+1231  &   07:50:50.4  &   12:31:34  &        2.04$\pm$0.03  &        2.69$\pm$0.07  &        3.18$\pm$0.15  &        3.90$\pm$0.15  &        4.54$\pm$0.08  &        ...  &        1.97  &        1.24  & 0.889000 & B, V \\
 J0754+4820  &   07:54:48.0  &   48:20:17  &        0.65$\pm$0.07  &        0.94$\pm$0.08  &        0.90$\pm$0.17  &        ...  &        0.69$\pm$0.09  &        ...  &        0.20  &        0.23  & 0.377142 & B \\
 J0836--2237  &    08:36:52.3  &  --22:37:01  &        0.39$\pm$0.06  &        0.48$\pm$0.08  &        0.80$\pm$0.12  &        ...  &        1.40$\pm$0.07  &        0.46$\pm$0.03  &        0.35$\pm$0.02  &        0.31$\pm$0.01  & 0.837000 & B \\
 J0841+7053  &   08:41:12.2  &   70:53:24  &        2.10$\pm$0.04  &        2.91$\pm$0.07  &        3.56$\pm$0.14  &        3.17$\pm$0.12  &        2.82$\pm$0.06  &        ...  &        1.75  &        2.34  & 2.172000 & B \\
 J0847--0659  &    08:47:52.6  &  --06:59:02  &        0.54$\pm$0.05  &        0.92$\pm$0.06  &        0.99$\pm$0.14  &        ...  &        0.91$\pm$0.06  &        0.73$\pm$0.04  &        0.47  &        0.44 & ... & B  \\
 J0854+2006  &   08:54:43.9  &   20:06:04  &        5.40$\pm$0.09  &        6.26$\pm$0.08  &        6.49$\pm$0.16  &        7.34$\pm$0.16  &        7.16$\pm$0.08  &        ...  &        3.41  &        2.91   & 0.306000 & B \\
 J0920+4441  &   09:20:58.8  &   44:41:13  &        1.43$\pm$0.06  &        1.39$\pm$0.07  &        1.81$\pm$0.15  &        2.15$\pm$0.15  &        2.29$\pm$0.08  &        ...  &        1.37  &        1.09   & 2.189910 & B \\
 J0923+2817  &   09:23:54.0  &   28:17:13  &        0.78$\pm$0.05  &        0.94$\pm$0.08  &        ...  &        1.44$\pm$0.16  &        1.18$\pm$0.07  &        ...  &        0.22  &        0.35  & 0.743902 & B \\
 J1102+7226  &   11:02:04.1  &   72:26:42  &        0.42$\pm$0.04  &        0.65$\pm$0.05  &        0.92$\pm$0.11  &        1.62$\pm$0.12  &        1.51$\pm$0.06  &        ...  &        0.37  &        0.86  & 1.460000 & B \\
 J1147+4000  &   11:47:01.0  &   40:00:36  &        0.59$\pm$0.04  &        0.82$\pm$0.06  &        0.91$\pm$0.11  &        1.31$\pm$0.12  &        1.14$\pm$0.05  &        ...  &        0.58  &        0.84  & 1.088000 & B \\
 J1152--0842  &    11:52:16.1  &  --08:42:40  &        0.75$\pm$0.05  &        1.13$\pm$0.09  &        1.25$\pm$0.21  &        ...  &        1.41$\pm$0.08  &        0.67$\pm$0.03  &        0.56  &        0.74  & 2.370000 & B \\
 J1153+4930  &   11:53:18.0  &   49:30:14  &        1.16$\pm$0.06  &        1.53$\pm$0.06  &        1.60$\pm$0.12  &        1.73$\pm$0.11  &        1.80$\pm$0.05  &        ...  &        0.44  &        0.17  & 0.333981 & B \\
 J1310+3221  &   13:10:26.2 &    32:21:50  &        1.30$\pm$0.06  &        1.56$\pm$0.09  &        2.27$\pm$0.18  &        2.96$\pm$0.16  &        3.58$\pm$0.08  &        ...  &        3.03  &        1.45   & 0.996000 & B, V \\
 J1332--0509  &    13:32:09.6  &  --05:09:14  &        0.84$\pm$0.04  &        1.00$\pm$0.09  &        ...  &        1.66$\pm$0.32  &        1.30$\pm$0.10  &        0.68$\pm$0.03  &        0.63  &        0.47 & 2.150000 & B  \\
 J1337--1256  &    13:37:40.1 &   --12:56:46  &        2.77$\pm$0.06  &        3.40$\pm$0.10  &        3.79$\pm$0.23  &        4.48$\pm$0.24  &        4.32$\pm$0.10  &        6.06$\pm$0.29  &        5.00  &        2.84  & 0.539000 & B, V\\
 J1457--3543  &    14:57:24.7  &  --35:43:19  &        0.51$\pm$0.04  &        0.76$\pm$0.07  &        0.84$\pm$0.16  &        1.59$\pm$0.17  &        1.18$\pm$0.08  &        0.90$\pm$0.05  &        0.88$\pm$0.04  &        0.93$\pm$0.05  & 1.424000 & B \\
 J1506+4237  &   15:06:52.1 &    42:37:26  &        0.41$\pm$0.04  &        0.60$\pm$0.07  &        ...  &        1.00$\pm$0.18  &        0.58$\pm$0.06  &        ... &        0.41  &        0.41 & 0.587000 & B \\
 J1516+0014  &   15:16:37.0  &   00:14:06  &        1.00$\pm$0.05  &        1.12$\pm$0.09  &        1.34$\pm$0.18  &        2.29$\pm$0.20  &        1.71$\pm$0.08  &        ...  &        0.96  &        1.59  & 0.052489 & B \\
 J1549+0237  &   15:49:28.1  &   02:37:34  &        1.14$\pm$0.04  &        1.29$\pm$0.08  &        1.50$\pm$0.17  &        2.22$\pm$0.16  &        1.78$\pm$0.07  &        ...  &        0.92  &        1.11  & 0.414414 & B, V \\
 J1553+1255  &   15:53:06.0  &   12:55:08  &        0.38$\pm$0.05  &        0.60$\pm$0.07  &        ...  &        1.57$\pm$0.23  &        0.96$\pm$0.07  &        ...  &        0.41  &        0.74  & 1.290000 & B \\
 J1604+5714  &   16:04:45.8  &   57:14:35  &        0.26$\pm$0.04  &        0.46$\pm$0.06  &        ...  &        1.02$\pm$0.07  &        0.91$\pm$0.03  &        ...  &        0.49  &        0.33  & 0.720000 & B \\
 J1635+3807  &   16:35:16.3  &   38:07:34  &        2.66$\pm$0.05  &        3.51$\pm$0.08  &        3.86$\pm$0.13  &        4.57$\pm$0.15  &        3.96$\pm$0.07  &        ...  &        2.40  &        3.22  & 1.813570 & B \\

\hline                                  
\end{tabular}
\medskip
\end{table}
\addtocounter{table}{-1}
\end{landscape}

\newpage

\begin{landscape}

~

\begin{table}
\caption{GPS source candidates in the ERCSC---{\it Continued}}
\begin{tabular}{M CrC CCC CCC CCN}        
\hline\hline               
\footnotesize

Name  & RA & {\hfill Dec\hfill}  & $S_{143~GHz}$ & $S_{100~GHz}$  & $S_{70~GHz}$ & $S_{44~GHz}$ & $S_{30~GHz}$ & $S_{20~GHz}$ & $S_{8.6~GHz}$ & $S_{4.8~GHz}$    & Redshift & Notes \\
 & &  & [Jy] & [Jy]  & [Jy] &  [Jy] & [Jy]  & [Jy] & [Jy] & [Jy]  &  & \\

\hline \\

 J1644--7716  &   16:44:32.4   &  --77:16:05  &        0.56$\pm$0.05  &        0.77$\pm$0.08  &        0.87$\pm$0.16  &        ...  &        0.90$\pm$0.08  &        0.40$\pm$0.02  &        0.24$\pm$0.01  &        0.25$\pm$0.01 & 0.042700 & B \\
 J1703--6213  &   17:03:39.4  &   --62:13:08  &        0.96$\pm$0.04  &        1.33$\pm$0.07  &        1.48$\pm$0.16  &        2.67$\pm$0.14  &        2.24$\pm$0.08  &        1.05$\pm$0.05  &        1.03$\pm$0.05  &        1.04$\pm$0.05  & ... & B \\
 J1727+4530  &  17:27:33.6 &     45:30:14 &        0.75$\pm$0.03  &        0.94$\pm$0.05  &        1.10$\pm$0.11  &        1.19$\pm$0.14  &        1.38$\pm$0.05  &        ...  &        1.36  &        0.94 & 0.717000 & B \\
 J1927+7358  &  19:27:51.6 &     73:58:37  &        2.42$\pm$0.04  &        3.28$\pm$0.05  &        3.82$\pm$0.10  &        4.94$\pm$0.11  &        5.01$\pm$0.05  &        ...  &        3.70  &        3.63 & 0.302100 & B \\
 J2005--1821  &   20:05:15.6  &   --18:21:04  &        0.46$\pm$0.05  &        0.62$\pm$0.08  &        0.98$\pm$0.17  &        ...  &        1.18$\pm$0.09  &        0.60$\pm$0.03  &        0.42$\pm$0.02  &        0.40$\pm$0.02   & 0.868000 &  \\
 J2006+6424  &  20:06:17.8  &    64:24:07  &        0.36$\pm$0.07  &        0.49$\pm$0.08  &        0.78$\pm$0.13  &        1.14$\pm$0.11  &        1.13$\pm$0.06  &        ...  &        0.96  &        0.72  & 1.574000 & B \\
 J2009+7229  &  20:09:58.1  &    72:29:06  &        0.49$\pm$0.04  &        0.86$\pm$0.06  &        1.24$\pm$0.10  &        0.90$\pm$0.10  &        0.68$\pm$0.04  &        ...  &        0.79  &        0.91  & ... & B \\
 J2035--6845  &   20:35:25.0  &   --68:45:04  &        0.51$\pm$0.06  &        0.65$\pm$0.08  &        0.90$\pm$0.14  &        1.35$\pm$0.17  &        1.02$\pm$0.07  &        0.47$\pm$0.02  &        0.37$\pm$0.02  &        0.41$\pm$0.02  & 1.084000  & B \\
 J2100--2932  &   21:00:57.1  &   --29:32:02 &        0.51$\pm$0.04  &        0.71$\pm$0.07  &        1.02$\pm$0.15  &        1.17$\pm$0.21  &        0.62$\pm$0.10  &        0.39$\pm$0.02  &        0.48$\pm$0.02  &        0.56$\pm$0.03  & 1.492000  & B  \\
 J2126--4607  &   21:26:36.5 &    --46:07:41  &        0.44$\pm$0.06  &        0.69$\pm$0.08  &        ...  &        1.24$\pm$0.17  &        1.04$\pm$0.07  &        0.55$\pm$0.03  &        0.64$\pm$0.03  &        0.69$\pm$0.03 & 1.670000  & B \\
 J2139+1424  &  21:39:08.6  &    14:24:14  &        0.39$\pm$0.06  &        0.73$\pm$0.08  &        1.30$\pm$0.17  &        2.48$\pm$0.13  &        2.64$\pm$0.06  &        ...  &        2.27  &        1.07  & 2.427000 & B \\
 J2147--7536  &   21:47:06.7  &   --75:36:40  &        2.12$\pm$0.04  &        2.48$\pm$0.06  &        3.09$\pm$0.14  &        3.37$\pm$0.12  &        2.93$\pm$0.06  &        0.49$\pm$0.03  &        0.86$\pm$0.04  &        1.01$\pm$0.05  & 1.139000 & B \\
 J2225+2119  &  22:25:36.0  &    21:19:23 &        1.10$\pm$0.05  &        1.19$\pm$0.08  &        1.28$\pm$0.12  &        2.16$\pm$0.15  &        1.87$\pm$0.06  &        ...  &        1.40  &        1.02  & 1.959000 & B \\
 J2230--3940  &   22:30:45.6  &   --39:40:19 &        0.45$\pm$0.06  &        0.72$\pm$0.07  &        1.06$\pm$0.14  &        ...  &        0.90$\pm$0.06  &        ...  &        0.53  &        0.56  & 0.318049 & B \\
 J2235--4836  &   22:35:01.7  &   --48:36:32  &        0.62$\pm$0.04  &        0.93$\pm$0.05  &        1.07$\pm$0.12  &        1.19$\pm$0.14  &        1.32$\pm$0.06  &        1.99$\pm$0.09  &        2.07$\pm$0.10  &        1.21$\pm$0.06  & 0.510000 & B  \\
 J2239--5701  &   22:39:16.3  &   --57:01:55 &        0.80$\pm$0.04  &        1.02$\pm$0.06  &        1.33$\pm$0.13  &        2.03$\pm$0.13  &        1.99$\pm$0.06  &        0.93$\pm$0.05  &        1.01$\pm$0.05  &        0.85$\pm$0.04 & ... & B \\
 J2253+1609  &  22:53:59.8  &    16:09:07  &       27.94$\pm$0.16  &       28.83$\pm$0.14  &       27.91$\pm$0.29  &       22.75$\pm$0.20  &       16.84$\pm$0.10  &        ...  &       10.38  &       14.47   & 0.859000 & B  \\

\hline                                 
\end{tabular}
\medskip

\begingroup
\rightskip36pt
\tablefoottext{a}{The 143 to 30\,GHz flux densities of these sources are from the ERCSC \citep{Planck2011-1.10}. The 20\,GHz flux density is from the AT20G catalogue \citep{murphy2010}. If a source is in the AT20G catalogue, its 8.6\,GHz and 4.8\,GHz flux densities are from the AT20G catalogue. If not, the 8.6 flux density (8.4 GHz to be accurate) and 4.8\,GHz flux densities are obtained from the CRATES catalogue \citep{healey2007}. Note that the CRATES catalogue does not provide flux uncertainties. The tabulated redshift values are from the NASA/IPAC Extragalactic Database (NED). } \\
\tablefoottext{b}{In the ``Notes'' column, ``B'' indicate sources in the Roma-BZCAT blazar catalogue \citep{massaro2008} and CGRaBS blazar catalogue \citep{healey2008}, ``V'' means these sources are found to be variable in \citet{torniainen2005}. } \\
\tablefoottext{c}{Source J2253+1609 is the well known quasar 3C454.3 that only shows GPS-like spectrum during strong flares \citep{rachen2010}. See also discussion in \S\,\ref{M-C-spec}.}
\endgroup
\end{table}
\end{landscape}

\subsubsection{Broad-peaked radio sources}\label{sec:broadpeak}

As noted above, many sources showing GPS-like features are radio sources with dominant compact jet emission, usually identified with blazars. At VLBI resolution, they generally appear as one-sided, compact, jet-like objects. The most likely explanation for the very flat, sometimes inverted, radio spectra of these sources is the overlay of synchrotron-self absorbed emission in a continuous jet (\citealt{marscher1977, marscher1985}). We may note at this point that the application of this mechanism to a standard Blandford-K\"onigl plasma jet \citep{blandford1979} yields an optically thick spectral index $\alpha$\,$\approx$\,$0.3{-}0.5$, depending on the electron spectral index \citep{marscher1985}.  The turnover to a completely optically thin spectral index $\alpha$\,$<$\,0, which is expected to happen somewhere in the GHz-THz regime from jet-size considerations, would then naturally produce a ``GPS type'' spectrum.  We may therefore conclude that GPS-type blazars agree better with the simplest model of compact, continuous jets than do ``typical'' blazars which show a low frequency spectral index  $\alpha$\,$\approx$\,0 \citep{Planck2011-6.3a}.

Most of these objects are known to be strongly variable, on time scales down to one day (usually on week-month scale in the millimeter-regime, however). Variability and one-sided VLBI morphology give strong arguments in favor of beaming, i.e., the emission is strongly Doppler boosted, making the measured flux of the source at a given frequency depend on the Doppler factor $D$ as $D^{3-\alpha}$. Therefore, small changes in the Doppler boosting, as  expected in helix-shaped or precessing jets, can lead to strong variability. The flux density changes would be expected to be achromatic, i.e., leaving the spectral shape unchanged. This can be distinguished from variability due to distortions in the jet, which is expected to be emphasised in a specific frequency range; such sources are discussed in \S\,\ref{M-C-spec}.          

\begin{figure*}
   \centering
   \begin{tabular}{cc}
\includegraphics[scale=0.5]{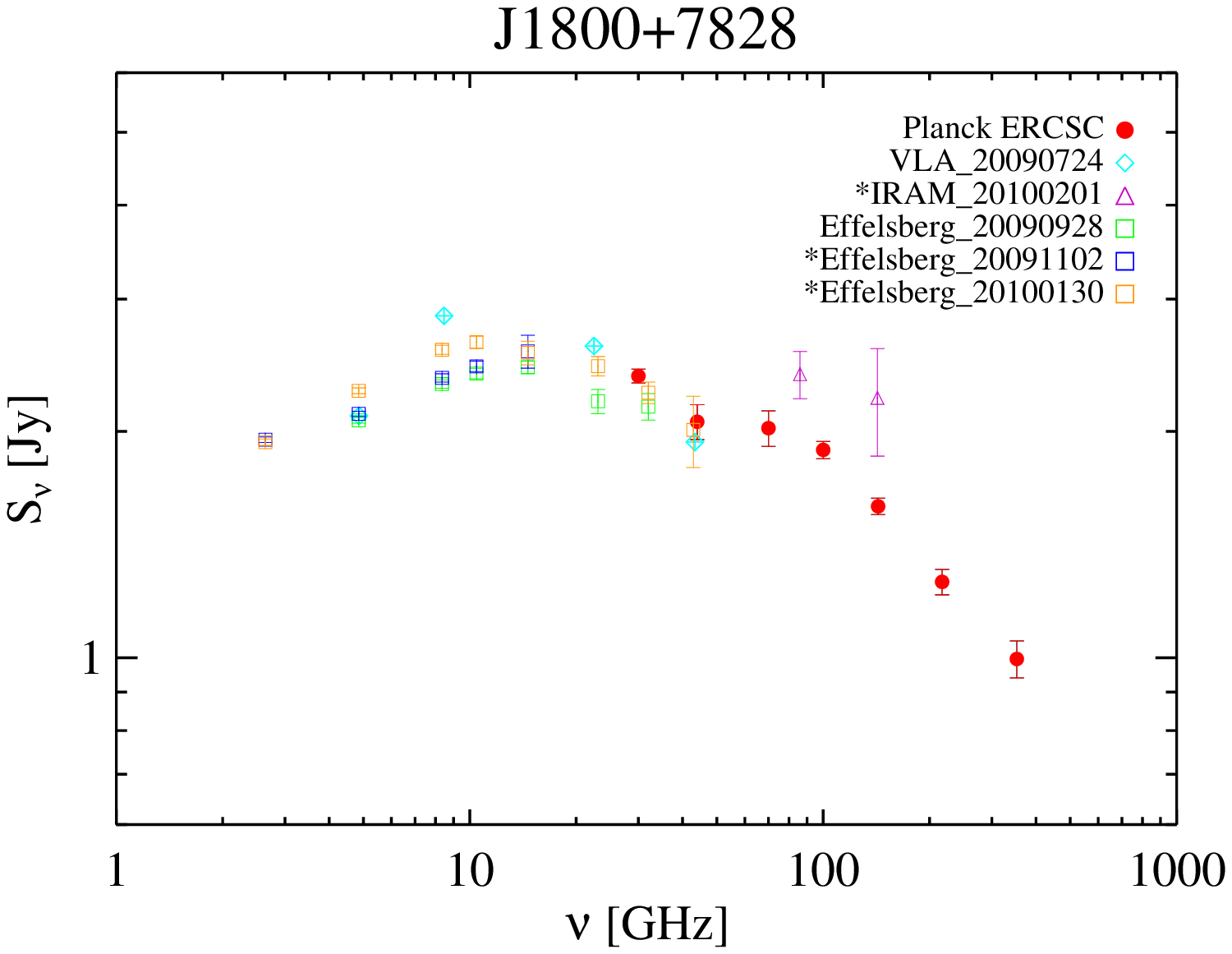} &
\includegraphics[scale=0.5]{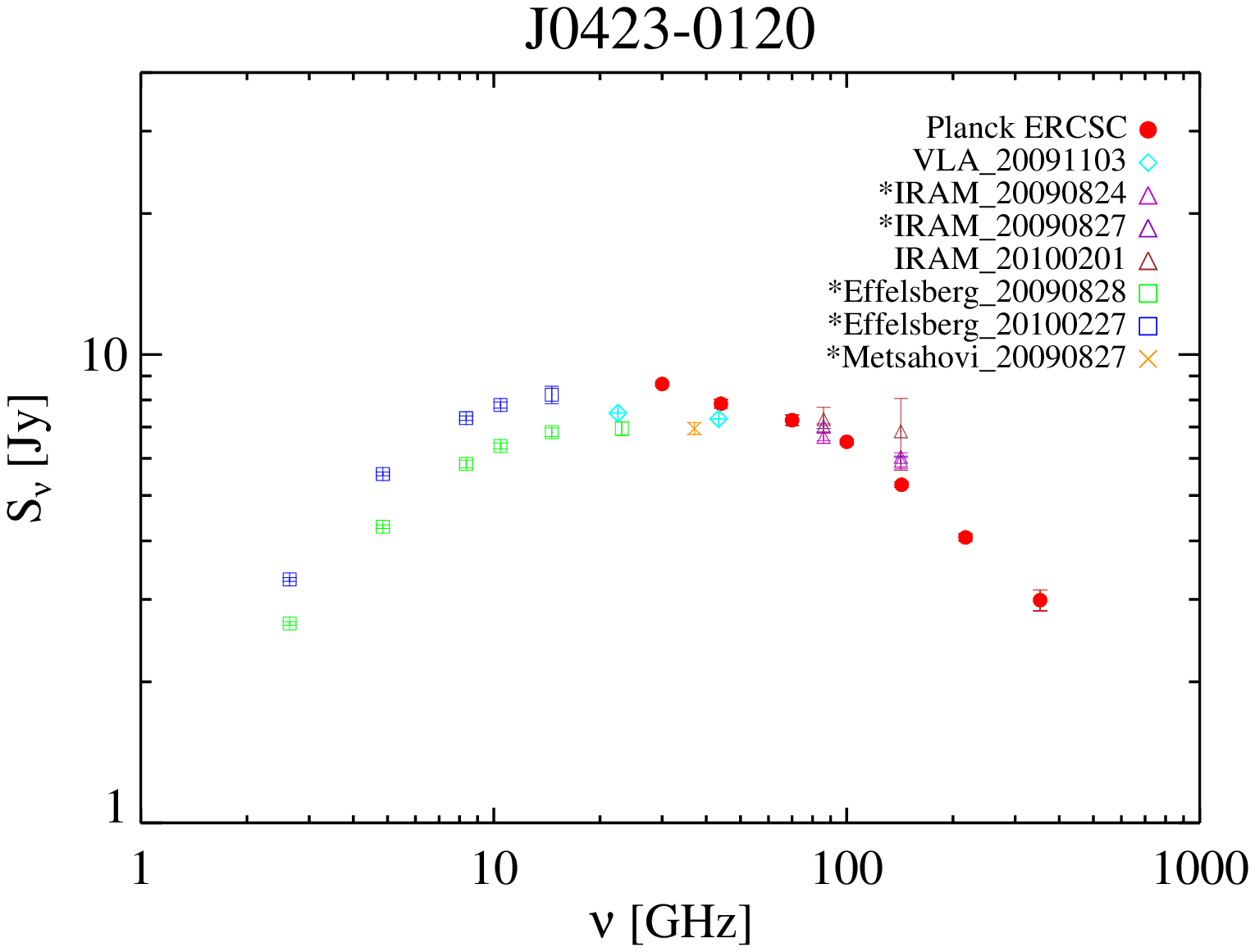}
\end{tabular}
   \caption{VLA, Mets\"ahovi, IRAM, Effelsberg and \Planck measurements of two known GPS blazars:
J1800+7828 (left) and J0423--0120 (right). While the former shows only small
signs of variability, the latter is clearly variable while preserving its
spectral shape, pointing to helical or pressing jet (see text). Asterisks
identify epochs within 10 days from \Planck observations at any of the LFI
channels. For both sources, the \Planck ERCSC data are a superposition of two
scans separated by 3 and 6 months, respectively.}
 \label{fig_1800+784}
    \end{figure*}

Figure~\ref{fig_1800+784} shows two examples of apparently achromatic variability: J1800+7828 shows a rather typical GPS blazar spectrum, with $\alpha$\,$\approx$\,0.3 below the peak frequency at about 10 GHz, and steepening to $\alpha$\,$\approx$\,$-0.17$ for $\nu$\,$\leq$\,100 GHz, with a further break to $\alpha$\,$\approx$\,$-0.5$ at still higher frequencies; The more dome-like spectrum of J0423--0120, with $\alpha$\,$\approx$\,1 at lower frequencies, points to a more uniform component producing the radiation. Supporting, ground-based, radio data from Effelsberg, Mets\"ahovi and the VLA further suggest significant, achromatic variability for this source, as would be expected from changes of the Doppler boosting in a helical or precessing jet. The high frequency spectrum has the same characteristics as J1800+7828. The temporarily very flat index $\alpha$\,$\approx$\,$-0.2$ after the peak may suggest an unusually flat electron spectrum, as discussed in \citet{Planck2011-6.3a}.

\subsection{Flat and multi-component spectra}\label{M-C-spec}

In our examination of radio sources in the ERCSC we found a large number with extended, flat, power law spectra.  For instance, the flux density of J2203+1725, as determined by \Planck and VLA observations, varies only between 0.99 and 1.18~Jy between frequencies of 4.9 and 100~GHz, and decreases only to 0.906~Jy at 143~GHz.  Many of these flat spectrum sources are the blazars discussed in  \citet{Planck2011-6.3a}. We also found several sources with prominently zig-zag or ``bumpy'' spectra. In some cases the variation in flux density from one \Planck band to the next was several times the associated errors. This could result from the superposition of emission from several components, as is probably the case in the examples shown in Figure \ref{fig_bumpy}. However, it is important to recall that \Planck\!\!'s multifrequency observations were not exactly simultaneous. A source could sweep through the \Planck beams at one frequency days before doing so at a neighboring frequency (depending on position, it takes 7-10 days for a source to drift entirely across the focal plane). Hence fast variability on time-scales of days can contribute to irregularities in its \Planck spectrum. We expect this effect to be smaller for the HFI frequencies, given the relatively closer packing in the focal plane. We discuss below the possible physical conditions for true multi-component sources, followed by examples of spectral artifacts caused by variability.

\subsubsection{Flat spectrum sources with a variable component}

Spectra resulting from superposed radiation of at least two components have become the standard model to explain flaring blazars; this is known as a ``shock-in-jet'' model 
\citep{marscher1985}. One component, the jet, produces a flat ($\alpha$\,$\sim$\,0) spectrum up to some break frequency, mostly between $\sim$10 and $\sim$100 GHz, above which the spectrum steepens to a typical index $\alpha$\,$\sim$\,$-0.7$. The second component, associated with an evolving shock, produces a self-absorbed synchrotron spectrum with a turnover frequency
$\sim$\,100~GHz. Below this turnover frequency, the spectrum is strongly inverted. As the shock evolves, the turnover frequency and the flux density of the shock component changes, and in some phases a detectable contribution of this component may not be present at all.

\begin{figure*}
   \centering
\includegraphics[scale=0.5]{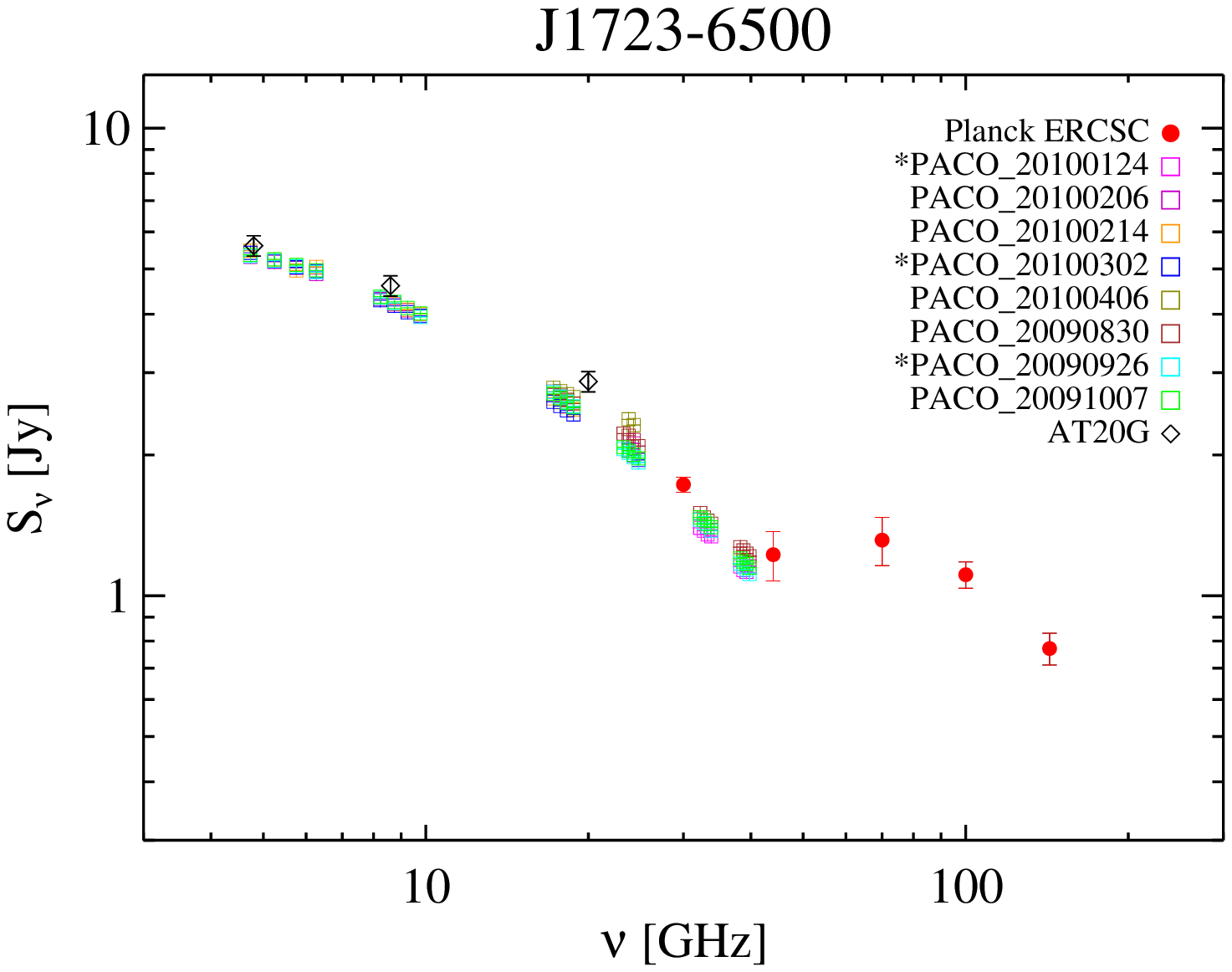}
\includegraphics[scale=0.5]{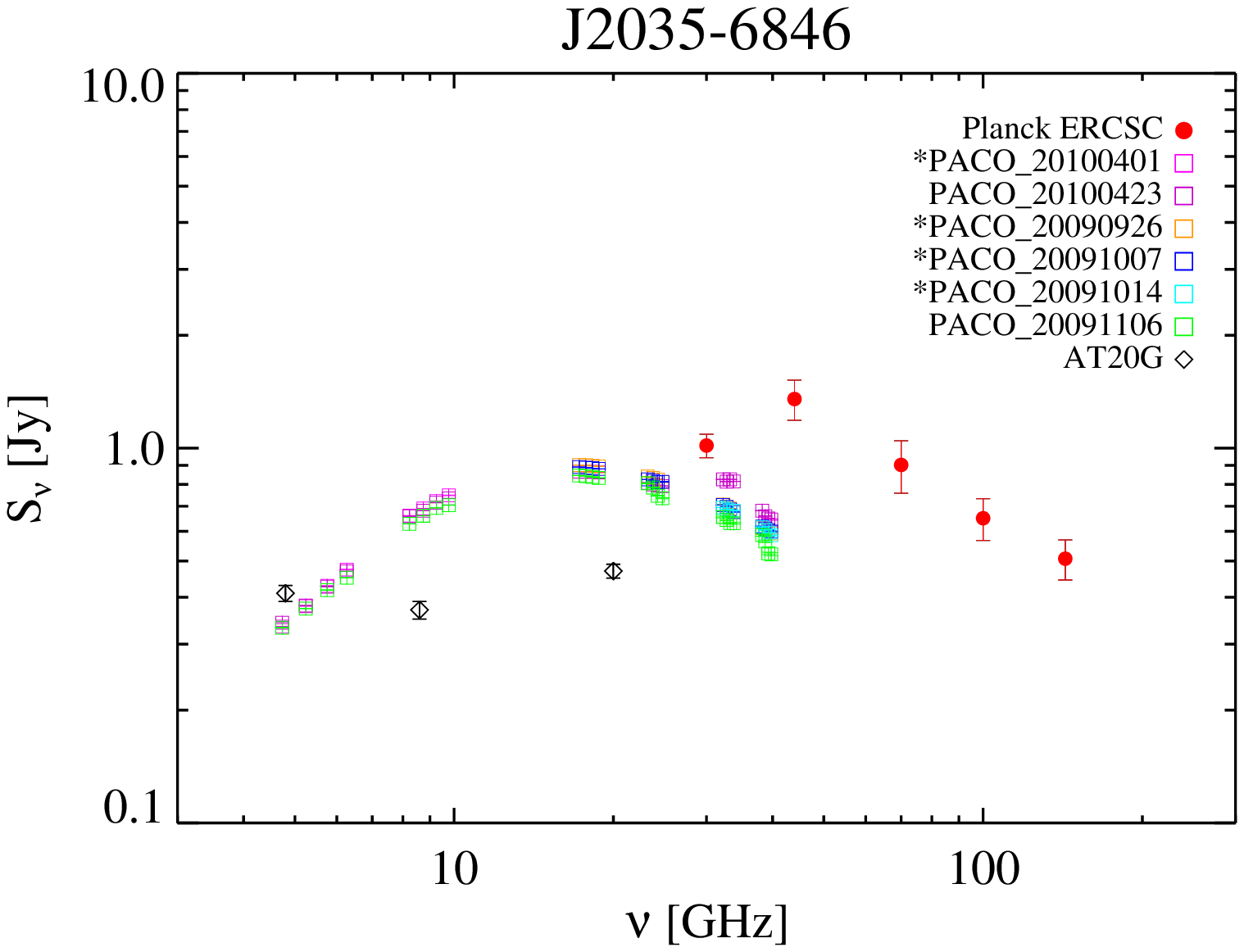}
   \caption{Two examples of clearly multi-component spectra. \Planck measurements are in filled circles; AT20G data are in open diamonds; PACO data, with their dense sampling in frequency space, are shown in open squares. Asterisks identify epochs within 10 days from \Planck observations. The multi-valued spectra in J2035--6846 are a clear sign that this source is variable on time scales less than two weeks (see text).}
              \label{fig_bumpy}
    \end{figure*}

An example of this class is J2253+1609, better known as 3C454.3. During an outburst in 2005/2006, the source showed a strong spectral peak around 300~GHz, evolving to lower frequencies over several months, while observations made in 2004 did not show this feature (see \citealt{Villata2007}, \citealt{Raiteri2008}, \citealt{Villata2009}, \citealt{rachen2010} and references therein). Incidentally, this source showed a similar outburst when it was first observed with \Planck \citep{Planck2011-6.3a}, which is why it is included in Table  \ref{table_candi_gps}.  Two other examples where Planck observations play a crucial role are shown in Figure~\ref{fig_bumpy}.  The sources shown in Figure~\ref{fig_hfp}, all identified with known blazars, present a very similar spectral shape. We therefore expect that their high-frequency-peaked spectrum is a temporary feature.

We emphasise that \Planck\!\!, which is expected to perform at least four full sky surveys, is a powerful tool to distinguish variable from permanent spectral features. By comparing catalogues derived from individual surveys made by \Planck\!\!, we will be able to resolve the questions raised here in the full frequency regime accessible to \Planck\!\!, even without reference to external data. Results from this research will be presented in future \Planck papers. 

\subsubsection{Artifacts in the ERCSC spectra of variable sources}\label{sec:artifacts}

\begin{figure*}
   \centering
\includegraphics[scale=0.5]{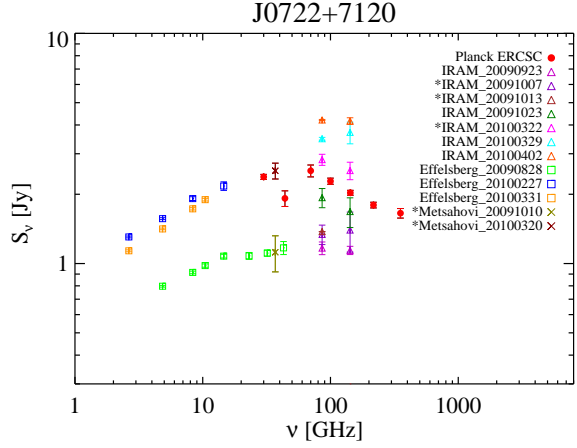}
   \caption{Based on ground-based and \Planck data, the
apparently anomalous 44 GHz point in the spectrum of the strongly variable
source J0722+7120 is shown to be due to the averaging of flux densities
described in \S\,\ref{sec:artifacts}. Asterisks identify epochs within 10 days from \Planck observations at any of the LFI
channels.}
              \label{fig_0716}
    \end{figure*}

In this subsection, we discuss some potential problems in interpreting ERCSC spectra for variable sources. As noted in \S\,\ref{sec:introduction}, 44\,GHz observations of a given source take place at two times separated by 7--10 days, for each scan. If the source is strongly variable on a time scale of order days, the 44\,GHz flux density contained in the ERCSC could be an awkward average. An example is the well-known, rapidly variable sources J0722+7120 \citep{Ostorero2006}, shown in Figure \ref{fig_0716}. Supporting observations, particularly the IRAM data at 86 and 143 GHz, show it to have varied on short time scales during the \Planck survey. The apparently anomalous 44\,GHz \Planck measurement is an average of the flux density in the source's high and low luminosity state. We show this case as a warning: spectral anomalies seen {\it only\/} at 44 GHz may be due to this effect.  

\begin{figure*}
   \centering
\includegraphics[scale=0.5]{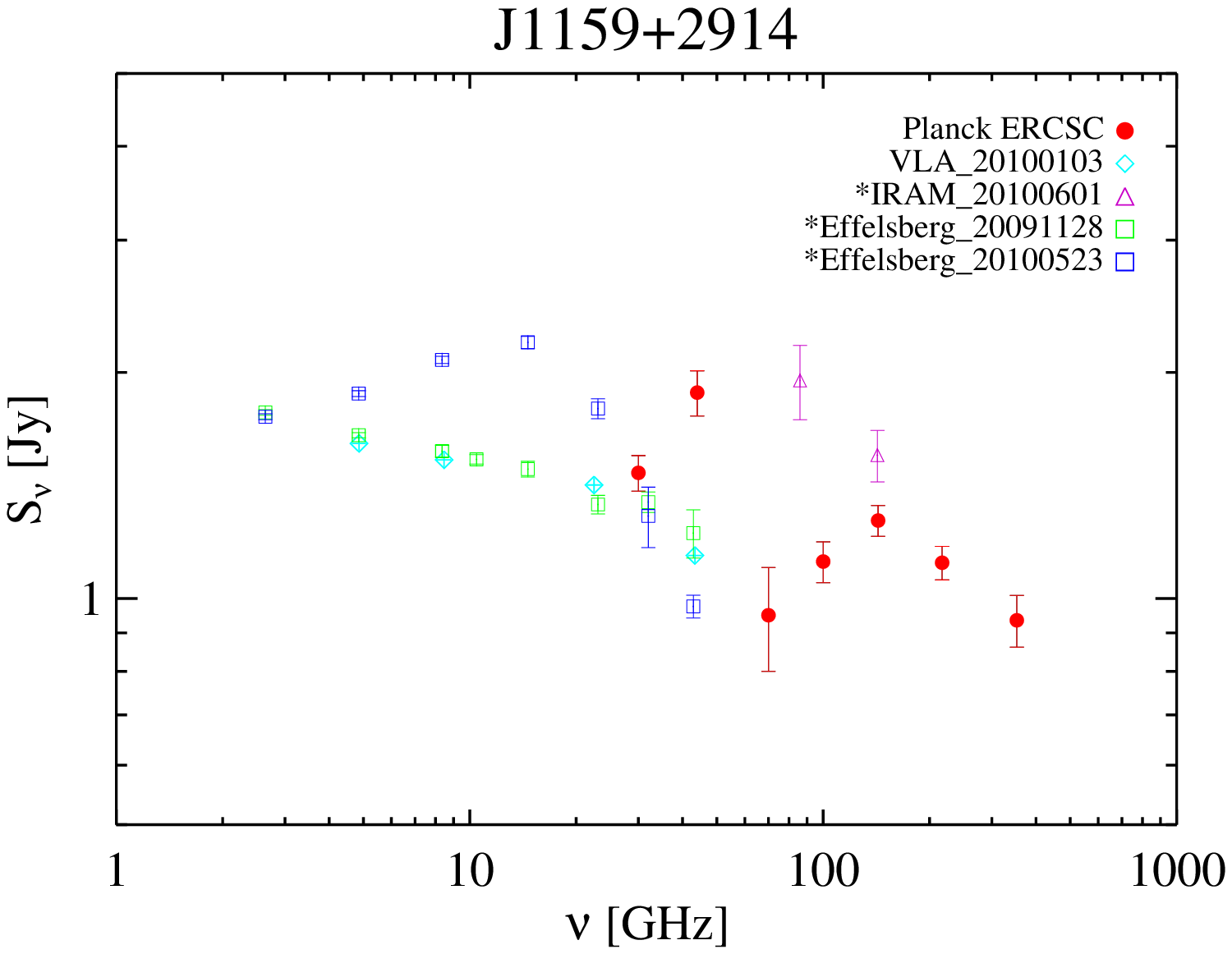}
   \caption{The zig-zag shape spectrum of the strongly variable
source J1159+2914. Asterisks identify epochs within 10 days from \Planck observations at any of the LFI
channels.}
              \label{fig_TON599}
    \end{figure*}

In sources with still faster variability, the effect of non-simultanity of the \Planck observations can produce even more bizarre effects. The ERCSC spectrum of the source J1159+2914, also known as the IDV blazar TON599, shows a strong zig-zag shape, dropping by a factor of two between 44 and 70 GHz, followed by another small bump (see Figure \ref{fig_TON599}). This source is known to show very fast variability, and the comparison with Effelsberg data, and IRAM 86\,GHz and 143\,GHz data taken around the time of the \Planck scans, suggests that it had a strong flare at a peak frequency of $\sim$50~GHz in the first days of June 2010. This flare must have started after 23~May (when the Effelsberg observations were made), and probably declined again on 7~June when the \Planck 100\,GHz point was taken. The ERCSC spectrum of the source is a superposition of a quite low state in December 2009, and the high, flaring state in June 2010, except for 30 and 70\,GHz measurements, which were made by \Planck for the second time shortly after the last day of data used for ERCSC, and are therefore not included in the average.

These examples make clear that a fair amount of care has to be taken when
intepreting \Planck ERCSC spectra for variable sources.

\subsection{Sources showing evidence of a spectral upturn at high frequency}

\begin{figure*}
   \centering
 \includegraphics[scale=0.35]{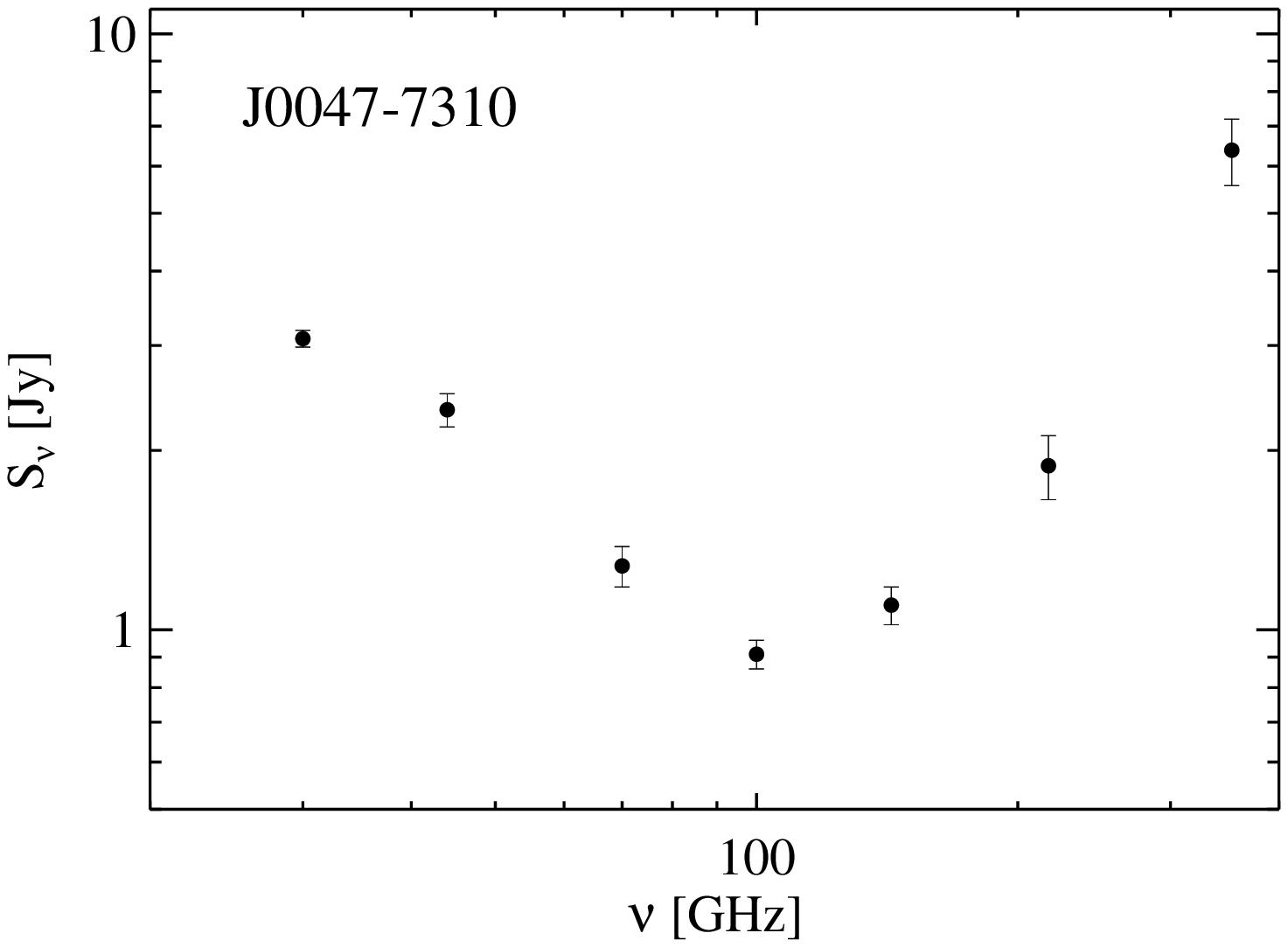}
  \includegraphics[scale=0.35]{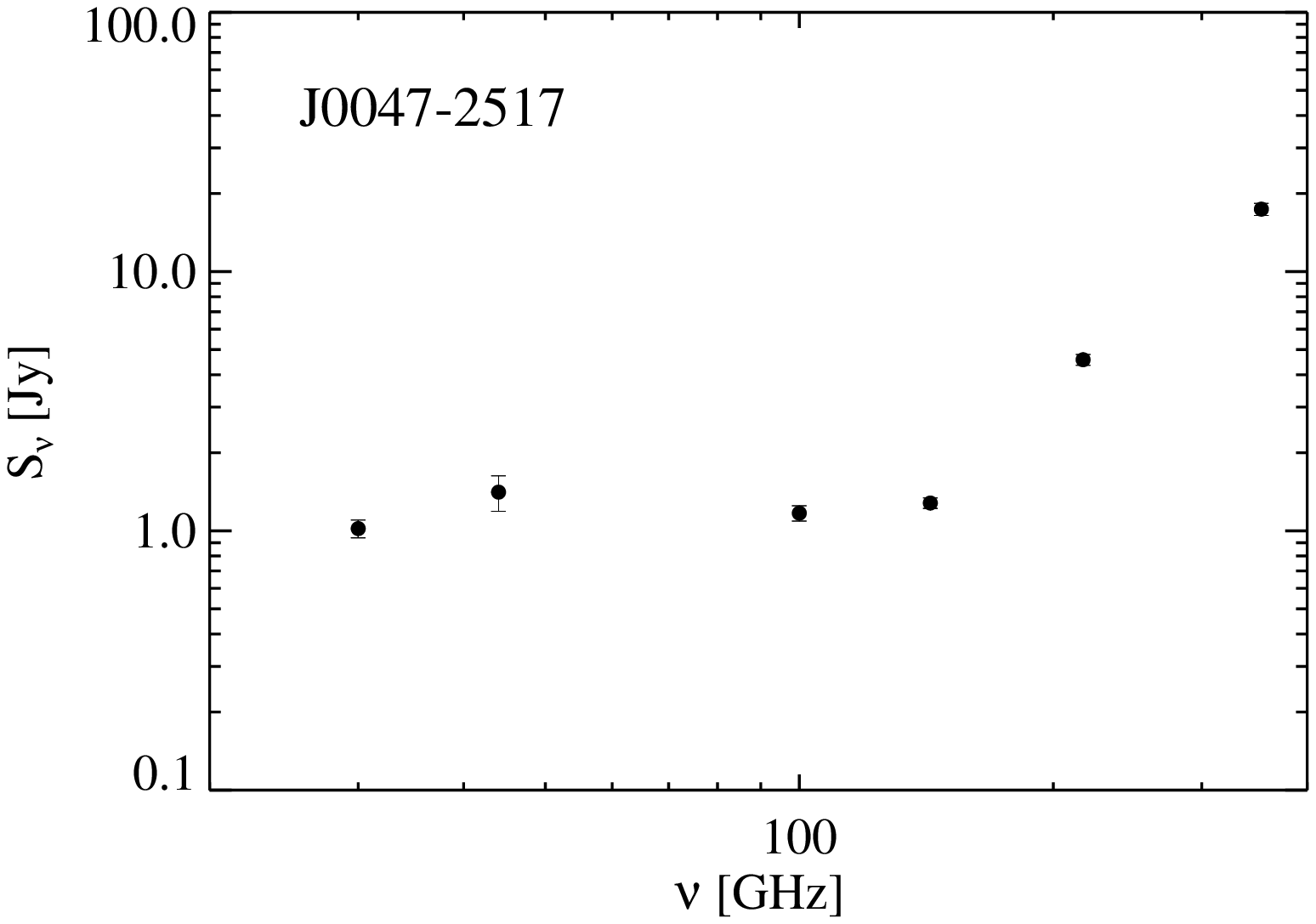}
 \includegraphics[scale=0.35]{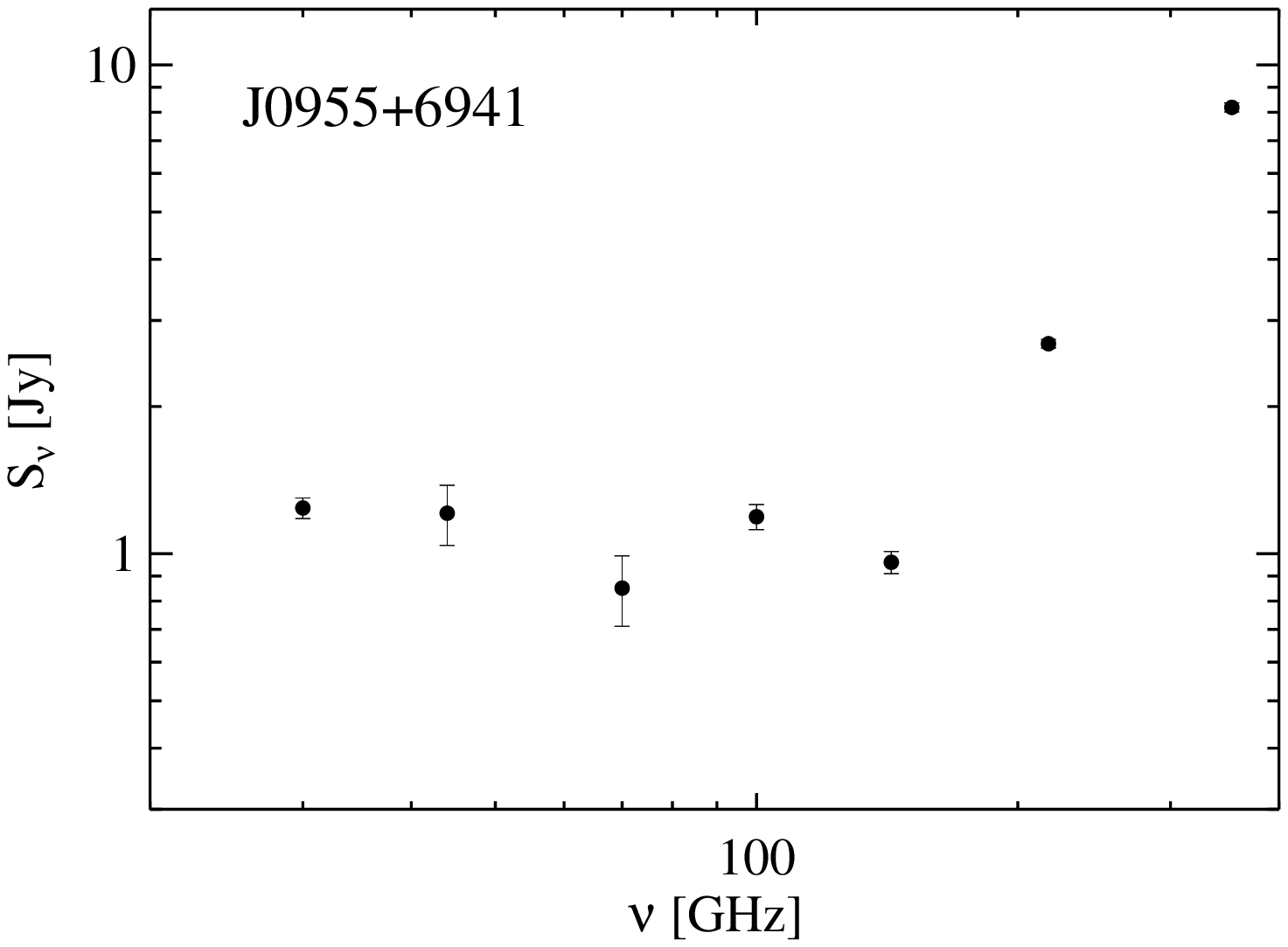}
   \caption{\Planck sources with an upturn in the spectrum. {\it Left:} J0047--7310 in the SMC. {\it Center:} \Planck spectrum of a familiar star-forming system, NGC253.  {\it Right:} The starburst galaxy M82.}
              \label{fig_upturn}
    \end{figure*}

In many radio sources (Arp 220 as an example), synchrotron emission is dominated at high frequencies by re-emission from warm dust.  As already noted, essentially no extragalactic \Planck sources show this pattern at frequencies below 143~GHz. Of all the sources at $|\,b\,|$\,$>$\,30$\degr$ that have \Planck measurements at 30 to 217 GHz, only eleven show evidence of a significant increase of flux density even from 143 GHz to 217 GHz; one of these is the nearby and well-studied star-forming galaxy NGC253 (Figure \ref{fig_upturn}). Of the remaining 10 sources, 7 are HII regions located in the Magellanic clouds, like J0047--7310 and J0048--7306; two are Galactic sources; and the remaining one is M82, a well-known starburst galaxy. The lack of upturn-spectrum sources in the ERCSC suggests that most of the sources detected by \Planck are flat spectrum sources like blazars, with high enough synchrotron luminosity to swamp dust reemission even up to 217~GHz.

We also note here the potential danger of apparently upturning spectra being produced by
source confusion. The low probability of source confusion discussed in \S\,\ref{sec:identification}
is valid only for the low frequency catalogues; at 545 and 857~GHz, in particular, the ERCSC
catalogues are much richer, and these frequencies are in the Rayleigh-Jeans part of
the spectra for most IR sources.  The apparently upturning spectra may result from confusion of a
radio source disappearing in the background with an IR source appearing within the large beam of \Planck\!\!. Careful checks with infrared catalogues are required to verify any case of a potentially
upturning spectrum.  Hence we restrict our attention in this paper to frequencies of 30--353~GHz, and generally to 30--217~GHz.

\subsection{ERCSC LFI sources with no plausible match in existing radio catalogues}\label{sec:unmatched}

Finally, we looked for sources with such extreme spectra that they have no counterparts in existing low-frequency, large area, radio surveys. This serves the dual purpose of looking for unusual or extreme sources and validating the ERCSC. We looked for  matches in the following catalogues: the WMAP 7 year catalogue \citep{gold2010}, the NEWPS catalogue \citep{massardi2009}, the 20\,GHz AT20G survey in the southern hemisphere \citep{murphy2010}, the CRATES catalogue in the northern hemisphere \citep{healey2007}, and the GB6 catalogue \citep{condon1994}.  To reduce the number of spurious identifications of the bright \Planck sources, a flux density cut of 0.3~Jy was applied to both the AT20G and GB6 catalogues. In the northern hemisphere, where the high frequency coverage is incomplete, we also looked at the 1.4\,GHz NVSS catalogue \citep{condon1998} with a 0.5 Jy flux density cut. We cross-correlated these catalogues with the ERCSC at 30~GHz, 44~GHz and 70~GHz using a search radius equal to 0.5~FWHM at each \Planck frequency channel. We have shown in \S\,\ref{sec:validation} that this search radius is unlikely to yield spurious matches. We also experimented with different flux cuts, but found that our adopted cuts are reasonable compromises between matching most sources and avoiding spurious matches. Significantly lower flux cuts result in spurious matches because the chance of random association given the large LFI beams goes up.  The unmatched sources listed in Table~\ref{table_unmatched} for example often have NVSS sources within the \Planck LFI beams, but the 1.4\,GHz flux densities of these NVSS sources are on the order of a few mJy; thus they are clearly unlikely to be associated with the ERCSC sources, whose flux denisities are on the order of 1 Jy or above.  

With these automatic matching steps, we found that among high Galactic latitude ($|\,b\,|$\,$>$\,5$\degr$) sources, 12 ERCSC sources at 30 GHz, 13 at 44 GHz and 26 at 70 GHz do not match with any known source. These represent respectively 2\,\%, 5\,\% and 8\,\% of the high Galactic latitude sources in each of these catalogues. However, there is a caveat in such automatic matching procedure. While the above matching radius is reasonable assuming that \Planck dominates the positional uncertainty, this is not necessarily the case, for example, when matching against the WMAP sources. The comparatively smaller beam sizes at 44 and 70 GHz of \Planck suggest that this procedure could miss genuine WMAP matches. 

\begin{figure*}
\centering
   \begin{tabular}{c}\vspace{0.08 in}
\includegraphics[scale=0.55]{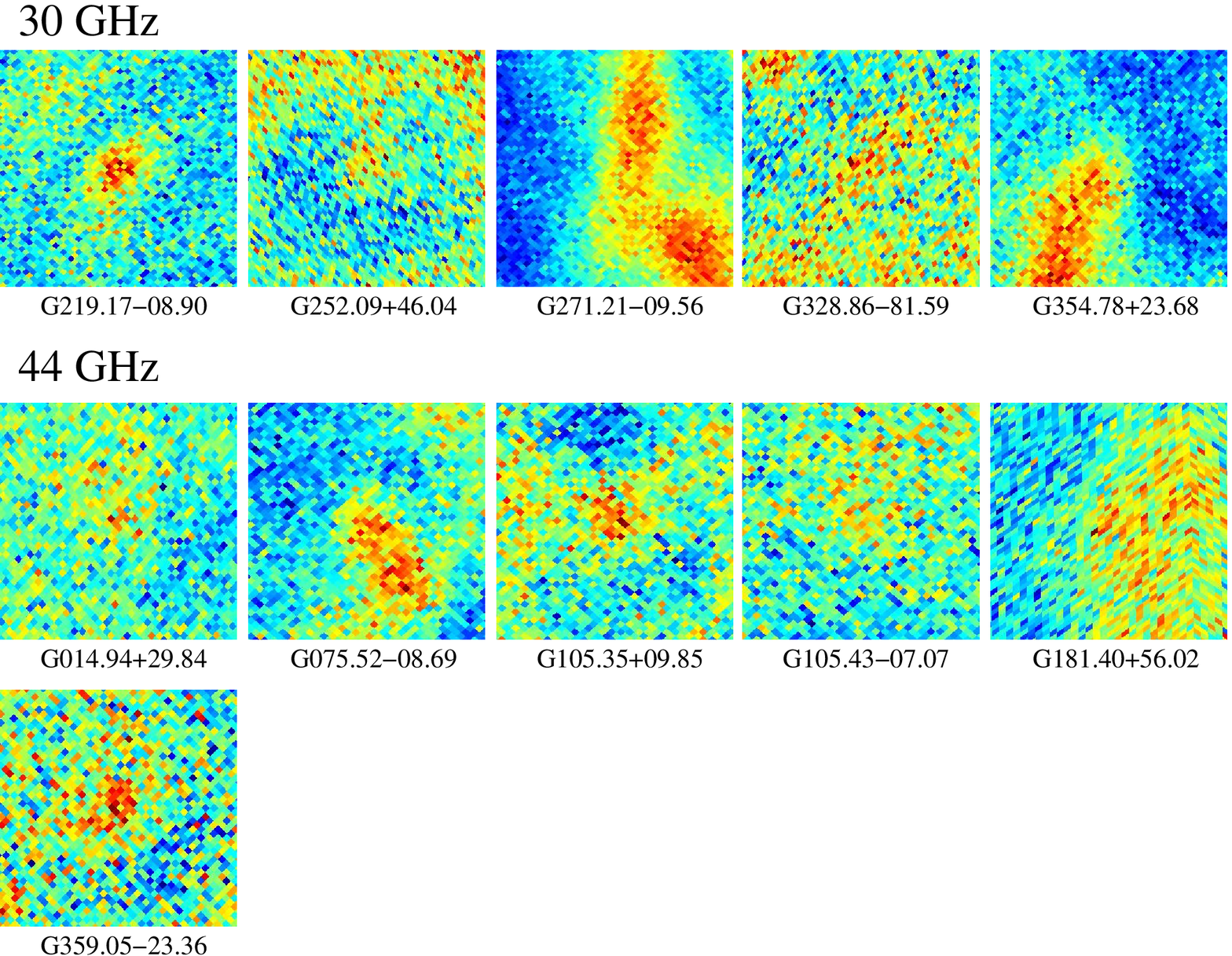} \\
\includegraphics[scale=0.55]{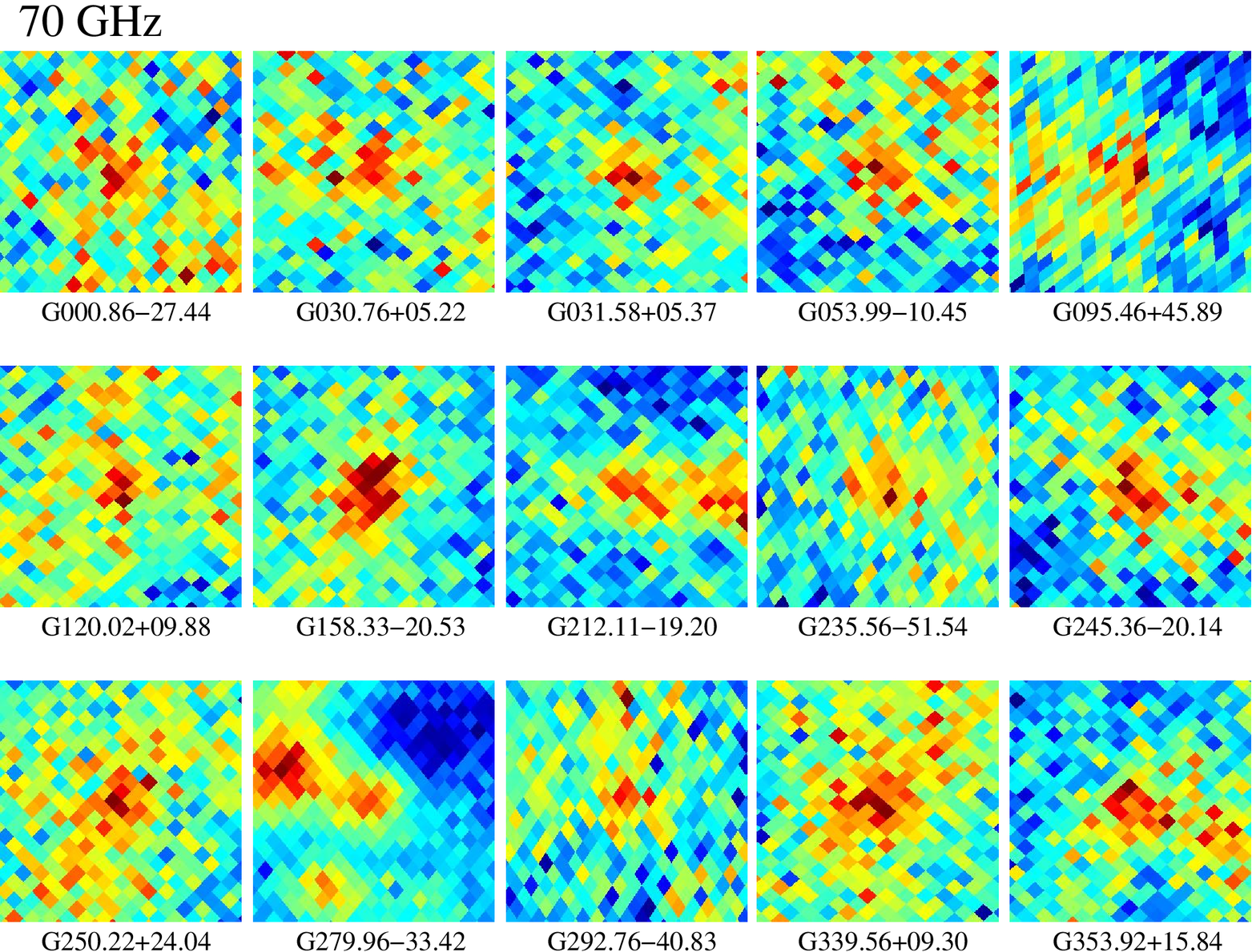}
  \end{tabular}
\caption{ERCSC sources with no plausible identifications in low frequency radio catalogues. } 
             \label{fig_unmatched}%
 \end{figure*}

Since sources that exist in only one \Planck channel are more likely spurious, we first looked for unmatched sources that exist in more than one \Planck band, and found a number of such cases among the 44 and 70\,GHz ``unmatched'' sources. We then looked in NED for each of these unmatched sources to see if they have counterparts in other radio catalogues that are not among those listed above. In particular, extended sources may appear in older, lower resolution surveys such as PMN \citep{griffith1993}, but not in newer, higher resolution surveys such as AT20G. Several such cases were found. We further associated a number of unmatched sources with Galactic objects such as PNe and SNR. After these checks, the number of  residual, apparently unmatched sources is 5 at 30~GHz, 6 at 44~GHz, and 15 at 70~GHz. These are listed in Table~\ref{table_unmatched}. Judging from the Galactic latitude, 2 of the 30\,GHz,  3 of the 44\,GHz and 4 of the 70\,GHz sources are likely associated with the Galaxy while one of the 70~GHz sources is likely in the LMC; these are flagged in Table~\ref{table_unmatched}.  The remaining sources are either exciting ``new'' sources or else spurious. To distinguish between these possibilities, we looked at the postage stamp images for all the sources that remain unidentified (see Figure~\ref{fig_unmatched}). In all cases, the postage stamps are 5~FWHM on a side.   A number of the sources at 44 and 70~GHz appear to be of low significance and hence could be spurious. Some are flagged in the ERCSC as possibly contaminated by CMB signals.  Others are at low Galactic latitude.  One apparent exception at 70 GHz is G158.33--20.53 (but see comments below). At 30~GHz, some sources are of high significance but appear extended, which, given the $\sim$30$'$ \Planck beam size, means they are likely to be either Milky Way objects or associated with very nearby galaxies (or could be CMB artifacts). An exception is G219.17--08.90 ($S_{\rm 30~GHz}$\,=\,1\,Jy) which appears to be real and point-like.  From NED, we find that there is a $S_{\rm 60~\mu m}$\,=\,68\,Jy IRAS source (IRAS06282--0935) near this position that seems to have been identified as IR cirrus \citep{strauss1992}. The most convincing 70 GHz source, G158.33--20.53, also has a counterpart, IRAS03259+3105, in NED, which again is identified as cirrus by \citet{strauss1992}.

We thus conclude that we do not identify any genuinely new population of sources among the ERCSC catalogues at the LFI frequencies. While confirming a match is generally much easier than claiming a source is ``unmatched", we can say that after the automated procedure described above, $>$\,90\,\% of the LFI sources had a reasonable match in an external radio catalogue. After cross-matching between the three LFI bands and searching in NED for plausible identifications, especially important in the case of extended radio sources, we are left with a total of 26 potentially ``unidentified'' LFI sources,  almost all of which can probably be explained by artifacts or extended Galactic structures.

\begin{table}
\caption{ERCSC sources at $|\,b\,|$\,$>$\,5\degr\ with no obvious match in external radio catalogues}             
\label{table_unmatched}     
\centering                          
\begin{tabular}{l l}       
\hline\hline                 
ERCSC Name & Notes${}^*$ \\  
 \hline
PLCKERC030 G219.17--08.90 & G \\
PLCKERC030 G252.09+46.04  & \\
PLCKERC030 G271.21--09.56   & G \\
PLCKERC030 G328.86--81.59  & \\
PLCKERC030 G354.78+23.68  & G \\
\hline
PLCKERC044 G014.94+29.84 & \\
PLCKERC044 G075.52--08.69 & G \\
PLCKERC044 G105.35+09.85 & G \\
PLCKERC044 G105.43--07.07  & G \\
PLCKERC044 G181.40+56.02  & \\
PLCKERC044 G359.05--23.36  & \\
\hline
PLCKERC070 G000.86--27.44    & G? \\
PLCKERC070 G030.76+05.22  & G \\
PLCKERC070 G031.58+05.37  & G? \\
PLCKERC070 G053.99--10.45   & G \\
PLCKERC070 G095.46+45.89  & \\ 
PLCKERC070 G120.02+09.88 & G \\
PLCKERC070 G158.33--20.53  & G? \\
PLCKERC070 G212.11--19.20  & \\
PLCKERC070 G235.56--51.54  & \\
PLCKERC070 G245.36--20.14  & \\
PLCKERC070 G250.22+24.04  & \\
PLCKERC070 G279.96--33.42 & L? \\
PLCKERC070 G292.76--40.83  & \\
PLCKERC070 G339.56+09.30 & G \\
PLCKERC070 G353.92+15.84  & G? \\
\hline                                  
\end{tabular}

\medskip
${}^*$ Here G stands for likely Galactic source. L stands for likely LMC source.

\end{table}

\section{Conclusions}

	We summarise in this section the primary conclusions to be drawn from \Planck\!\!'s study of extragalactic radio sources. We stick closely to the observational results, and provide comments on the fit between these observational results and current theories of the physics of extragalactic radio sources.

\subsection{Overall properties}

	\Planck has demonstrated that the high frequency counts (at least for frequencies $\leq$\,143 GHz) of extragalactic sources are dominated at the bright end by synchrotron emitters, not dusty galaxies. This finding is in agreement with conclusions reached by the South Pole Telescope (SPT) and Atacama Cosmology Telescope (ACT) teams (\citealt{vieira2010, marriage2010}, respectively) from the properties of sources at much lower flux densities. An inference from this result is that the cores of extragalactic sources, and not extended structure, dominate the high frequency emission, as suggested by earlier work at frequencies below most of the \Planck bands (see recent discussions in \citealt{lin2009} and \citealt{murphy2010}). The conclusion that the core dominates the high frequency emission is supported by the close agreement between \Planck and VLA flux densities (\S\,\ref{sec:validation}), implying that the emission is found within the VLA beam of order arcseconds in size.

	The emerging dominance of emission from a flat spectrum core implies that extrapolation of flux densities and/or counts of sources to frequencies $\geq$\,30\,GHz cannot be reliably made from low frequency catalogues, where the emission is dominated by lobes. This has been recognised for some time (see \citet{lin2009} and \citet{murphy2010}); \Planck strongly confirms this conclusion. If we look at the SEDs of the sources, including all components, the spectra grow flatter as frequency increases and the steep-spectrum lobes fade away. On the other hand, \Planck observations allow us to follow the SEDs of a statistically significant sample of sources to much higher frequencies than can generally be employed in ground-based observations. As noted in \S\,\ref{sec:validation} (and described in more detail in \citealt{Planck2011-6.1}), \Planck provides clear evidence of a spectral steepening in the radio and millimeter wave emission from extragalactic cores; the steepening sets in at frequencies above $\sim$44--70\,GHz.  The observed spectral steepening in turn means that radio sources contribute less foreground noise to increasingly sensitive searches for small angular scale anisotropies in the CMB than earlier models had suggested (e.g., \citealt{dezotti2005}). This finding, too, is consonant with results from SPT and ACT.

\subsection{Properties of individual sources}

\Planck allows us for the first time to investigate sub-mm spectra beyond 200~GHz, which is the limit of most ground-based monitoring programs. In this regime, the spectra of radio sources are usually expected to be optically thin synchrotron emission, and our results largely confirm that this part of the spectrum is represented by a single power law. The vast majority of extragalactic sources in the ERCSC lists at 30--100 GHz are flat spectrum radio sources of the sort discussed in detail in \citet{Planck2011-6.3a}, with a scattering of bright steep spectrum sources strong enough to register in the lowest frequency bands of \Planck\!\!. A small fraction ($\lsim$\,10\,\%; Table~\ref{table_candi_gps}) show peaked or convex spectra, but even these sources are mostly identified as known blazars. As blazars are known to be variable, care has to be taken in the interpretation of their spectra. For most sources, ERCSC flux densities represent the average over two scans, and even during one scan, spectral artifacts may occur due to the fact that it requires 7--10 days for the \Planck focal plane to cross a point source. 

In contrast, most known examples of compact, newborn radio galaxies (CSOs), which were originally thought to be the dominant class of sources with Gigahertz-peaked spectra, are mostly too faint to be detected by {\it Planck}.  A new population of bright, very compact, high-peaked CSOs has not been found. Likewise, very few of the extragalactic radio sources found by \Planck show evidence of the sharp spectral upturn expected from dust reemission at high frequencies. Only NGC253 shows a clear upturn at a frequency $\leq$\,143 GHz; sources that show dust reemission dominating at 217\,GHz are in many cases Galactic sources, in the Magellanic clouds, or nearby known star-forming galaxies. 

Although we have not investigated in detail all the sources with no obvious matches in lower frequency radio catalogues (\S\,\ref{sec:unmatched}), let alone every extragalactic source in the ERCSC, we find no convincing evidence for the emergence of a new and unexpected population of sources.  The 26 ERCSC sources with no match in radio catalogues appear to be a heterogeneous mixture of conventional radio sources, many of them Galactic.  
	
\subsection{Future observations and analysis}

Two sets of future observations will help clarify the status of some of the
sources listed in \S\,\ref{sec:peaked} and \S\,\ref{M-C-spec}. Careful monitoring and/or VLBI observations of
sources with peaked spectra would determine whether they are highly
compact radio galaxies (CSOs), compact jet objects with stable GPS type
spectra, or are instead merely flaring sources with temporarily convex spectra. In fact,
our results seem to suggest most extreme spectral features seen in \Planck
sources may be associated with flares. With at least four full sky
surveys, \Planck is a powerful tool to distinguish these possibilities, by deriving catalogues from individual, six-month, sky surveys and comparing flux densities of all bright point sources at six-month intervals. Further ground-based observations, now underway at
frequencies below and overlapping the \Planck frequency bands, will support this
effort.

Likewise, for the non-variable extreme radio sources like CSOs, the addition of
more surveys to the \Planck maps will allow us to extract deeper catalogues, with the
potential to find some more of these usually faint objects. 

In the case of Galactic sources, the role of CO emission lines is important. The situation for extragalactic sources is more complicated, because the CO lines redshift in and out of the \Planck bands. We have underway a study to look for the influence of CO emission in the ERCSC spectra. We expect the effect to be small, since most of the ERCSC sources are extremely bright, non-thermal emitters, with strong continua. There is no obvious evidence in the SEDs we have examined for the presence of CO emission. In this regard, we again warn readers to be careful in interpreting anomalous 44\,GHz observations; a spectral bump at 44 GHz is not necessarily evidence for redshifted CO.

We used ground-based observation data for both validation purposes, specifically cross-calibration between ground-based instruments and \Planck\!\!, and in the study of the spectral and variability properties of extreme radio sources. Most of these observation programs, like the \Planck mission itself, are still ongoing.


\begin{acknowledgements}
A description of the Planck Collaboration and a list of its members can be found at \url{http://www.rssd.esa.int/index.php?project}\\\url{=PLANCK&page=Planck_Collaboration}. This paper makes use of observations obtained at the Very Large Array (VLA) which is an instrument of the National Radio Astronomy Observatory (NRAO). The NRAO is a facility of the National Science Foundation operated under cooperative agreement by Associated Universities, Inc. This research also makes use of observations with the 100\,m telescope of the Max Planck Institut f\"ur Radioastronomie (MPIfR), the 30\,m telescope of Institut de Radioastronomie Millim{\'e}trique (IRAM), the Australia Telescope Compact Array (ATCA) and the 13.7\,m telescope of the Mets\"ahovi Radio Observatory. The Mets\"ahovi observing project is supported by the Academy of Finland (grant numbers 212656, 210338 and 121148). We acknowledge the use of the NASA/IPAC Extragalactic Database (NED) which is operated by the Jet Propulsion Laboratory, California Institute of Technology, under contract with the National Aeronautics and Space Administration. The Planck Collaboration acknowledges the support of: ESA; CNES and CNRS/INSU-IN2P3-INP (France); ASI, CNR and INAF (Italy); NASA and DoE (USA); STFC and UKSA (UK); CSIC, MICINN and JA (Spain); Tekes, AoF and CSC (Finland); DLR and MPG (Germany); CSA (Canada); DTU Space (Denmark); SER/SSO (Switzerland); RCN (Norway); SFI (Ireland); FCT/MCTES (Portugal); and DEISA (EU).
\end{acknowledgements}

\bibliographystyle{aa}

\bibliography{Planck_bib,ers_refs}

\listofobjects

\raggedright
\end{document}